\documentclass[PAPER, atlasdraft=false, UKenglish, cernpreprint, texmf, orcidlogo]{atlasdoc}

\usepackage{todonotes}
\DeclareBoolOption[true]{LANGEDIT}
\LEcontact{Chip Brock brockr@msu.edu}
\newcommand*{\LEinline}[1]%
{\todo[inline, color=yellow!30, size=\footnotesize]{#1}}

\usepackage{atlaspackage}
\usepackage{atlasbiblatex}

\usepackage{atlasphysics}

\usepackage{multirow}
\usepackage{graphicx}
\newcolumntype{~}{>{\global\let\currentrowstyle\relax}}
\newcolumntype{^}{>{\currentrowstyle}}
\newcommand{\rowstyle}[1]{\gdef\currentrowstyle{#1}%
#1\ignorespaces
}

\graphicspath{{logos/}{figures/}}

\usepackage{ANA-TOPQ-2023-02-PAPER-defs}

\AtlasTitle{Measurement of high-mass $t\bar{t}\ell^{+}\ell^{-}$ production and lepton flavour universality-inspired effective field theory interpretations at $\sqrt{s}=13$~\TeV with the ATLAS detector}

\AtlasAbstract{%
Measurements of $t\bar{t}\ell^{+}\ell^{-}$ production in the region of high dilepton invariant mass with effective field theory (EFT) interpretations are presented. They are performed using final states with three isolated leptons (electrons or muons) and are based on $\sqrt{s} = 13$~\TeV proton--proton collision data with an integrated luminosity of $140\,\mathrm{fb}^{-1}$, recorded from $2015$ to $2018$ with the ATLAS detector at the Large Hadron Collider. Measurements of the \ttll signal strength and cross-section upper-limits are performed inclusively in lepton flavour and separately for electrons and muons. The study also aims to probe anomalous four-fermion interactions including to test for possible lepton flavor universality violation. No significant deviations from the Standard Model predictions are observed and the measurements are interpreted through the EFT formalism to provide new constraints on the relevant operators.
}

\AtlasRefCode{TOPQ-2023-02}

\PreprintIdNumber{CERN-EP-2025-053}

\AtlasJournalRef{\EPJC 85 (2025) 1434}
\AtlasDOI{10.1140/epjc/s10052-025-14695-9}

\hypersetup{pdftitle={ATLAS document},pdfauthor={The ATLAS Collaboration}}

\begin{document}

\maketitle

\tableofcontents

\section{Introduction}
\label{sec:intro}

The top quark as the heaviest fundamental particle has a unique role
in testing the Standard Model (SM) and electroweak (EW) symmetry breaking.
The study of rare processes involving the production of top quarks and
heavy bosons is interesting in its own right as a test of the SM electroweak sector at high energy.
This also offers a window on to couplings between top quarks and bosons that cannot be
directly tested elsewhere in particle collisions. The production of a
top--antitop quark pair in association with a $Z$ boson, \ttZ, is an example of such a process. One of
the cleanest signatures of this process is where the $Z$ boson
decays leptonically, $\ttZ \to \ttbar \ell^+\ell^-$. Measurements of this process, where the $Z$ boson is on-shell, $O(10)$~\GeV from the $Z$ mass, have recently been performed at the Large Hadron Collider (LHC)~\cite{Evans:2008zzb} by ATLAS and CMS~\cite{TOPQ-2020-20,CMS-TOP-21-001,CMS-TOP-23-004} inclusively and at the differential level. This process gives direct sensitivity to the $t$--$Z$ coupling, which is important as deviations from the SM could
be a sign of new physics effects in the EW symmetry breaking
mechanism.  Several beyond the SM (BSM) models predict such
deviations~\cite{Aguilar-Saavedra:2009xmz,Aguilar-Saavedra:2013qpa,Perelstein:2005ka,Chivukula:1992ap,Chivukula:1994mn,PhysRevD.52.5374,PhysRevD.55.5848,PhysRevD.56.402}.%

\begin{figure}[t!]
\centering
\subfloat[]{\includegraphics[width=0.45\linewidth]{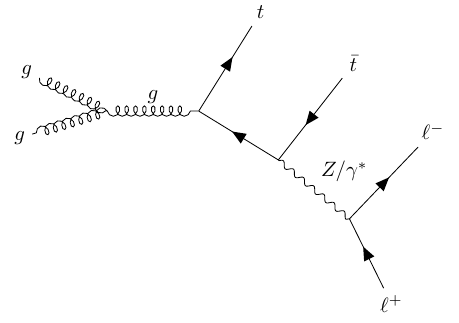}}
\subfloat[]{\includegraphics[width=0.45\linewidth]{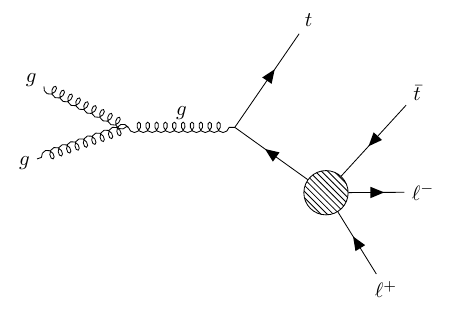}}
\caption{Representative Feynman diagrams for \ttll (a) in the SM and (b) with an effective vertex.}
\label{fig:Intro:LFU_feyn}
\end{figure}

Off-shell \ttll production, where the opposite-sign
same-flavour lepton pair originates from the $Z/\gamma^*$ decay, has
not yet been measured by ATLAS. This process is also highly
interesting as it gives sensitivity to the effective \ttll coupling. It is
possible to more generically probe possible BSM contributions to these
kinds of couplings in the context of effective field theories
(EFTs)~\cite{EFT:Bylund,Afik:2021xmi}. Figure~\ref{fig:Intro:LFU_feyn}
shows Feynman diagrams for \ttll production in the SM and
a possible effective vertex that affects the coupling of
interest. The \ttll type of coupling is poorly constrained in
existing measurements probing EFT couplings relating to the top
quark~\cite{CMS-TOP-22-006}. High-mass \ttll events are particularly
sensitive to such couplings, the impact of the relevant EFT operators
strongly increasing with energy~\cite{Afik:2021xmi}, motivating a dedicated
measurement.

Furthermore, some of the most intriguing evidence for new physics has
come from anomalies observed in lepton flavour universality (LFU)
tests in $B$-hadron decays into a $D^*$ and lepton-neutrino pair~\cite{BaBar:2012obs,BaBar:2013mob,Belle:2015qfa,Belle:2016dyj,Belle:2017ilt,Belle:2019rba,LHCb:2023zxo,LHCb:2023uiv}. These
measurements have highlighted the importance of probing LFU in
general, and there is a commonality in couplings which means anomalies
in the $B$-sector could also be manifest in the top quark
sector~\cite{Bissmann:2020mfi,Fuentes-Martin:2020lea,Grunwald:2023nli,Bartocci:2023nvp,Grunwald:2023nli,Allwicher:2023shc,Atkinson:2024hqp}.

Therefore, the target of this analysis is to make dedicated
measurements of \ttll, $(\ell=e,\mu)$, with a focus on the high dilepton invariant
mass region. This is achieved by selecting events with three isolated
leptons with additional requirements on jet activity to provide a
region pure in \ttll production. The three-lepton selection provides a balance between high purity, worsened with a two-lepton selection, and reasonable data statistics, worsened by a four-lepton selection. Many aspects of the analysis are
built upon the previous ATLAS measurement of \ttZ
production~\cite{TOPQ-2020-20} with several extensions to deal with
the relatively larger background contributions away from the $Z$ boson
mass peak. Measurements of the \ttll signal strength, $\mu(\ttll) = \sigma(\textrm{obs})/\sigma(\textrm{SM})$, and cross-section upper limits are
made inclusively in lepton flavour and separately for electrons
and muons.  These measurements are used to search for new physics
entering via four-fermion top quark--lepton EFT operators in the
SMEFT~\cite{Brivio:2020onw} formalism, with the \texttt{top} flavour assumptions. The relevant operators,
considering only those of dimension-6, are shown in
Table~\ref{tab:smeft-operators} and the effect of these operators is
parameterised as a function of their corresponding Wilson Coefficients
(WCs) assuming a mass scale, $\Lambda = 1\ \tev$, that characterizes the beyond the SM dynamics. The operators, and WCs, can be treated as flavour-inclusive (no
differentiation between lepton flavours), or flavour-split (with
individual operators per lepton flavour). Both interpretations are
considered in this measurement. Additionally, to probe any possible
LFU violation that could be present, an analysis of the
flavour-relative WCs, the difference between the flavour-split
electron and muons WCs, is tested for the first time at the LHC. The
final outcome is measurements of the relevant WCs in
flavour-inclusive, flavour-split and flavour-relative scenarios
looking for deviations from the SM and setting limits on possible new
physics.

\begin{table}[!t]
\footnotesize
\caption{Definitions are given of the relevant dimension-6 SMEFT operators. The convention for notation used follows that of Ref.~\cite{Brivio:2020onw}, where relevant $SU(2)_L$ indices are represented by $k$ and $p, r$ represent flavour indices. The \texttt{top} flavour assumptions are used. Relevant for this measurement is that each of the seven top--lepton operators have corresponding three flavour-diagonal Wilson Coefficients (WCs) denoted $c_{X,i}$ with $i=1,2,3$ an index running over the lepton generations, for a given operator, $X$. The handedness, left (L) and right (R), of the quarks and leptons in the interaction vertices is given.}
\label{tab:smeft-operators}
\def\arraystretch{1.7}
\begin{center}
\begin{tabular}{lll}
\toprule
Operator & Definition & Description \\
\midrule
$O_{te}$         & $(\bar e_p \gamma_\mu e_r)(\bar t \gamma^\mu t)$          & R-handed leptons and R-handed quarks in the $t\bar{t}\ell^{+}\ell^{-}$ vertex \\
$O_{Qe}$         & $(\bar Q \gamma_\mu Q)(\bar e_p \gamma^\mu e_r)$           & R-handed leptons and L-handed quarks in the $t\bar{t}\ell^{+}\ell^{-}$ and $b\bar{b}\ell^{+}\ell^{-}$ vertices \\
$O_{tl}$         & $(\bar l_p \gamma_\mu l_r)(\bar t \gamma^\mu t)$          & L-handed leptons and R-handed quarks in the $t\bar{t}\ell^{+}\ell^{-}$ vertex \\
$O_{Ql}^{1}$   & $(\bar l_p \gamma_\mu l_r)(\bar Q \gamma^\mu Q)$          & L-handed leptons and L-handed quarks in the $t\bar{t}\ell^{+}\ell^{-}$ and $b\bar{b}\ell^{+}\ell^{-}$ vertices, weak-singlet \\
$O_{Ql}^{3}$   & $(\bar l_p \sigma^i\gamma_\mu l_r)(\bar Q\sigma^i\gamma^\mu Q)$ &  L-handed leptons and L-handed quarks in the $t\bar{t}\ell^{+}\ell^{-}$ and $b\bar{b}\ell^{+}\ell^{-}$ vertices, weak-triplet \\
$O_{leQt}^{1}$   & $(\bar l_p^j e_r) \epsilon_{jk} (\bar Q^k t)$ & Mixed L-/R-handed quarks and leptons in the $t\bar{t}\ell^{+}\ell^{-}$, $t\bar{b}\ell^{+}\ell^{-}$ and $b\bar{b}\ell^{+}\ell^{-}$ vertices, scalar \\
$O_{leQt}^{3}$   & $(\bar l_p^j \sigma_{\mu\nu} e_r) \epsilon_{jk} (\bar Q^k \sigma^{\mu\nu} t)$ & Mixed L-/R-handed quarks and leptons in the $t\bar{t}\ell^{+}\ell^{-}$, $t\bar{b}\ell\nu$, $\bar{t}b\ell\nu$ and $b\bar{b}\ell^{+}\ell^{-}$ vertices, tensor \\

\bottomrule
\end{tabular}
\end{center}
\end{table}

The paper is structured as follows:
Section~\ref{sec:detector} outlines the ATLAS detector, Section~\ref{sec:samples} provides an overview of the data
and MC samples used in the measurements, and Section~\ref{sec:objects}
details the object reconstruction and selection.
Section~\ref{sec:analysisstrategy} describes the signal region definitions and provides a description of the strategy to
estimate backgrounds, an overview of the systematic
uncertainties is given in Section~\ref{sec:systematics}, the results
are presented in Section~\ref{sec:result} and conclusions are given
in Section~\ref{sec:conclusion}.


%

%
\section{ATLAS detector}
\label{sec:detector}


The ATLAS detector~\cite{PERF-2007-01} at the LHC covers nearly the entire solid angle around the collision point.\footnote{\AtlasCoordFootnote}
It consists of an inner tracking detector surrounded by a thin superconducting solenoid, electromagnetic and hadronic calorimeters,
and a muon spectrometer incorporating three large superconducting air-core toroidal magnets.

The inner-detector system (ID) is immersed in a \qty{2}{\tesla} axial magnetic field
and provides charged-particle tracking in the range of \(|\eta| < 2.5\).
The high-granularity silicon pixel detector covers the vertex region and typically provides four measurements per track,
the first hit generally being in the insertable B-layer (IBL) installed before Run~2~\cite{ATLAS-TDR-19,PIX-2018-001}.
It is followed by the SemiConductor Tracker (SCT), which usually provides eight measurements per track.
These silicon detectors are complemented by the transition radiation tracker (TRT),
which enables radially extended track reconstruction up to \(|\eta| = 2.0\).
The TRT also provides electron identification information
based on the fraction of hits (typically 30 in total) above a higher energy-deposit threshold corresponding to transition radiation.

The calorimeter system covers the pseudorapidity range \(|\eta| < 4.9\).
Within the region \(|\eta|< 3.2\), electromagnetic calorimetry is provided by barrel and
endcap high-granularity lead/liquid-argon (LAr) calorimeters,
with an additional thin LAr presampler covering \(|\eta| < 1.8\)
to correct for energy loss in material upstream of the calorimeters.
Hadronic calorimetry is provided by the steel/scintillator-tile calorimeter,
segmented into three barrel structures within \(|\eta| < 1.7\), and two copper/LAr hadronic endcap calorimeters.
The solid angle coverage is completed with forward copper/LAr and tungsten/LAr calorimeter modules
optimised for electromagnetic and hadronic energy measurements respectively.

The muon spectrometer (MS) comprises separate trigger and
high-precision tracking chambers measuring the deflection of muons in a magnetic field generated by the superconducting air-core toroidal magnets.
The field integral of the toroids ranges between \num{2.0} and \qty{6.0}{\tesla\metre}
across most of the detector.
Three layers of precision chambers, each consisting of layers of monitored drift tubes, cover the region \(|\eta| < 2.7\),
complemented by cathode-strip chambers in the forward region, where the background is highest.
The muon trigger system covers the range \(|\eta| < 2.4\) with resistive-plate chambers in the barrel, and thin-gap chambers in the endcap regions.

The luminosity is measured mainly by the LUCID--2~\cite{LUCID2} detector that records Cherenkov light produced in the quartz windows of photomultipliers located close to the beampipe.

Events are selected by the first-level trigger system implemented in custom hardware,
followed by selections made by algorithms implemented in software in the high-level trigger~\cite{TRIG-2016-01}.
The first-level trigger accepts events from the \qty{40}{\MHz} bunch crossings at a rate below \qty{100}{\kHz},
which the high-level trigger further reduces in order to record complete events to disk at about \qty{1.25}{\kHz}.

A software suite~\cite{SOFT-2022-02} is used in data simulation, in the reconstruction
and analysis of real and simulated data, in detector operations, and in the trigger and data acquisition
systems of the experiment.


%

%
\section{Data and simulated event samples}
\label{sec:samples}

%

The full \RunTwo data sample, collected in \pp collisions during the years 2015--2018 and corresponding to a luminosity of \lumi~\ifb\ is used.
Only events from time periods in which LHC beams were stable and all ATLAS detectors were operational are selected~\cite{DAPR-2018-01}.
The uncertainty in the combined 2015--2018 integrated luminosity is 0.83\% \cite{DAPR-2021-01}, obtained using the LUCID-2 detector~\cite{LUCID2} for the primary luminosity measurements, complemented by measurements using the inner detector and calorimeters
The average number of interactions per bunch crossing, \avgmu, ranges from \num{0.5} to approximately $80$ and has a mean of \num{33.7}.

The data were collected using a combination of single-electron and
single-muon triggers, with requirements on the lepton candidates to
maintain high efficiency across the full momentum range while
controlling the trigger rates~\cite{TRIG-2016-01,TRIG-2018-05,TRIG-2018-01,TRIG-2019-03}. For electrons the trigger thresholds
were \pT = 26, 60 and 140~\GeV, whereas for muons the thresholds
were \pT = 26 and 50~\GeV. Isolation requirements were applied to the
triggers with the lowest \pT thresholds of 26~\GeV for electrons
and muons.

Samples of simulated events are produced using Monte Carlo (MC)
techniques to model the different SM processes. The
estimate of the \ttll signal and most considered backgrounds is
based upon these simulations. More details of the event generation are
given in the following paragraphs and are summarised in Table~\ref{tab:MCSamples},
with the samples in black, or in parentheses, indicating those used as the
nominal configurations, or to estimate the systematic uncertainties, respectively.
Several data-driven corrections are applied to
the estimated backgrounds and detailed in Section~\ref{sec:analysisstrategy}.

After generating events for each process of interest, the
detector response was modelled by either a full simulation of the
ATLAS detector based on \GEANT~\cite{SOFT-2010-01,Agostinelli:2002hh}
or a fast simulation~(\textsc{AtlFast~2}) relying on parameterised
showers in the
calorimeter~\cite{ATL-SOFT-PUB-2018-002}.
The latter was only used for the modelling of some rare background processes %
and the modelling of %
processes with alternative generators (see Section~\ref{sec:systematics}).
Additional simulated $pp$ collisions, generated with \Pythia[8.186]~\cite{Sjostrand:2007gs} using
parameter values from the A3 tune~\cite{ATL-PHYS-PUB-2011-014},
are overlaid to model the effects of both the in-time and
out-of-time \pileup. Events are weighted such that the \pileup distribution reflect the number of additional interactions observed in
data. All simulated events are processed using the same
reconstruction algorithms and analysis chain as the data. Corrections are applied so that the object reconstruction and
identification efficiencies, energy scales and energy resolutions
in simulation match those determined from data, with uncertainties described in Section~\ref{sec:systematics}.

\begin{table}[htb]
\begin{center}
\caption{
The configurations used for event generation of signal and background processes.
The samples in black indicate those used as the
nominal configurations, those in grey and parentheses are used to estimate the systematic uncertainties.
}
\label{tab:MCSamples}
\vspace{0.25cm}
\small
\setlength\tabcolsep{1.5pt}
\begin{tabular}{~l^l^l^l^l^l}
\hline\hline
Process & Generator & ME order & Parton shower & PDF & Tune  \\
\hline
\ttll                  & MG5\_aMC & \textsc{NLO} & \PYTHIA[8]  &  \NNPDF[3.0nlo] &  A14 \\
\rowstyle{\color{gray}}
& (MG5\_aMC) & (NLO) &  (\HERWIG[7])  & (\NNPDF[3.0nlo]) &   (H7-UE-MMHT)\\
\rowstyle{\color{gray}}
& (MG5\_aMC) & (NLO) &  (\PYTHIA[8])  & (\NNPDF[3.0nlo]) &   (\textsc{A14 Var3c})\\
$\ttbar W$             & \textsc{Sherpa 2.2.10} & \textsc{MePs@Nlo} & \textsc{Sherpa} &  \NNPDF[3.0nnlo] & \textsc{Sherpa} default \\
\rowstyle{\color{gray}}
& (MG5\_aMC)  & (FxFx NLO) & (\PYTHIA[8]) & (\NNPDF[3.0nlo]) & (A14)   \\
\rowstyle{\color{gray}}
& (\POWHEG)  & (NLO) & (\PYTHIA[8]) & (\NNPDF[3.0nlo]) & (A14)   \\
\rowstyle{\color{gray}}
& (\POWHEG)  & (NLO) & (\HERWIG[7]) & (\NNPDF[3.0nlo]) & (H7-UE-MMHT)   \\
$\ttbar W$ (EW)        & \textsc{Sherpa 2.2.10} & LO & \textsc{Sherpa} &  \NNPDF[3.0nnlo] & \textsc{Sherpa} default \\ %
\rowstyle{\color{gray}}
& (MG5\_aMC)  & (LO) & (\PYTHIA[8]) & (\NNPDF[3.0nlo]) & (A14)   \\
\ttH                   & \textsc{MG5\_aMC}  & NLO & \PYTHIA[8]\ & \NNPDF[3.0nlo] & A14 \\
$t\bar t t\bar t$      & MG5\_aMC & NLO & \PYTHIA[8] & \NNPDF[3.1nlo] & A14 \\
$\ttbar$               & \textsc{Powheg-BOX} & NLO & \PYTHIA[8] & \NNPDF[3.0nlo] & A14  \\
\rowstyle{\color{gray}}
& (\textsc{Powheg-BOX}) & NLO & (\HERWIG[7]) & (\NNPDF[3.0nlo]) &  (H7-UE-MMHT) \\
\rowstyle{\color{gray}}
& (\textsc{Powheg-BOX} $2\hdamp$) & NLO & (\PYTHIA[8]) & (\NNPDF[3.0nlo]) &  (A14) \\
\rowstyle{\color{gray}}
& (\textsc{Powheg-BOX}) & NLO & (\PYTHIA[8] $p_{\textrm{T,hard}}=1$) & (\NNPDF[3.0nlo]) &  (A14) \\
$t\bar t t$            & MG5\_aMC & LO & \PYTHIA[8] & \NNPDF[2.3lo] & A14 \\
$V$, $VV$, $qqVV$,    & \textsc{Sherpa} 2.2.2(1)  & \textsc{MePs@Nlo} & \textsc{Sherpa} & \NNPDF[3.0nnlo] & \textsc{Sherpa} default  \\
$VVV$    &   &  &  &  &   \\
$Z \to \ell^{+}\ell^{-}$         & \textsc{Sherpa} 2.2.1 & \textsc{MePs@Nlo} & \textsc{Sherpa} & \NNPDF[3.0nnlo] & \textsc{Sherpa} default \\ %
$Z\gamma$ & Sherpa 2.2.4 & \textsc{MePs@Nlo} & \textsc{Sherpa} & \NNPDF[3.0nnlo] & \textsc{Sherpa} default \\
$VH$ & \PYTHIA[8] & LO & \PYTHIA[8] & \NNPDF[2.3lo] & A14 \\
$t\bar{t}WW$ & MG5\_aMC & LO & \PYTHIA[8] & \NNPDF[2.3lo] & A14 \\
$t\bar{t}\gamma$ & MG5\_aMC & \textsc{NLO} & \PYTHIA[8]  &  \NNPDF[3.0nlo] &  A14 \\
$t\bar{t}Z(\to qq,\nu\nu)$ & MG5\_aMC & \textsc{NLO} & \PYTHIA[8]  &  \NNPDF[3.0nlo] &  A14 \\
\tZq, $tWZ$ & MG5\_aMC & \textsc{NLO} & \PYTHIA[8] & \NNPDF[2.3lo] & A14 \\ \\
\hline\hline
\end{tabular}
\end{center}
\end{table}

The SM signal process, associated production of a top--antitop
pair with a pair of same-flavour opposite-charge leptons, is modelled
using the \MGNLO[2.8.1]~\cite{Alwall:2014hca} (MG5\_aMC) generator
that provides matrix elements (MEs) at next-to-leading-order~(NLO) in the
strong coupling constant \alphas with
the \nnpdfnlo~\cite{Ball:2014uwa} parton distribution function (PDF)
set.  The functional form of the renormalization and factorization
scales (\muR, \muF) is set to the default scale
$0.5 \times \sum_i \sqrt{m^2_i+p^2_{\textrm{T},i}}$, where the sum runs over
all the particles generated from the matrix element calculation.  Top
quarks are decayed at leading-order (LO)
using \MADSPIN~\cite{Frixione:2007zp,Artoisenet:2012st} to preserve
all spin correlations.  The top-quark mass is set to 172.5~\GeV, in
this sample and for all other samples
involving top-quarks.  The events are interfaced
with \PYTHIA[8.244]~\cite{Sjostrand:2014zea} for the parton shower and
its hadronization, using the \tune{A14} set of tuned
parameters~\cite{ATL-PHYS-PUB-2014-021} and the \nnpdflo PDF set.  The
decays of bottom and charm hadrons are simulated using
the \evtgen~1.7.0 package~\cite{Lange:2001uf}, as are all other
samples in the measurement showered using \PYTHIA.  The
(on-shell) \ttZ cross-section is calculated at NLO quantum
chromodynamics (QCD) and NLO EW accuracy using \MGNLO and reported in
Ref.~\cite{deFlorian:2016spz}.  The $\gamma^{*}$ contribution and the
$Z/\gamma^{*}$ interference are taken into account with an off-shell
(down to $m(\ell^{+}\ell^{-})=5~\GeV$) correction calculated in
Ref.~\cite{ATL-PHYS-PUB-2016-005}.  The predicted value at 13~\TeV is
$0.88^{+0.09}_{-0.10}$~pb, where the uncertainties arise from
variations of renormalization and factorization scales, PDF uncertainties,
and \alphas variations.

For top-quark--antiquark pair production with a \Wboson boson (\ttW),
the \SHERPA[2.2.10]~\cite{Bothmann:2019yzt,Gleisberg:2008ta} generator and default parton shower are used at
NLO accuracy in QCD, with multi-leg merging of up to one additional
parton at NLO and up to two additional partons at LO (\tune{MEPS@NLO}
setup~\cite{Hoeche:2011fd,Hoeche:2012yf,Catani:2001cc,Hoeche:2009rj} with a merging scale of 30~\GeV). The \NNPDF[3.1nnlo]~\cite{NNPDF:2017mvq} PDF set is used. Additionally a LO QCD sample
is also generated with \SHERPA[2.2.10] but for the electroweak production of the $\ttW j$ final state.
The cross-sections of these QCD and EW \ttW samples
are scaled separately to the predictions of
Ref.~\cite{Frederix:2021agh}. %

Events featuring the production of a \ttbar pair in association with a
SM Higgs boson with a mass of 125~\GeV\ (\ttH) are simulated using NLO
matrix elements in \MGNLO[2.6.0] with the \nnpdfnlo PDF set.  The
samples are showered with \PYTHIA[8.230] using the \tune{A14} tune.

The production of single top quarks in association with a
$Z/\gamma^{*}$ in the dileptonic decay mode, including off-shell
effects down to $m(\ell^{+}\ell^{-})=5~\GeV$, \tZq, is modelled using
the \MGNLO[2.9.5] generator at NLO. The production of single top
quarks in association with a \Wboson and a \Zboson boson (\tWZ) is
modelled using \MGNLO[2.2.2] at NLO~\cite{Demartin2017}, both using the \nnpdfnlo PDF
set.  The \tZq events are interfaced
with \PYTHIA[8.245] and the \tWZ events with \PYTHIA[8.212], using
the \tune{A14} tune and the \nnpdflo PDF set.  The \tZq sample is
simulated in the four-flavour scheme (therefore including an
additional $b$-quark in the final state) and normalised to a cross
section obtained in the five-flavour scheme.

The possible overlap between events in the \ttZ sample and \tWZ sample,
which can occur when a NLO real emission is included in the latter, must be removed.
The \tWZ sample follows the ``diagram removal 1'' (DR1) scheme
described in Ref.~\cite{Demartin2017} and ignores any Feynman diagrams
containing two resonant top quarks for the overlap removal.

The background production of \ttbar events is modelled using
the \POWHEGBOX[2] generator~\cite{Alioli:2010xd} at NLO with
the \nnpdfnlo PDF set and the damping factor $h_\text{damp}$ set to
1.5 times the top quark mass.  The events are interfaced with \PYTHIA[8.230]
using the \tune{A14} tune and the \nnpdflo PDF set.

Diboson plus jets processes featuring the production of three charged leptons and one neutrino or four charged leptons (denoted by $WZ$+jets or $ZZ$+jets, respectively) are simulated using the \SHERPA[2.2.2] generator.
Multiple matrix elements are matched and merged with the \SHERPA parton shower based on the Catani--Seymour dipole factorisation scheme~\cite{Gleisberg:2008fv,Schumann:2007mg} using the \tune{MEPS@NLO} prescription.
The virtual QCD correction for matrix elements at NLO accuracy are provided by the \openloops library~\cite{Cascioli:2011va,Denner:2016kdg,Buccioni:2019sur}.
The production of \Wboson and \Zboson bosons with multiple jets ($V+$jets) is simulated similarly but with the \SHERPA[2.2.1] generator.

Smaller backgrounds featuring Higgs boson production in association with
a \Wboson or \Zboson ($VH$), four top-quark production (\tttt), triple
top-quark production (\ttt), the production of a \ttbar pair with
two \Wboson bosons~(\ttWW), the production of a \ttbar pair with two neutrinos or two jets and processes with three heavy gauge
bosons ($WWW$, $WWZ$, $WZZ$ and $ZZZ$) are simulated with the settings
shown in Table~\ref{tab:MCSamples}.

Additional MC samples encapsulating EFT effects are produced at LO in
QCD for the \ttll process, using the \MADGRAPH[2.9.3] generator
and \PYTHIA[8.245] parton shower (with the default \tune{A14} tune
settings). They include the \texttt{SMEFTsim 3.0}
model \cite{Brivio:2020onw} as the Universal Feynman Output (UFO) model, in the $M_W$ electroweak input
scheme~\cite{Brivio:2021yjb} with the \texttt{top} flavour
restrictions (5FS).
The nominal events are generated according to the
SM, and the \MADGRAPH reweighting module~\cite{Mattelaer:2016gcx} is used to compute a large
number of alternative event weights corresponding to the inclusion of
dimension-6 EFT vertices and propagators in the production Feynman
diagrams.
%
%


%

%
\section{Object identification and reconstruction}
\label{sec:objects}

Two categories of leptons are defined in the analysis: a looser category referred to as \emph{pre-selected} leptons used for object overlap removal, described later in this section, and to define background-enriched regions; and a tighter category of \emph{signal} leptons used in the selection requirements for most regions, including the signal regions.

Electron candidates are reconstructed from clusters of energy deposits in the electromagnetic calorimeter
that are matched to a track in the ID. They are required to satisfy $\pT>7$~\GeV, $|\eta|<2.47$ and
a likelihood-based electron identification~\cite{EGAM-2018-01} is required.
Further requirements for pre-selected electrons and signal electrons used in the analysis are summarised in the following.
For pre-selected electrons, the \texttt{LooseAndBLayerLH} identification is used, while for the signal electrons used in one of the analysis regions, \texttt{MediumLH} is required.
Electrons in the LAr transition region ($1.37 < \abseta < 1.52$) are rejected and requirements on the transverse ($d_{0}$) and longitudinal ($z_{0}$) impact parameter are applied to mitigate the contribution from electrons from reducible backgrounds and pile-up. To suppress electrons with incorrect charge assignment, a boosted decision tree (BDT) discriminant based on calorimeter and tracking quantities is applied and finally a procedure to resolve ambiguities between overlapping reconstructed objects is applied.
The electron energy scale is calibrated using data~\cite{EGAM-2018-01,EGAM-2021-02}. The associated scale factors~(SFs) for electron reconstruction, identification, and isolation are applied in MC, to correct for the efficiency differences between data and simulation~\cite{EGAM-2021-01}.
The associated scale factors~(SFs) for electron reconstruction, identification, and isolation are applied in MC, to correct for the efficiency differences between data and simulation.%
The selection criteria above largely suppress the contribution from non-prompt leptons arising from hadron decays that contain bottom- or charm-quarks (referred to as ``heavy-flavour (HF) non-prompt leptons'').  However, as this measurement is particularly sensitive to such backgrounds, signal electrons have a further selection requirement from a non-prompt lepton BDT discriminant~\cite{HIGG-2017-02}, based on isolation and lifetime information about a track-based jet that matches the selected light lepton. %

Muon candidates are reconstructed from MS tracks matched to ID tracks
in the pseudorapidity range of $|\eta|<2.5$ and are required to have
$\pt > 7$~\GeV.  Both the pre-selected and signal muons are required to
satisfy \texttt{Medium} identification requirements defined in
Refs.~\cite{MUON-2022-01,MUON-2018-03}. Selections on
the longitudinal and transverse impact parameters are applied as for the electrons and also, as with
electrons, the pre-selected muons have no isolation requirements,
while the signal muons selected for the analysis need to satisfy
the non-prompt lepton BDT isolation working point requirements.  The associated SFs for
identification and isolation are applied as multiplicative factors to
the MC event weight, to correct for the efficiency differences between
data and MC.

Jets are reconstructed using the \antikt jet algorithm~\cite{Cacciari:2008gp} as implemented in the \textsc{FastJet} package~\cite{Fastjet} with the distance parameter $R$ set to $0.4$. Topological clusters are used~\cite{ATL-PHYS-PUB-2015-036} along with ID tracks as input to the particle flow reconstruction algorithm~\cite{PERF-2015-09, }. The algorithm removes calorimeter energy deposits due to charged hadrons and uses more precise measurements of their momenta from tracks instead. Jets are calibrated~\cite{JETM-2018-05} with the \texttt{EMPFlow} scheme applying the jet area pile-up corrections.
The jets are kept only if they have $\pt > 25$~\GeV\ and are inside a pseudorapidity range of $\abseta < 2.5$.
The \tool{JetVertexTagger}~(JVT)~\cite{ATLAS-CONF-2014-018} is employed to mitigate pile-up effects, and specifically rejects any selected jets after the overlap removal procedure with $\pt < 60$~\GeV, $\abseta < 2.4$ and for which $\JVT < 0.5$ (\texttt{Tight} working point).

Jets containing a $b$-hadron, referred to as $b$-jets, are identified with the \texttt{DL1r} $b$-tagging algorithm~\cite{FTAG-2018-01}.
A working point (WP) corresponding to an $85\%$ $b$-tagging efficiency is used in most regions for pre-selections.
Exclusive bins in the $b$-tagging discriminant corresponding to different $b$-tagged jet identification efficiencies are also used, as pseudo-continuous $b$-tagging (PCBT).
This allows the possible use of different calibrated $b$-tagging WPs in defining regions.

The missing transverse momentum $\vec{p}_{\mathrm{T}}^{\mathrm{miss}}$
(with magnitude $\met$) is defined as the negative vector sum of the
$\pt$ of all selected and calibrated objects in the event, including a
term to account for the momenta of soft particles that are not
associated with any of the selected objects~\cite{JETM-2020-03}.  This
soft term is calculated from inner-detector tracks matched to the
primary vertex, which makes it more resilient to contamination
from \pileup interactions.

After the identification, overlaps between objects are resolved to avoid double-counting of physics objects.
Pre-selected leptons and jets are used for this procedure.
The sequence of operations that are performed to solve these ambiguities, referred to as the overlap removal (OR) procedure, is applied in the following steps. First, the lowest \pT of any electron pair found with a shared track or an overlapping calorimeter cluster is removed. Next, any muon matched with significant calorimeter energy deposits and found to share a track with an electron is removed, then any subsequent electron found to share a track with a muon is removed. Any jet found within a $\dR$ of 0.2 of an electron is removed, then any electron found within a $\dR$ of 0.4 of a jet is removed. Any jet with fewer than three tracks associated with it and found within a $\dR$ of 0.2 of a muon is removed. Any jet with fewer than three tracks associated with it that has a muon inner-detector track ghost-associated~\cite{FTAG-2018-01,Cacciari:2008gn} to it, is removed. Finally, any muon later found within a $\dR$ of 0.4 of a jet is removed.


%

%
\section{Analysis strategy}
\label{sec:analysisstrategy}


Regions enriched in the signal of \ttll at high invariant mass of the
$\ell^+\ell^-$ pair are defined to obtain sensitivity to the targeted
four-fermion EFT operators. The three-lepton final state is utilised due
to the high \ttll purity that can be obtained.

Several aspects of the analysis are built on a
previous ATLAS measurement of \ttZ production~\cite{TOPQ-2020-20}. The
main changes are to deal with the relatively larger background
contributions away from the Z boson mass peak. These changes are the use of a
tighter lepton isolation requirement and the addition of new control
regions defined to target the relevant backgrounds.

Background contamination in the signal regions (SRs) is due to contributions
from several SM processes which can feature the same final state
particles as the signal process. These are broadly split into two
categories based on the source of the reconstructed leptons:
backgrounds where the three final state leptons are prompt (discussed
in Section~\ref{sec:PromptBackgrounds}), and those where at least one
of the leptons is labelled as \emph{fake} as it originates from a non-prompt source (described in
Section~\ref{sec:NonPromptBackgrounds}).

\begin{figure}[t!]
\centering
\includegraphics[width=0.5\linewidth]{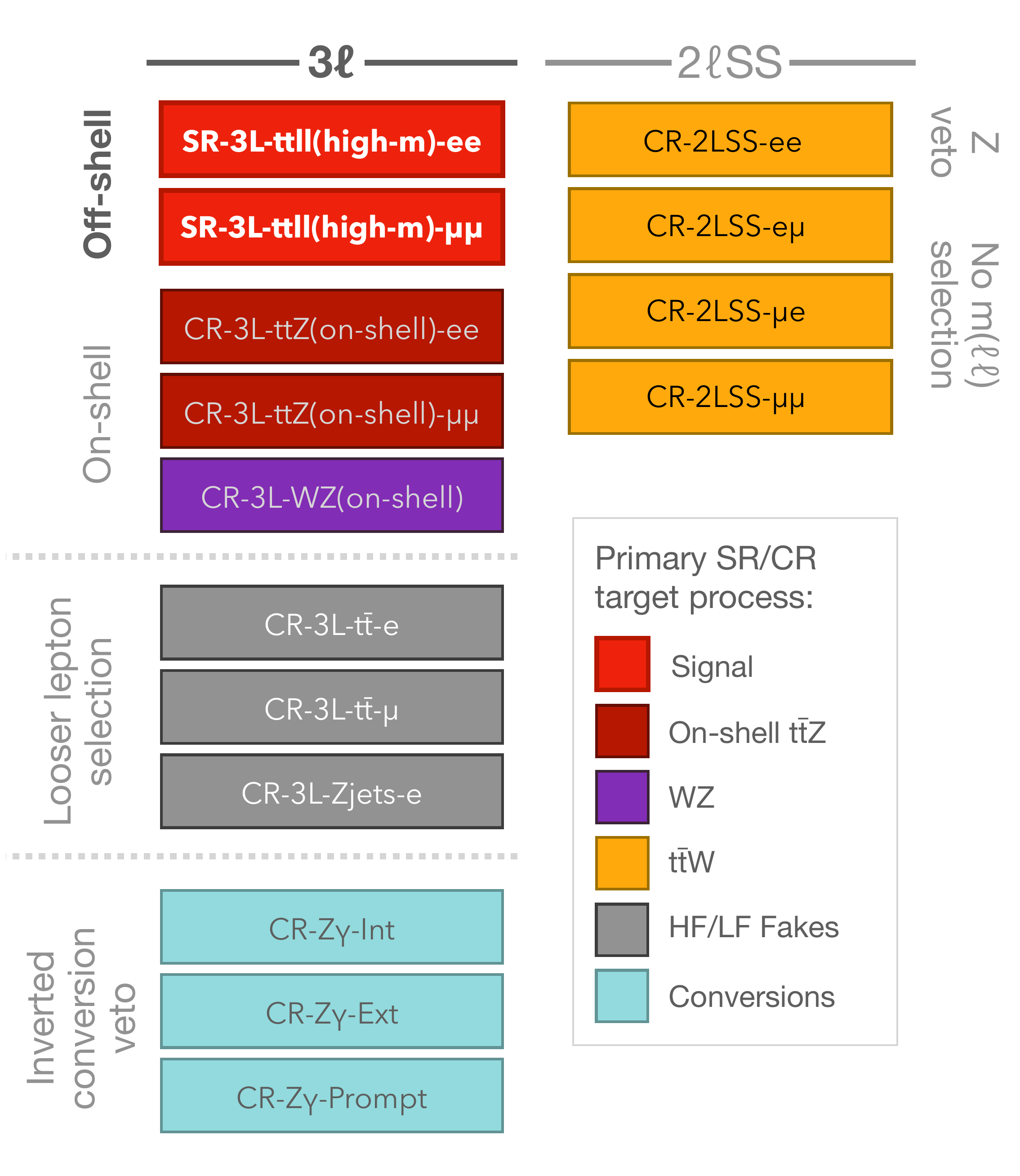}\\
\caption{Schematic diagram of the signal and control regions in the fits. Labelled are their main defining features and the signal or background process they are primarily targetting.}
\label{fig:RegionDefs:Schematic}
\end{figure}

Dedicated control regions (CRs) enriched in the dominant backgrounds are
defined to constrain these backgrounds from data. There are in total
thirteen of these CRs.
In eight of them only the overall normalisation enters into the fit: three that target the normalisation of the events with a fake electron from photon conversions (F-e-IntConv, F-e-ExtConv);
two to constrain events with non-prompt fake electrons and muons
originating from heavy-flavour hadrons (F-e-HF, F-m-HF); and three
targeting on-shell \ttZ and diboson (VV) production. There are five further CRs which are binned in a dedicated observable: one to
constrain electrons from light-flavour hadrons (F-e-LF); and four primarily sensitive to \ttW, \ttH and electrons with misidentified
charge (charge-flip) in addition to the previous
backgrounds. The charge-flip background is
estimated by using a data-driven approach in the 2LSS CRs and all other backgrounds are
estimated purely from simulation. An overview of the signal and CRs is provided
schematically in Figure~\ref{fig:RegionDefs:Schematic} and comparisons to data are provided in Figure~\ref{fig:RegionDefs:Summary_post} (c).

The binned yields (templates) of some simulated processes, namely on-shell \ttZ and the
\ttW, $VV$, F-e-HF, F-m-HF, F-e-LF, F-e-IntConv, F-e-ExtConv backgrounds are
adjusted via normalisation factors that are determined by performing a
likelihood fit to data across these dedicated CRs
simultaneously with the signal regions.

\subsection{Event categorisation}
\label{sec:eventcat}
In the following, simulated events are categorised according to different
criteria. If all leptons in the events are categorised as prompt, the
event is classified according to its process (rare processes are
grouped  as \enquote{Other}, including \tttt, \ttt, $VH$,
$WW$, $VVV$, \ttWW, $t\bar{t}qq$ and $t\bar{t}\nu\nu$).  If one lepton
is categorised as non-prompt, the event is classified according to the
type of the non-prompt lepton.

Leptons are categorised as prompt if generator-level information, using a $\Delta R$ matching between generator- and detector-level information,
identifies them as electrons or muons originating from a $\tau$-lepton or $W$, $Z$ or $H$ boson. Otherwise the leptons
are categorised as fake and attributed to one of the
following categories using their generator-level information. Heavy flavour
fakes are those identified as originating from a strange, charmed or
bottom hadron. Light
flavour fakes are similarly those originating from a light baryon or
meson, for example from pions.  Charge-flip fakes are prompt electrons with
mis-identified charge.  Conversion electrons are those from photon
conversion and are further categorised into two sub-categories;
internal photon conversions, from $\gamma^{*}\to e^{+}e^{-}$, and external conversions, photon
conversions in the detector material.  Finally, conversion
muons are those originating from a muon internal photon conversion.

\subsection{Signal selection}

Events entering the signal region are required to contain exactly three
signal leptons with transverse momentum greater than
\SI{27}{\giga\eV}, \SI{20}{\giga\eV} and \SI{15}{\giga\eV} and least
one opposite-sign (charge) same-flavour (OSSF) lepton pair. Additionally
all OSSF pairs are required to have a minimal invariant mass of
\SI{10}{\giga\eV} in order to remove contributions from low mass
resonances. Finally at least three jets including at least one
$b$-tagged jet at the \SI{85}{\percent} efficiency $b$-tagging working
point are required.  The mass of the OSSF pair closest to the $Z$ mass
(referred to as $m(\ell^{+}\ell^{-})$ in the following) has to be above
\SI{101.2}{\giga\eV}, targeting specifically the high-mass \ttll region.
Additionally none of the leptons should be flagged as conversion candidates as described in Subsection~\ref{sec:NonPromptBackgrounds}.

To increase the sensitivity to the flavour specific EFT WCs,
the SR is split into two regions corresponding to \ttee and \ttmumu
by requiring at least two electrons or two muons respectively.

The resulting SRs (summarised in Table~\ref{tab:3lSRCR} and shown in
Figure~\ref{fig:RegionDefs:SRs}) with the event yields for each
process detailed in Table~\ref{tab:RegionsDefs:SRyields} are then
binned in the $m(\ell^{+}\ell^{-})$ distribution. This variable is found to
give the highest sensitivity to the four-fermion EFT operators. As the
impact of these EFT operators grows with energy, an optimised binning
in the highest $m(\ell^{+}\ell^{-})$ bins increases the sensitivity.

\begin{table}[t!]
\begin{center}
\caption{
Summary of the selection criteria for the 3L SRs and CRs. The neural network used for the splitting the \ttZ CR and the $WZ+$jets CR is introduced in Section~\ref{sec:PromptBackgrounds}.
}
\label{tab:3lSRCR}
\vspace{0.25cm}
\small
\setlength\tabcolsep{1.5pt}
\begin{tabular}{c|c|c}
\hline\hline
\ttll & \ttZ & $WZ+$jets \\
\hline
\multicolumn{3}{c}{exactly 3 signal leptons}\\
\multicolumn{3}{c}{$\pT(\ell_{1})>\SI{27}{\giga\eV}$, $\pT(\ell_{2})>\SI{20}{\giga\eV}$, $\pT(\ell_{3})>\SI{15}{\giga\eV}$}\\
\multicolumn{3}{c}{$\geq1$ OSSF pair}\\
\multicolumn{3}{c}{$m(\ell^{+}\ell^{-})>\SI{10}{\giga\eV}$ for all OSSF pairs}\\
\multicolumn{3}{c}{$N_{\mathrm{jets}}\geq 3$ and $N_{b-\mathrm{tagged\,jets}}\geq 1$ at the \SI{85}{\percent} efficiency WP}\\
\multicolumn{3}{c}{none of the leptons passes the conversion cuts}\\
\hline
$m(\ell^{+}\ell^{-}) > m_Z+\SI{10}{\giga\eV}$ & \multicolumn{2}{c}{$|m(\ell^{+}\ell^{-})-m_Z|\leq\SI{10}{\giga\eV}$}\\
for $m(\ell^{+}\ell^{-})$ closest to $m_Z$ & \multicolumn{2}{c}{for $m(\ell^{+}\ell^{-})$ closest to $m_Z$}\\
\hline
--- & $DNN(VV)<0.22$ & $DNN(VV)\geq0.22$\\
--- & $DNN(\tZq)<0.43$ & ---\\
\hline\hline
SR-3L-ttll(high-m)-ee & CR-3L-ttZ(on-shell)-ee & \multirow{ 2}{*}{CR-3L-WZ(on-shell)} \\
SR-3L-ttll(high-m)-$\mu\mu$ & CR-3L-ttZ(on-shell)-$\mu\mu$ & \\
\hline\hline
\end{tabular}
\end{center}
\end{table}

\begin{table}[t!]
\caption{Expected yields from MC of the different processes in the \ttee and \ttmumu SRs, SR-3L-ttll(high-m)-ee and SR-3L-ttll(high-m)-$\mu\mu$, and are compared to the observed yields in data. The indicated uncertainties include those from the limited dataset size as well as all other systematic uncertainties, discussed in Section~\ref{sec:systematics}, before the fit to data.}
\begin{center}
\begin{tabular}{lS[table-format=2.1]@{\(\;\pm\;\)}S[table-format=1.1]@{\,} c S[table-format=2.1]@{\(\;\pm\;\)}S[table-format=1.1]@{\,}}
\hline
\hline
& \multicolumn{5}{c}{SR-3L-ttll(high-m)} \\
Process & \multicolumn{2}{c}{ee} & & \multicolumn{2}{c}{$\mu\mu$} \\
\hline
\ttll & \num{19.4} & \num{ 1.4} & & \num{27.4} & \num{1.2} \\
\ttW & \num{19.9} & \num{ 1.9} & & \num{27.5} & \num{3.5} \\
\ttH & \num{8.6} & \num{ 0.8} & & \num{11.9} & \num{1.1} \\
$WZ$ & \num{12} & \num{ 5} & & \num{20} & \num{8} \\
$ZZ$ & \num{1.4} & \num{ 0.6} & & \num{2.4} & \num{1.0} \\
\tZq, \tWZ & \num{4.2} & \num{ 0.8} & & \num{6.5} & \num{1.2} \\
F-e-ExtConv & \num{1.2} & \num{ 1.1} & & \num{0.67} & \num{0.04} \\
F-e-IntConv & \num{4.9} & \num{ 1.1} & & \num{6.2} & \num{2.6} \\
F-e-HF & \num{2.1} & \num{ 1.5} & & \num{4.5} & \num{0.7} \\
F-e-LF & \num{2.4} & \num{ 0.7} & & \num{2.0} & \num{0.5} \\
F-m-HF & \num{5.8} & \num{ 1.5} & & \num{4.2} & \num{0.5} \\
Other (fakes) & \num{4.4} & \num{ 2.1} & & \num{6} & \num{4} \\
Other (non-fakes) & \num{4.6} & \num{ 1.6} & & \num{5.7} & \num{1.9} \\
\hline
Total & \num{91} & \num{ 7} & & \num{124} & \num{9}\\
\hline
Data & \multicolumn{2}{c}{89}  & & \multicolumn{2}{c}{128}  \\
\hline
\end{tabular}
\end{center}

\label{tab:RegionsDefs:SRyields}
\end{table}

\subsection{Prompt backgrounds}
\label{sec:PromptBackgrounds}

The dominant background in the SRs is from the $t\bar{t}W$ process, for which the normalisation is treated as a free floating parameter in the fit mainly constrained by dedicated CRs (detailed in Table~\ref{tab:2lSSCR}).
These regions require events to contain exactly two same-sign (SS) leptons that do not fulfil the conversion criteria and satisfy $\pT>\SI{27}{\giga\eV}$ and $\pT>\SI{20}{\giga\eV}$ respectively. Additionally at least three jets are required, where at least one is $b$-tagged at the \SI{85}{\percent} efficiency WP. Events are split into $2\mu$, $2e$, $e\mu$ and $\mu e$ regions, where for the latter two regions the order defines the flavour of the leading and subleading \pT leptons respectively. For the $2e$ regions events with an invariant mass of the two leptons around the $Z$ peak are vetoed to suppress backgrounds containing charge-flip leptons. The 2LSS CRs are binned in leading lepton $\pT$ as this leads to additional EFT sensitivity from high-mass \ttll events where one lepton is not reconstructed, with the $\pT$ of the leading lepton characterising the energy of the event well.

\begin{table}[!t]
\begin{center}
\caption{
Summary of the selection criteria for the 2LSS CRs.
}
\label{tab:2lSSCR}
\vspace{0.25cm}
\small
\setlength\tabcolsep{1.5pt}
\begin{tabular}{c|c|c|c}
\hline\hline
2$\mu$ & 2$e$ & $e\mu$ & $\mu e$ \\
\hline
\multicolumn{4}{c}{exactly 2 signal leptons}\\
\multicolumn{4}{c}{$\pT(\ell_{1})>\SI{27}{\giga\eV}$, $\pT(\ell_{2})>\SI{20}{\giga\eV}$}\\
\multicolumn{4}{c}{SS lepton pair}\\
\multicolumn{4}{c}{$N_{\mathrm{jets}}\geq 3$ and $N_{b-\mathrm{tagged\,jets}}\geq 1$ at the \SI{85}{\percent} efficiency WP}\\
\multicolumn{4}{c}{none of the leptons passes the conversion requirements}\\
\hline
2 muons & 2 electrons & $\pt(e)>\pt(\mu)$ & $\pt(\mu)>\pt(e)$ \\
& $|m(\ell^{+}\ell^{-})-m_Z|>\SI{10}{\giga\eV}$ & & \\
\hline\hline
CR-2LSS-$\mu\mu$ & CR-2LSS-ee & CR-2LSS-e$\mu$ & CR-2LSS-$\mu$e \\
\hline\hline
\end{tabular}
\end{center}
\end{table}

Additionally two CRs targeting contributions from the \ttZ and the $WZ+$jets processes are obtained by using the SR selection requirements, but requiring the invariant mass of the OSSF pair within the $Z$ peak,  $\SI{81.2}{\giga\eV} < m(\ell^{+}\ell^{-}) < \SI{101.2}{\giga\eV}$. The \ttZ and the $WZ+$jets region are split by a requirement on a Deep Neural Network (DNN) output score, classifying how likely it is that an event is a diboson event. This DNN is the same as that used in the on-shell \ttZ analysis~\cite{TOPQ-2020-20}. It corresponds to a 3-class DNN discrimination among \ttZ, \tZq and diboson events and the requirements used for splitting the regions were defined by optimising the signal sensitivity. In this analysis, the \tZq region is disregarded. As in the \ttZ analysis, events entering the \ttZ CR are required to satisfy the $DNN(\tZq)<0.43$ and $DNN(VV)<0.22$ requirements, while events entering the $WZ+$jets CR need to satisfy the $DNN(VV) \ge 0.22$ requirement. Compared with the \ttZ analysis, the tighter requirement of a $b$-tagged jet at the \SI{60}{\percent} tagging working point is removed in the $WZ+$jet CR (while still requiring at least one $b$-tagged jet at \SI{85}{\percent}), in order to get a $WZ+$jets CR targetting all flavour contributions instead of a specific $WZ+b$ CR to better match the composition in the high-mass SR. Similar to the SR, the \ttZ region is split into \ttee and \ttmumu. All three of these CRs only contain one bin, with the selections summarised in Table~\ref{tab:3lSRCR}.

\subsection{Non-prompt backgrounds}
\label{sec:NonPromptBackgrounds}

This measurement is sensitive to events containing non-prompt fake
leptons as they constitute a reducible background in the two same-sign
lepton and three-lepton regions. Even after dedicated object selection
criteria to reduce them, the large \ttbar cross-section relative to
the rare top-quark processes probed in this analysis means that this
background can be significant.

The SRs have too  few MC events to use for the fake lepton
backgrounds. The default estimations, using signal leptons, are replaced by those produced
with the looser preselection lepton criteria applied.
The distribution with the preselection lepton requirement is then
scaled to the integral of the distribution with the signal selection,
with the shape of the template in $m(\ell^{+}\ell^{-})$ agreeing very well
between both of the distributions. This method significantly reduces uncertainty
in the fake templates due to limited MC events.

Three regions target the HF and LF fakes: the events are required to
contain exactly three leptons, two signal leptons and one lepton that satisfies the preselection lepton definition but fails the signal lepton definition, with
transverse momentum greater than \SI{27}{\giga\eV}, \SI{20}{\giga\eV}
and \SI{15}{\giga\eV} and the sum of charges of plus or minus
one. Additionally, there have to be at least three jets, with at least
one of them $b$-tagged at an efficiency working point
of \SI{85}{\percent}.
Events are split into three regions based on requirements on the
leptons. If the event contains an OSSF pair, has
$\met<\SI{80}{\giga\eV}$ and the non-signal lepton is an electron it enters the \lllzeLF region. If the event does not have an OSSF pair, depending on whether the non-signal lepton is an electron or muon it enters the \lllttbareHF or \lllttbarmuHF region respectively.
While the \ttbar-e/$\mu$ CRs only contain one bin, the Zjets-e CR is
binned in the transverse mass of the lepton--\met system, $m_{\textrm{T}}(W)$. The latter
binning is taken from the on-shell \ttZ analysis~\cite{TOPQ-2020-20} and
is found to give a good discrimination of the LF electron fakes.

To create regions with relatively high event counts of
conversion fakes and high purity a $Z(\rightarrow\mu\mu\gamma)$ CR is
implemented. The events are required to contain three tight leptons,
of which two are an opposite sign muon pair and one is an
electron. The \met must be smaller than \SI{50}{\giga\eV} and the
event should contain no $b$-tagged jets at a WP of
\SI{85}{\percent}. Additionally the invariant mass of the two muons is
required to be above \SI{20}{\giga\eV} but outside the Z boson peak defined
as \SI{10}{\giga\eV} around the $Z$ mass. The trileptonic invariant
mass is then required to lie inside this $Z$ mass window.
Events are added to the CRs if the electron has a
reconstructed conversion vertex~\cite{EGAM-2018-01}. If the conversion radius
$>\SI{20}{\milli\m}$ and $m(\ell^{+}\ell^{-})<\SI{100}{\mega\eV}$ (where the mass is calculated relative to the conversion
vertex) the event enters the external conversion CR,
\lllzyext. If the conversion radius $<\SI{20}{\milli\m}$ and $m(\ell^{+}\ell^{-})<\SI{100}{\mega\eV}$ (where the mass is calculated relative to the primary vertex) it enters the internal
conversion CR, \lllzyint.  Additionally to these two CRs
selecting conversions, the prompt $Z\gamma$ region is added as a CR,
where, similar to the SR and the other prompt background CRs, events
with conversion like leptons are vetoed.
This prompt region emulates a selection closer to the SR and the 2LSS CRs as it contains events that fail  to satisfy the conversion rejection criteria.
All three of these
conversion CRs only contain one bin.  As the conversion variables
cannot be computed for muons, the CRs do not include conversion muons
but only conversion electrons. As the physics process between internal
electron and muon conversions is the same, the computed normalisation
factor is applied to electrons and muons.

\begin{figure}[t!]
\centering
\subfloat[]{\includegraphics[height=0.4\linewidth]{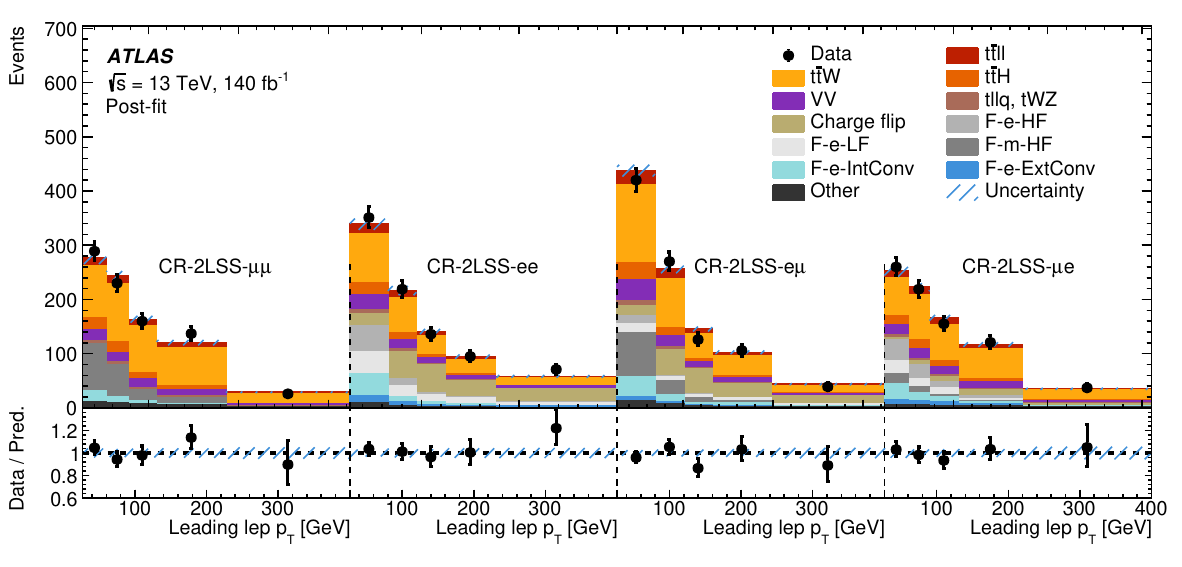}}\\
\subfloat[]{\includegraphics[height=0.4\linewidth]{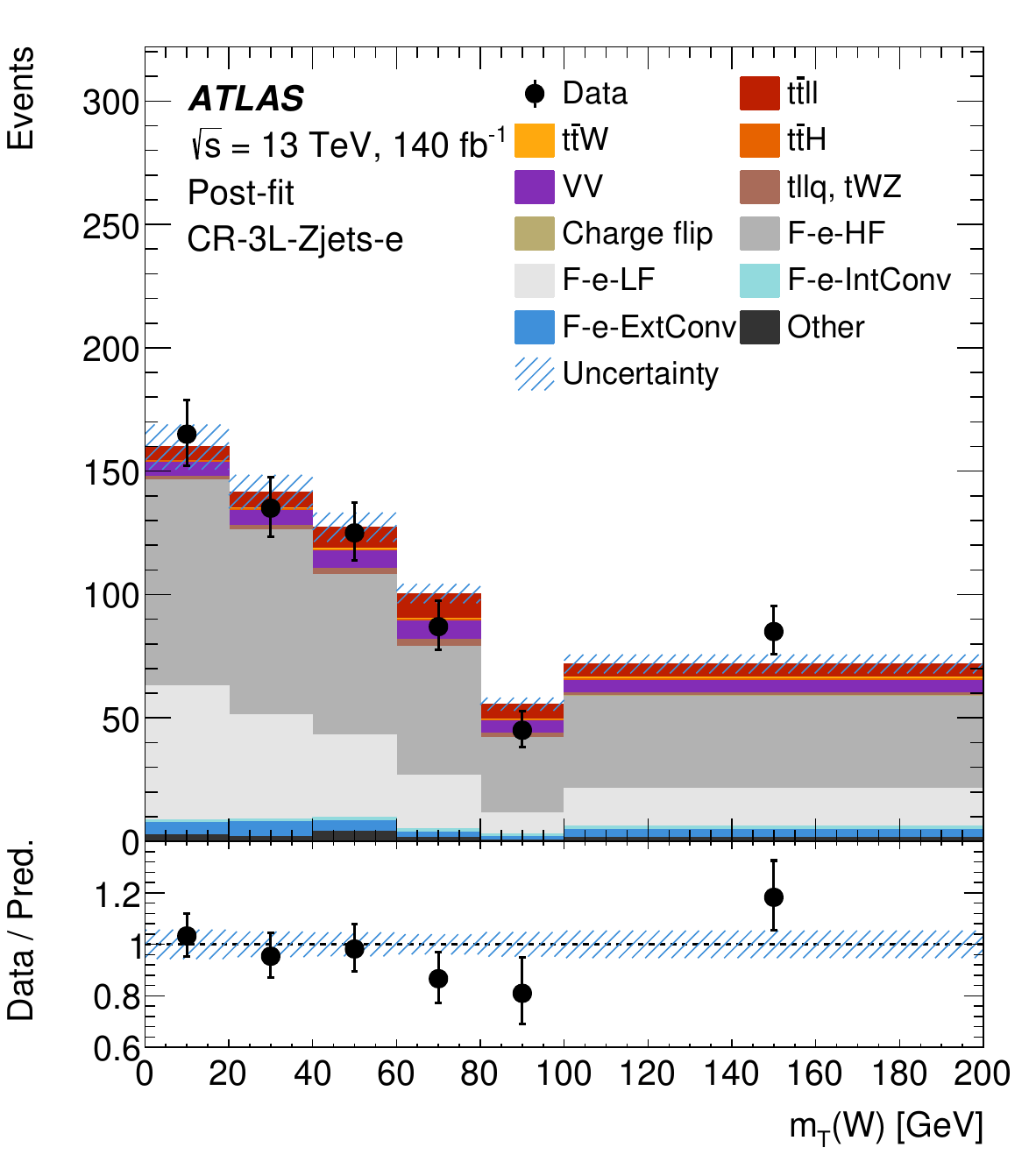}}
\subfloat[]{\includegraphics[height=0.4\linewidth]{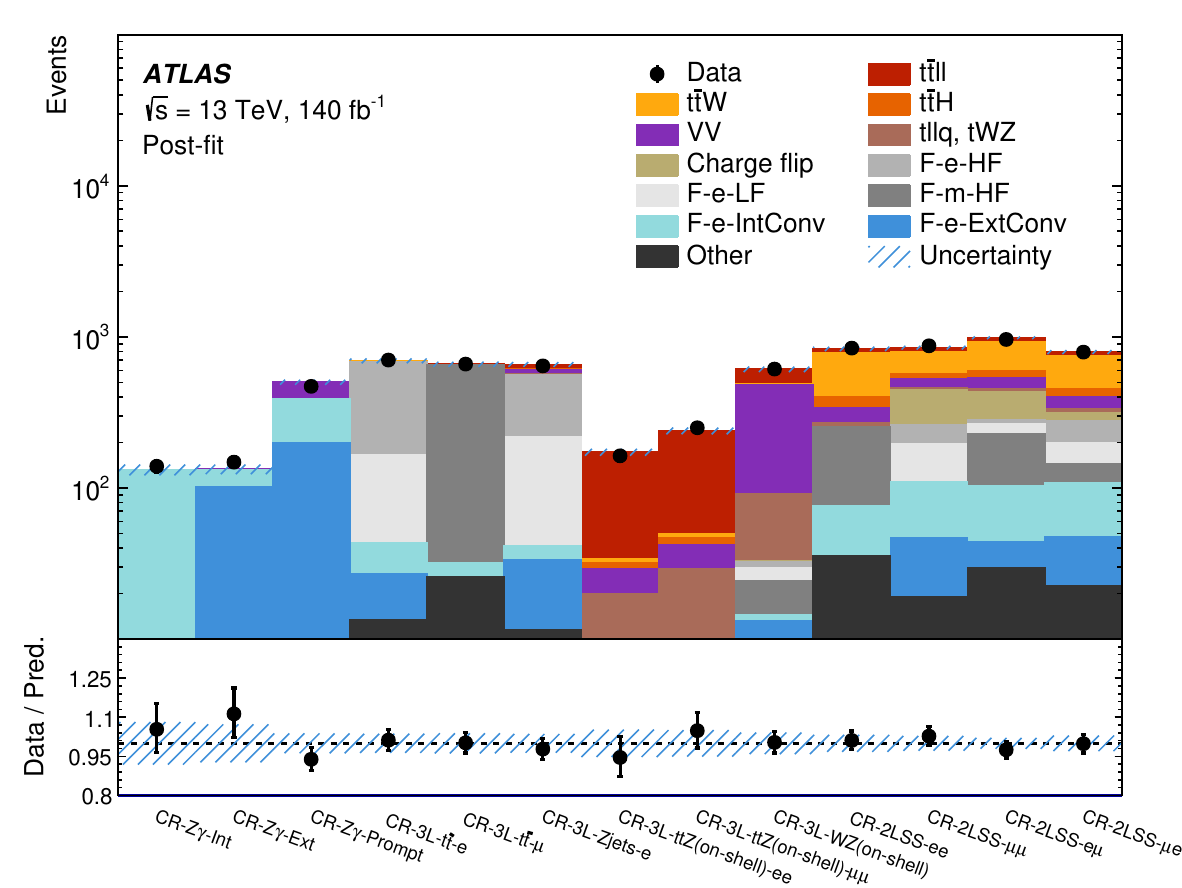}}
\caption{Event yields as a function of the relevant kinematic variables in (a) the CR-2LSS CRs, (b) the CR-3L-Zjets-e CR, and (c) inclusively in all CRs considered in the analysis after the fit to data. The uncertainty band includes the statistical as well as the systematical components. In figures (a) and (b) the highest bin is inclusive and contains events above the bin edge.}
\label{fig:RegionDefs:Summary_post}
\end{figure}

Finally, a data-driven charge-flip estimate is used in the 2LSS CRs. Charge-flip rates are derived for electrons in $\pT$ and $\eta$ bins using two $Z\rightarrow ee$ regions, one with opposite-sign electrons and one with same-sign electrons. The regions select events with exactly two signal electrons, at most one jet and no $b$-tagged jet. Additionally a $Z$ peak region is defined with $\SI{71}{\giga\eV} < m(e^{+}e^{-}) < \SI{101}{\giga\eV}$, where a background subtraction is performed, assuming linear backgrounds, by subtracting the event yields from two side-band regions ($\SI{61}{\giga\eV} < m(e^{+}e^{-}) < \SI{71}{\giga\eV}$ and $\SI{101}{\giga\eV} < m(e^{+}e^{-}) < \SI{111}{\giga\eV}$). The charge-flip rates are then obtained by maximising a likelihood function where the charge-flip rates in each lepton $\pT$ and $\eta$ bin are the parameters of interest. The charge-flip estimate in the 2LSS CR are then estimated by applying these rates to a region with identical selection except with the $SS$ switched to an $OS$ one.

Distributions for all the CRs considered in
the analysis are shown, after the fit to data, in
Figure~\ref{fig:RegionDefs:Summary_post}.


%

%
\section{Systematic uncertainties}
\label{sec:systematics}

The systematic uncertainties described in the following are separated into detector systematic uncertainties, which are related to the reconstruction and identification of physics objects in the detector, theory uncertainties related to the modelling of the different processes and uncertainties associated with the estimate of the fake-lepton background.

\subsection{Detector-related uncertainties}

The luminosity estimate has a relative uncertainty of 0.83\%, derived following the prescriptions described in Ref.~\cite{DAPR-2021-01} and is applied to all processes determined from MC simulations.

An uncertainty related to the SFs used for MC to account for differences in pile-up distributions between MC and data is applied.
This uncertainty is obtained by re-scaling the \avgmu value in data by $1/0.99$ and $1/1.07$, around the central SF of $1/1.03$.
The uncertainty in the jet energy scale (JES) is derived combining information from test-beam data, LHC collision data and simulation~\cite{JETM-2018-05}.
The jet energy resolution (JER) is  measured separately for data and MC using two in situ techniques~\cite{JETM-2018-05}.
A systematic uncertainty is defined as the quadratic difference between the jet energy resolutions for data and simulation.
To determine this the energy of jets in the simulation is smeared by this residual difference, and the changes in the normalisation and shape of the final discriminant are compared with the default prediction.
Uncertainties associated with the jet vertex tagging~(JVT) are applied and account for the choice of MC generator and residual contamination from pile-up jets after pile-up suppression.

The effects of uncertainties in efficiencies for the heavy-flavour identification of jets by the $b$-tagging algorithm were evaluated.
These efficiencies are measured from data and depend on the jet flavour and kinematic quantities of the jets.
The SFs and their uncertainties are applied to each jet in the simulation depending on its flavour, \pt and $\eta$~\cite{FTAG-2018-01}.
Additionally, flavour tagging extrapolation uncertainties for jets with $\pT$ above the calibration range are considered.

Uncertainties associated with the electron efficiency SFs are considered for the reconstruction and trigger uncertainties (observed to be largely sub-dominant) and the electron identification and isolation SFs.
The accuracy of the momentum scale and resolution of electrons in simulation is checked using reconstructed distributions of the $Z \rightarrow e^{+}e^{-}$ and $J/\psi \rightarrow e^{+}e^{-}$ masses.
Small discrepancies are observed between data and simulation, and corrections for the energy scale and resolution are applied. A detailed description can be found in Refs.~\cite{EGAM-2018-01}.%

SFs are also applied to correct for the difference between data and simulation in the muon identification, isolation and trigger efficiency, as well as track-to-vertex association~\cite{MUON-2018-03} and associated uncertainties in these SFs are considered.
Momentum scale and resolution corrections are applied for muons in the simulation.
Uncertainties in both the momentum scale and resolutions in the MS and the ID are considered and varied separately.
Additional uncertainties are considered to account for the charge-dependent scale correction applied on data.
A more detailed description can be found in Refs.~\cite{MUON-2018-03}.

Uncertainties are applied to the scale and resolution of the soft track component of the missing transverse energy.
They are derived from the agreement between data and MC of the \pt balance between the hard and soft \met components.

\subsection{Modelling uncertainties}

To obtain the uncertainties in \ttll related to missing higher-order effects, \muR and \muF are simultaneously varied up and down by a factor of two and compared with the nominal predictions.
Uncertainties on the PDF are evaluated following the recommended \tool{PDF4LHC} prescription~\cite{Butterworth:2015oua}.
The uncertainties associated with the parton showering algorithm and the underlying event model\footnote{In the following only, referred to  as ``parton shower uncertainty''.} are evaluated by comparing the nominal samples, generated with \MGNLO interfaced to \PYTHIA[8], to equivalent samples interfaced to \HERWIG[7]~\cite{Bahr:2008pv,Bellm:2015jjp} instead.
Uncertainties related to the modelling of initial-state radiation are obtained from the up and down \tune{Var3c} variations of the \PYTHIA \tune{A14} tune in dedicated alternative samples.

Uncertainties in \ttW due to missing higher-order QCD corrections in the modelling are considered by including simultaneous variations of \muR and \muF by a factor of two up and down. %
Matrix element and parton shower uncertainties are estimated comparing the nominal \SHERPA prediction with \MGNLO interfaced with \PYTHIA[8] using the FxFx merging prescription~\cite{Frederix:2012ps} for the QCD part and with \MGNLO{}+\PYTHIA[8] for the EW part. An additional parton shower uncertainty is considered using the relative difference between \POWHEG{}+\PYTHIA[8] and \POWHEG{}+\HERWIG[7] predictions.
Finally a normalisation-only uncertainty in the relative yields of the EW sample within the QCD+EW $t\bar{t}W$ templates is applied by varying the EW fraction between the values of the Sherpa and the MG5$\_$aMC@NLO prediction.
Uncertainties in the modelling of the $WZ \rightarrow \ell\ell\ell\nu$ ($WZ$+jets) and $ZZ \rightarrow \ell\ell\ell\ell$ ($ZZ$+jets) processes related to the matrix element matching scale, resummation scale and recoil scheme are estimated by using alternative ``truth-level'' samples.
The \muR and \muF scale and PDF uncertainties are evaluated following the same prescription as used for the \ttll sample.
In addition, normalisation uncertainties are applied to the different additional jet flavour components: light quarks and gluons ($l$), charm quarks ($c$) and bottom quarks ($b$). An uncertainty of $30\%$ is applied to \WZl and \WZc and of $50\%$ to the \WZb components of the $WZ+$jets background. Similarly, normalisation uncertainties of $10\%$, $30\%$ and $50\%$ are applied to the \ZZl, \ZZc and \ZZb components of the $ZZ+$jets background.

A cross-section normalisation uncertainty of $14\%$ is assigned to the \tZq process, based on the dedicated ATLAS measurement presented in Ref.~\cite{TOPQ-2018-01}.
Uncertainties in the parton shower model, initial-state radiation, scale and PDF variations are obtained in a similar way to the \ttll sample.
For the \tWZ background process, a flat \SI{30}{\percent} normalisation uncertainty is added. No parton shower uncertainty is considered, but instead the difference relative  to an alternative diagram removal scheme (DR2)~\cite{Demartin2017} is treated as a modelling uncertainty. The DR2 scheme, additionally to the nominal DR scheme, considers the interference terms (at the level of squared amplitudes) between single- and double-resonant \ttbar production.
As with the processes described above, PDF and scale uncertainties are taken into account. %

Only theory uncertainties in the normalisation of the cross-section of the \ttH process are considered. These follow the scale and PDF uncertainties prescribed in Ref.~\cite{deFlorian:2016spz}.
The \ttbar process is the dominant contribution to the fake backgrounds. While the normalisation of the fake backgrounds is obtained in the template fit, shape systematic uncertainties are considered for \ttbar separately for each of the fake background contributions. The parton shower uncertainty is estimated by using the alternative \POWHEG{}+\HERWIG[7] sample. Additionally uncertainties in the value of \hdamp are covered by using an alternative sample with twice the value of the default \hdamp (set to $1.5m_t$). Finally, a sample with the \POWHEG{}-specific parameter in \PYTHIA[8] set to $p_{\textrm{T,hard}}=1$ (compared with the nominal sample with $p_{\textrm{T,hard}}=0$) is considered for the matrix element uncertainty.

For other minor background processes, such as $HV$, $VVV$, $\ttWW$ or multi-top-quark (\ttt) production, a common conservative overall normalisation uncertainty of $50\%$ is applied. These background components contribute to about $2.5\%$ of the total event yields in the signal regions.
For \tttt, a normalisation uncertainty of $+70\%$ $-15\%$ is applied based on the measurement in Ref.~\cite{TOPQ-2021-08}. Additionally, a parton shower uncertainty is obtained similarly to the \ttll sample. %

\subsection{Non-prompt background uncertainties}

A normalisation uncertainty of \SI{30}{\percent} is applied to the simulation-based charge-flip templates in the $3\ell$ regions, based on the largest uncertainties seen in the data-driven estimate, and a conservative \SI{50}{\percent} uncertainty is assigned to the fake leptons that do not fall into the categories defined in Section~\ref{sec:eventcat}.

An uncertainty is applied to the estimate of the fake templates in the SRs, consisting of the difference between the distribution of the tight and loose selection (taking half of the difference between the event yield predicted by the tight and the loose selection in every bin and mirroring it).

Three uncertainties are considered for the data-driven charge-flip estimate: an uncertainty due to the limited event numbers of the charge-flip rate extraction samples, an uncertainty due to the choice of binning used to derived those rates and an uncertainty in the extrapolation from a $Z$ boson-enriched region to a \ttbar-enriched region.
%

%
%
%
%
%
%
%
%


%

%
%
%
%
%

%
\section{Results}
\label{sec:result}


Results are presented in two interpretations, one with a focus on
measurement of the SM \ttll process in the signal region, and one
using the EFT formalism to probe for signs of BSM physics.  In both of these
cases maximum-likelihood fits of the signal and background predictions
in the signal and control regions to the event yields in data are
performed.  The likelihood
function ${\cal L}(\vec{s},\vec{\lambda},\vec{\theta})$ is constructed
as a product of Poisson probability terms over all bins considered in
the measurement, and depends on the signal, $\vec{s}$, the
normalisation factors for several backgrounds (see
Section~\ref{sec:analysisstrategy}), $\vec{\lambda}$, and a set of
nuisance parameters (NP), $\vec{\theta}$, encoding systematic
uncertainties in the signal and background
expectations~\cite{Cowan:2010js,Cranmer:2012sba}. Systematic uncertainties can
impact the estimated signal and background rates, the migration of
events between categories, and the shape of the fitted
distributions. Both $\vec{s}$ and $\vec{\lambda}$ are treated as free
parameters in the likelihood fit. The NPs $\vec{\theta}$ allow
variations of the expectations for signal and background according to
the systematic uncertainties, subject to Gaussian-shape constraints in the
likelihood fit.  Their fitted values represent the deviations from the
nominal expectations that are needed to provide the best fit to the
data.  Statistical uncertainties in each bin due to the limited size
of the simulated event samples are taken into account with dedicated
parameters, using the Beeston--Barlow ``lite''
technique~\cite{Barlow:1993dm}.

\subsection{SM results}

Measurements are reported of the SM \ttll signal strength,
$\mu(\ttll)$, inclusive in lepton flavour and split into separate
electron and muon channels. The signal strength is defined as the
ratio of the \ttll yield in the SR measured in the data divided by the
predicted yield at detector level in the same SR.  These provide a
direct indication of the agreement of the data with the SM expectation
before moving on to consider an interpretation searching for BSM
contributions.
The percentage of simulated \ttll events passing the detector-level SR selection which have generated
$m(\ell^+\ell^-)>m_Z+10$~\GeV is approximately 60\%.

However, for BSM signals that contribute primarily at high mass the
cross-section for the full invariant mass range,
$m(\ell^+\ell^-)>m_Z+10$~\GeV, has limited use, therefore
model-independent visible cross-section upper limits for a BSM
contribution on top of SM \ttll are provided for the highest
$m(\ell^+\ell^-)$ bins ($>550$~\GeV) at the 95\% confidence level
(CL)~\cite{Cowan:2010js}. This is equivalent to fitting the highest
bin in each of the signal regions shown in
Figure~\ref{fig:RegionDefs:SRs}. The visible cross-section upper limit
is defined by dividing the upper limit on the number of signal events
in the high mass bin by the integrated luminosity.

The results are shown in Table~\ref{tab:Res:SM}.  The
measured signal strengths are all compatible with the SM with a
combined observed (expected) \ttll significance of $2.9\sigma$ ($2.8\sigma$). The combined observed (expected) 95\% CL cross-section
upper limit for the high mass region, $m(\ell^+\ell^-)>550$~\GeV, is 30.8~ab (30.5~ab).
For the $\ttbar{}e^{+}e^{-}$ channel the observed (expected) significance is  $1.4\sigma$ ($1.7\sigma$) and the 95\% CL cross-section
upper limit for the high mass region is 26.9~ab (25.4~ab).
For the \ttbar{}$\mu^{+}\mu^{-}$ channel the observed (expected) significance is  $2.5\sigma$ ($2.0\sigma$) and the 95\% CL cross-section
upper limit for the high mass region is 26.0~ab (27.5~ab).

\begin{table}[t!]
\caption{Results of the measurement of the SM \ttll signal strength and significance for the signal region with $m(\ell^+\ell^-)>m_Z+10$~\GeV. Also the expected and observed 95\% confidence level (CL) upper limits on a possible BSM signal for the very high mass region with $m(\ell^+\ell^-)>550$~\GeV are shown.}
\begin{center}
\renewcommand{\arraystretch}{1.5}
\begin{tabular}{l c c c c}
\hline
&       \multicolumn{2}{c}{$m(\ell^+\ell^-)>m_Z+10$~\GeV} &  \multicolumn{2}{c}{$m(\ell^+\ell^-)>550$~\GeV} \\
\hline
& $\mu(\ttll)$ & Significance & Exp. 95\% CL upper limit  & Obs. 95\% CL upper limit \\
\hline
\hline
\ttbar{}$e^{+}e^{-}$     & $0.8^{+0.6}_{-0.6}$ & $1.4\sigma$ & 25.4~ab & 26.9~ab \\
\ttbar{}$\mu^{+}\mu^{-}$ & $1.1^{+0.5}_{-0.5}$ & $2.5\sigma$ & 27.5~ab & 26.0~ab \\
\hline
\ttll                    & $1.0^{+0.4}_{-0.4}$ & $2.9\sigma$ & 30.5~ab & 30.8~ab \\
\hline
\end{tabular}
\end{center}

\label{tab:Res:SM}
\end{table}

\subsection{EFT interpretation fit}

The focus of EFT interpretation of the analysis is the measurement of the WCs of the
six four-fermion EFT operators outlined in Section~\ref{sec:intro}:
$O_{tl}$, $O_{te}$, $O_{Qe}$, $O_{Ql}^{-}$ ($O_{Ql}^{1}-O_{Ql}^{3}$),
$O_{leQt}^{1}$ and $O_{leQt}^{3}$. These are considered in flavour-inclusive and flavour-split scenarios, the latter denoted by 11 and 22
indices ($\ell^+\ell^- = ee,\mu\mu = 11,22$). Fits are performed
individually to a single operator at a time (other WCs fixed to
0). A dedicated flavour-relative fit configuration is used to test
for LFU-violating signals by testing the
difference between electron and muon WCs for each operator,
e.g. $\Delta(c_{tl(22)}-c_{tl(11)})$.
The primary focus is on results relating to the single operator fits due to a lack of sensitivity to multi-operator fits, which is explained in more detail in Section~\ref{sec:multiEFT}.

The effect of the EFT operators on the \ttll process is parameterised
bin by bin with a second order polynomial function parameterised by
the relative change in yield from each operator due to interference
with the SM at $\mathcal{O}(\Lambda^{-2})$ (linear) and pure EFT
contributions (quadratic, including cross-terms where relevant) at
$\mathcal{O}(\Lambda^{-4})$. The EFT contribution approximates higher
order QCD effects by scaling the SM EFT prediction at LO in QCD to the
SM prediction at NLO in QCD bin-by-bin. Once the EFT parameterisation
is obtained, a profile-likelihood fit is performed, where the
individual WCs are treated as free floating in the fit to data. Maximum-likelihood fits are performed with
the signal parameterised by EFT WCs for each of the scenarios
described.

The EFT contributions are only applied to the \ttll signal process,
sensitivity to \tZq was also investigated but found to be negligible
and is therefore neglected. EFT contributions are considered in the
on-shell \ttZ process but these contribute very little to the overall
sensitivity. This allows the overall \ttll normalisation, $\lambda_{\ttZ}$, to be controlled, primarily from the on-shell CRs, in the fit to data.

Figure~\ref{fig:RegionDefs:SRs} shows the signal regions after the fit
to data for the $c_{tl}$ WC (other WCs fixed to 0).  As expected from
the SM fit results, there are no significant deviations from the
expected background predictions after the fit to data, the fitted
normalisation factors are: $\lambda_{\ttZ}=1.09\pm 0.10$,
$\lambda_{\ttW}=1.18\pm 0.11$, $\lambda_{WZ}=0.92\pm 0.26$,
$\lambda_{\textrm{IntCo}}=1.04\pm 0.12$,
$\lambda_{\textrm{ExtCo}}=1.22\pm 0.15$,
$\lambda_{\textrm{HF}_{\mu}}=1.03\pm 0.11$,
$\lambda_{\textrm{HF}_{e}}=0.89\pm 0.19$,
$\lambda_{\textrm{LF}_{e}}=1.84\pm 0.82$. No significant changes in
the results for the background normalisation factors are observed in
the fits to other combinations of WCs. The \ttW normalisation factor
is higher than unity but compatible with measurements made
previously in ATLAS in a similar phase space~\cite{TOPQ-2019-30}.

\begin{figure}[!t]
\centering
\subfloat[]{\includegraphics[width=0.45\linewidth]{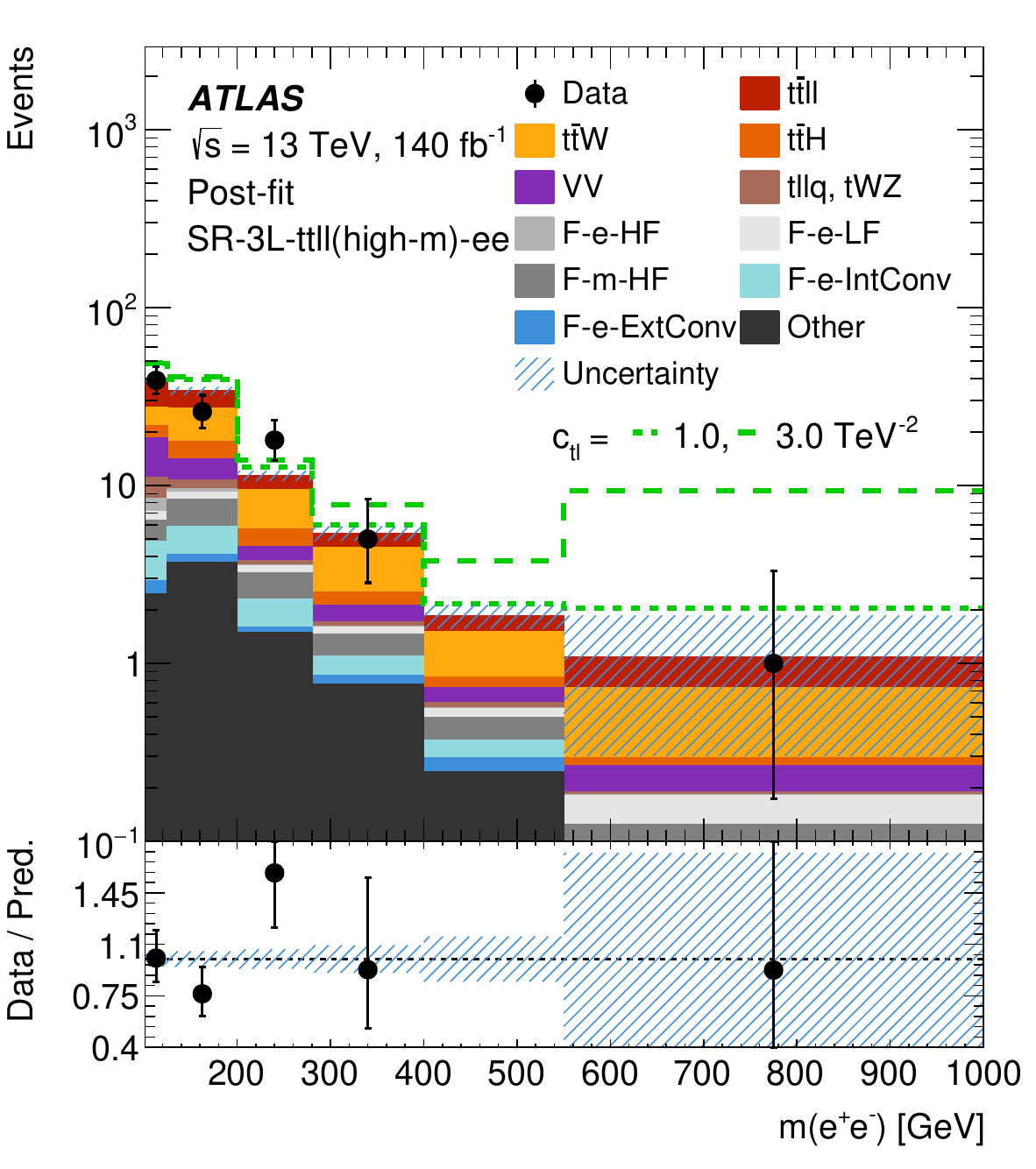}}
\subfloat[]{\includegraphics[width=0.45\linewidth]{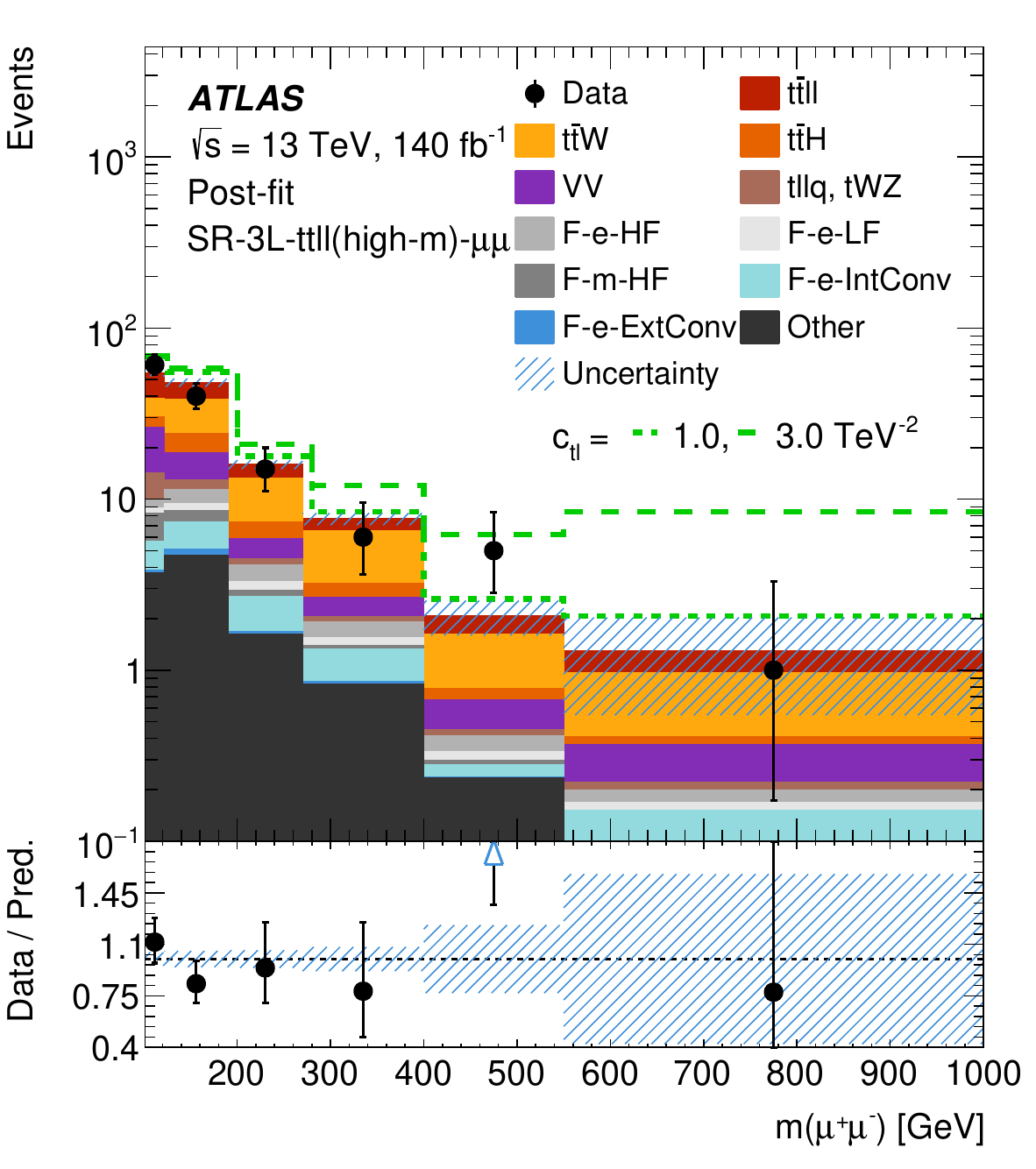}}
\caption{Distribution of (a) the \ttee and (b) the \ttmumu SRs binned in the dilepton invariant mass closest to the $Z$ mass after the fit to data where the $c_{tl}$ WC is the primary parameter of interest. The uncertainty bands include statistical and systematical uncertainties. In both figures the highest bin is inclusive and contains events above the bin edge. The green dotted and dashed lines show the total expectation in the presence of an EFT signal corresponding to $c_{tl} = 1.0$ and $c_{tl} = 3.0$ respectively. The triangle symbol in the ratio panel indicates that a data point is above the limit of the frame.}
\label{fig:RegionDefs:SRs}
\end{figure}

The data are in good agreement with the SM expectation within the
quoted uncertainty which is dominated by the limited size of the
data samples. For the fit to the $c_{tl}$ WC (other WCs
fixed to 0) the statistical uncertainty due to limited data set size
is $^{+1.63}_{-0.60}\ \TeV^{-2}$ compared with the total systematic uncertainty of $^{+0.41}_{-0.28}\ \TeV^{-2}$.
The largest systematic uncertainties
are due to the SM \ttll and \ttW modelling, and the limited
size of MC samples in some of the high $m(\ell^+\ell^-)$ bins.

The EFT WC values are fitted and further confirm compatibility
with the SM as reported in the following.

\subsection{EFT flavour-inclusive fit results}

The results for the various flavour-inclusive fit configurations
described above are shown in Tables~\ref{tab:Res:EFT_flavincl_linquad}
and~\ref{tab:Res:EFT_flavincl_lin} for cases where all EFT effects
are considered, $\mathcal{O}(\Lambda^{-4})$, and where only SM
interference effects of the EFT operators are considered, $\mathcal{O}(\Lambda^{-2})$, respectively. In the latter case the
$O_{leQt}^{1}$ and $O_{leQt}^{3}$ operators are not included as they
have no equivalent SM interaction so there is no interference with the SM and sensitivity only
arises from the pure EFT contribution.

Table~\ref{tab:Res:EFT_flavincl_linquad} shows the best fit values,
68\% and 95\% confidence intervals (CIs) from the fit to data
including $\mathcal{O}(\Lambda^{-4})$ EFT contributions. All the
fitted WC values are compatible with the SM. The dominant uncertainty
in all cases is due to the limited data size. Constraints on the
$O_{leQt}^{3}$ operator are tighter than for other operators, this is
expected due to the definition of this operator.
These limits improve upon previous
constraints on these operators from the CMS Collaboration~\cite{CMS-TOP-22-006}.

In the case of EFT paramerisations
where the (quadratic) pure-EFT $\mathcal{O}(\Lambda^{-4})$ terms are
dominant a breakdown of a key axiom of Wilks'
theorem~\cite{Wilks:1938dza} can occur, the profile-likelihood ratio (PLR) may not follow
the expected $\chi^2$-distribution~\cite{Bernlochner:2022oiw}. This can lead to coverage in the
test statistic different to what is expected in the asymptotic case. The true coverage can be determined with toy
pseudodata and adjusting where to read off the CI accordingly. Therefore, additionally
to the non-coverage adjusted results (Asymp. CI), results
are also given where the 68\% and 95\% CIs are coverage-adjusted
(Pseud. CI) for the fits to data where only a single WC at a
time is a free parameter in the fit. This adjustment is calculated by
generating toy pseudodata and taking the value of the 68\% and 95\% percentiles
of the fitted pseudodata test statistics. Scans of the fitted negative
log-likelihood value as a function of the WC value are shown in
Figure~\ref{fig:Results:coverage} for $c_{tl}$ and $c_{Qe}$. The solid
line shows the negative log-likelihood as a function of the WC, the crosses and plusses show the
values of the 68\% and 95\% coverage calculated from the fits to
pseudodata, compared with the coverage value in the asymptotic assumption of a profile likelihood ratio, $-\Delta\ln{L}=0.5$ and $-\Delta\ln{L}=1.96$. The confidence intervals calculated with the coverage from pseudodata are quoted by interpolating between the markers and giving the
intersection with the likelihood scan. The percentage value
written in text is the actual coverage value at the WC value
corresponding to the nominal $-\Delta\ln{L}=0.5$ and $-\Delta\ln{L}=1.96$ points.
The $\mathcal{O}(\Lambda^{-4})$
CIs calculated from pseudodata and the corresponding 95\% confidence level limits on the new physics scale, $\Lambda$, are also summarised in Figure~\ref{fig:Results:EFT_summary_incl}.

\begin{table}[t!]
\caption{Results of the fit to data for each WC fitted individually considering $\mathcal{O}(\Lambda^{-4})$ effects. The best fit value, 68\% and 95\% confidence intervals for the asymptotic assumption (Asymp. CI) and the coverage calculated from pseudodata (Pseud. CI) are shown.}
\begin{center}
\renewcommand{\arraystretch}{1.2}
\begin{tabular}{l c c c c c}
\hline
& \multicolumn{4}{c}{$\mathcal{O}(\Lambda^{-4}) [\TeV^{-2}]$} \\
& Best fit & 68\% Asymp. CI & 68\% Pseud. CI & 95\% Asymp. CI  & 95\% Pseud. CI\\
\hline     \hline
$c_{tl}$ & $-0.44$ & $[-1.10,1.24]$ &  $[-1.17,1.31]$ & $[-1.65,1.80]$ & $[-1.71,1.86]$\\
$c_{te}$ &  $-0.45$ & $[-1.05,1.18]$ & $[-1.21,1.41]$ & $[-1.57,1.79]$& $[-1.64,1.96]$\\
$c_{Qe}$ &  $-0.54$ & $[-1.23,1.19]$ & $[-1.32,1.32]$ & $[-1.82,1.83]$ & $[-1.87,1.88]$ \\
$c_{Ql}^{-}$ & $0.69$ & $[-0.98,1.41]$ & $[-1.15,1.74]$ & $[-1.57,2.03]$ & $[-1.66,2.33]$ \\
$c_{leQt}^{1}$ & $-0.61$ & $[-1.51,1.50]$ & $[-1.55,1.54]$ & $[-2.27,2.27]$ & $[-2.32,2.33]$\\
$c_{leQt}^{3}$ & $-0.12$ & $[-0.26,0.26]$ & $[-0.27,0.27]$ & $[-0.37,0.37]$ & $[-0.38,0.38]$\\
\hline
\end{tabular} \\
\end{center}
\label{tab:Res:EFT_flavincl_linquad}
\end{table}

\begin{table}[b!]
\caption{Results of the fit to data for each WC fitted individually considering only $\mathcal{O}(\Lambda^{-2})$ effects. The best fit value and 68\% and 95\% confidence intervals from the asymptotic assumption (Asymp.) are shown.}
\begin{center}
\renewcommand{\arraystretch}{1.2}
\begin{tabular}{l c c c}
\hline
& \multicolumn{3}{c}{$\mathcal{O}(\Lambda^{-2}) [\TeV^{-2}]$}\\
& Best fit value & 68\% Asymp. CI  & 95\% Asymp. CI \\
\hline     \hline
$c_{tl}$ &  $-1.21$ & $[-8.23,4.22]$ & $[-11.83,8.62]$ \\
$c_{te}$ &  $-1.42$ & $[-4.45,1.09]$ & $[-7.94,3.10]$ \\
$c_{Qe}$ &  $-11.65$ & $[-25.75,0.98]$ & $[-40.97,11.53]$ \\
$c_{Ql}^{-}$ &  $-0.16$ & $[-2.66,1.91]$ & $[-5.58,3.67]$ \\
\hline
\end{tabular}
\end{center}

\label{tab:Res:EFT_flavincl_lin}
\end{table}

\begin{figure}[!t]
\centering
\subfloat[]{\includegraphics[width=0.48\linewidth]{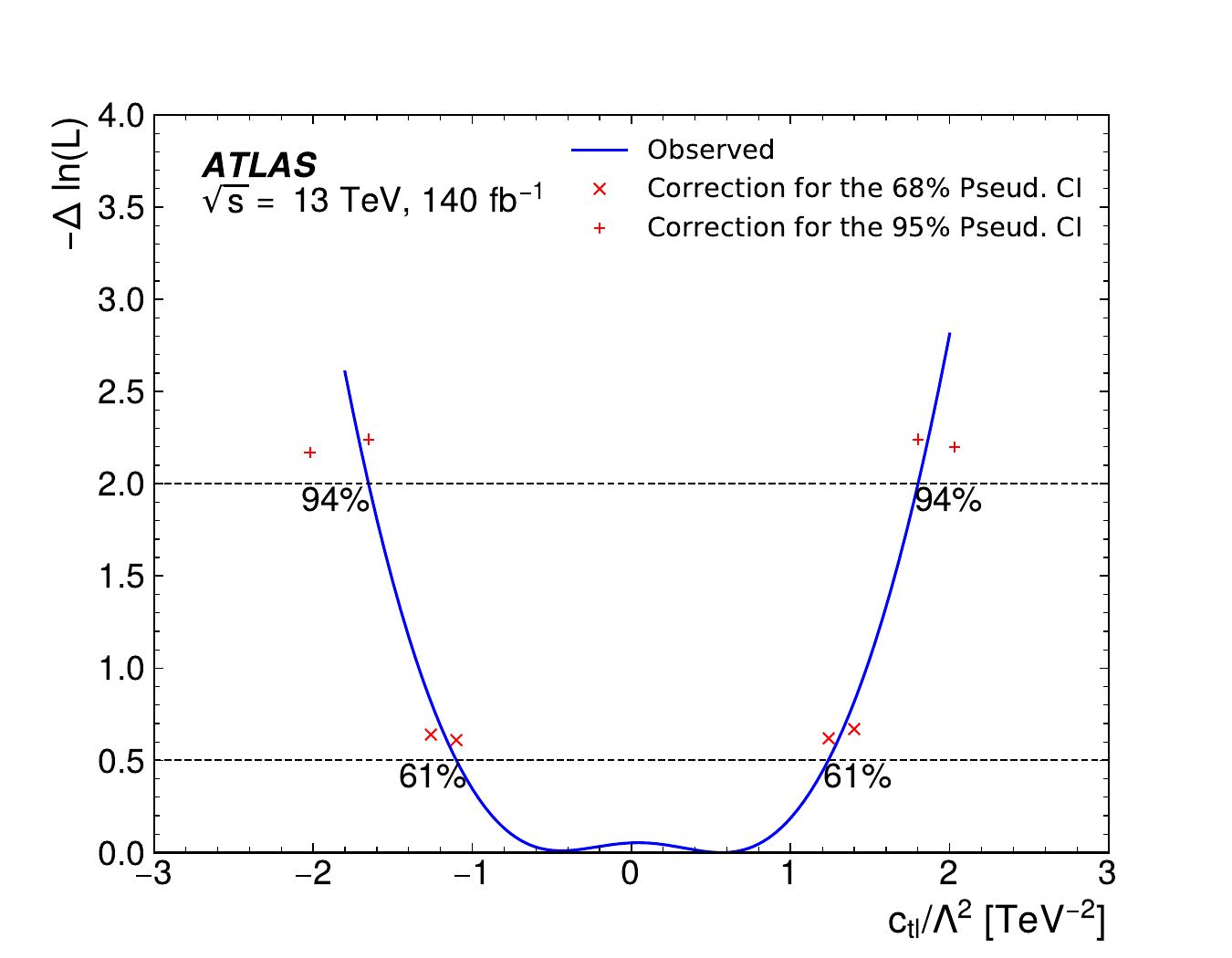}}
\subfloat[]{\includegraphics[width=0.48\linewidth]{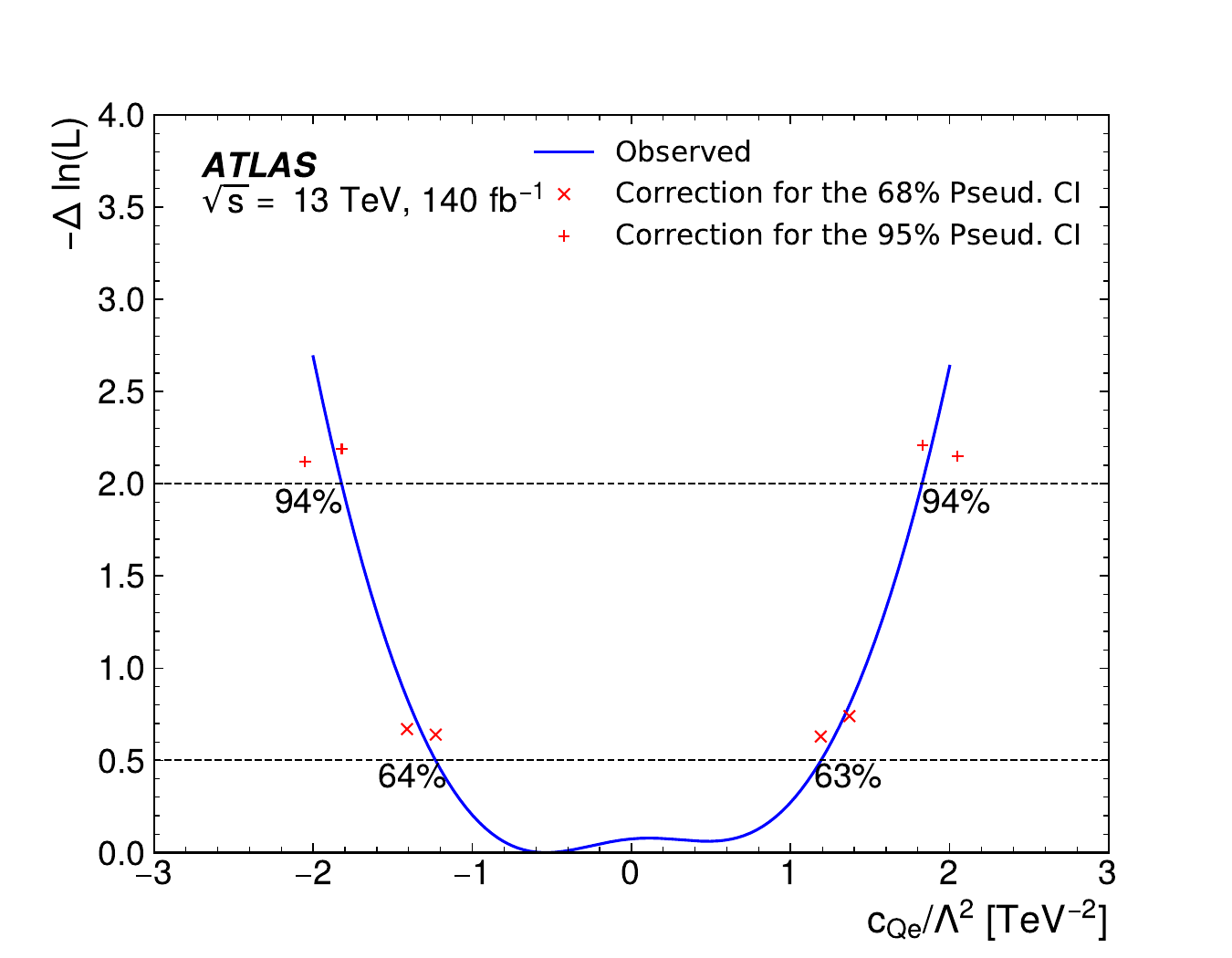}} \\
\caption{Scans of the fitted negative log-likelihood value as a function of the WC value for (a) $c_{tl}$ and (b) $c_{Qe}$. The solid line shows the standard profile likelihood ratio approach, the cross and plus markers show the values of the 68\% and 95\% coverage, respectively, calculated from the fits to pseudodata (Pseud.).}
\label{fig:Results:coverage}
\end{figure}

Table~\ref{tab:Res:EFT_flavincl_lin} shows results of the fit to data
considering only $\mathcal{O}(\Lambda^{-2})$ EFT--SM interference
contributions. As expected given that no significant deviation from
the expectation is observed in the data, the fitted WC values are
compatible with the SM. There is only limited sensitivity to this
contribution alone, as the energy growth that gives the best
sensitivity at high mass comes primarily from
$\mathcal{O}(\Lambda^{-4})$ contributions.

\begin{figure}[!htbp]
\centering
\subfloat[]{\includegraphics[width=0.49\linewidth]{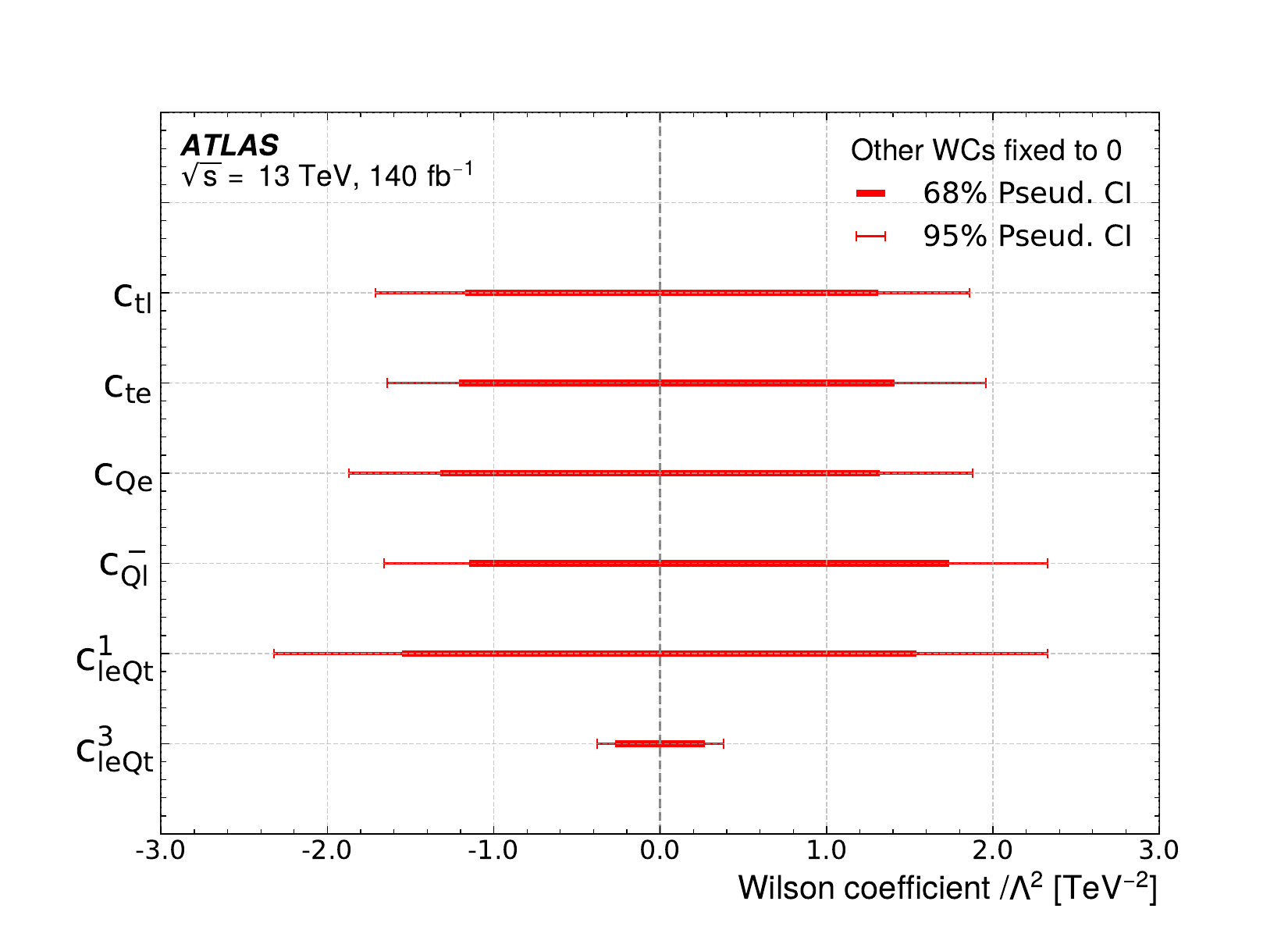}}
\subfloat[]{\includegraphics[width=0.49\linewidth]{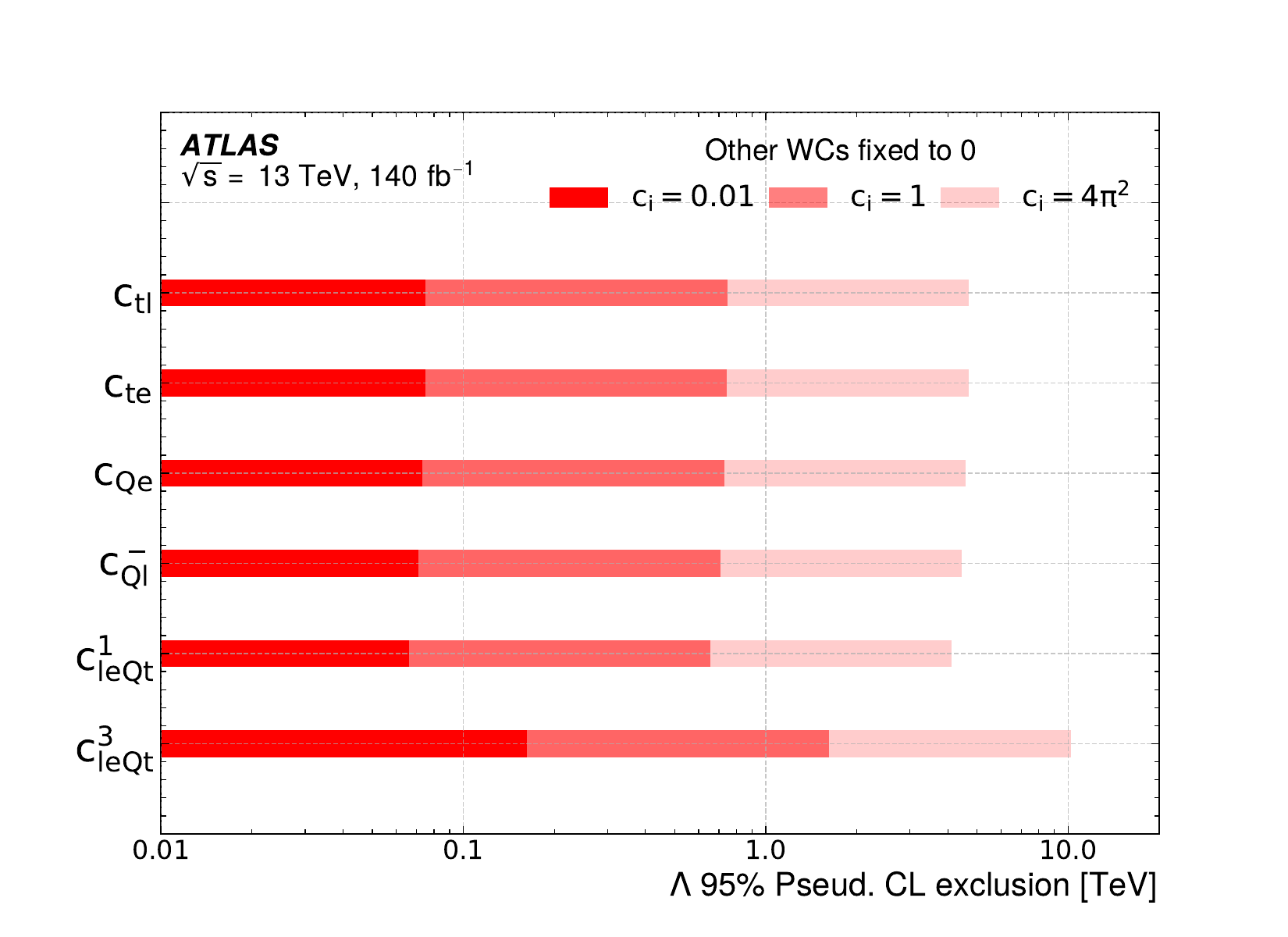}}
\caption{Confidence intervals on the WCs for a fit to data with only one WC constrained at a time for (a) the flavour-inclusive WCs and (b) the corresponding 95\% exclusion limits on new physics scale $\Lambda$ for different assumptions on the value of the (dimensionless) coupling. The results have the coverage calculated from pseudodata (Pseud.).}
\label{fig:Results:EFT_summary_incl}
\end{figure}

It is important to consider the range of EFT validity for measurements which are
sensitive to the pure EFT contribution. As suggested by the LHC EFT
Working Group~\cite{Brivio:2022pyi} clipping can be applied by
removing events above a particular $m(\ell^+\ell^-)$
value. Figure~\ref{fig:Results:ctl_mllcutoff} shows the results for
the $c_{tl}$ and $c_{Qe}$ WCs as a function of different values of the
high \mll cut-off used in the analysis. The clipping values are chosen
to be \SI{550}{\giga\eV}, \SI{800}{\giga\eV} and \SI{1}{\tera\eV}. It
can be seen that the WC contraints degrade, as expected, when lower
cut-offs are applied but that they are rather stable for \SI{1}{\tera\eV}
and above. This gives confidence that the nominal result without
clipping can be trusted and does not suffer from issues
related to the region of validity of the EFT model.  Therefore, the
default EFT results are obtained without applying any \mll cut-off.

\begin{figure}[!t]
\centering
\subfloat[]{\includegraphics[width=0.48\linewidth]{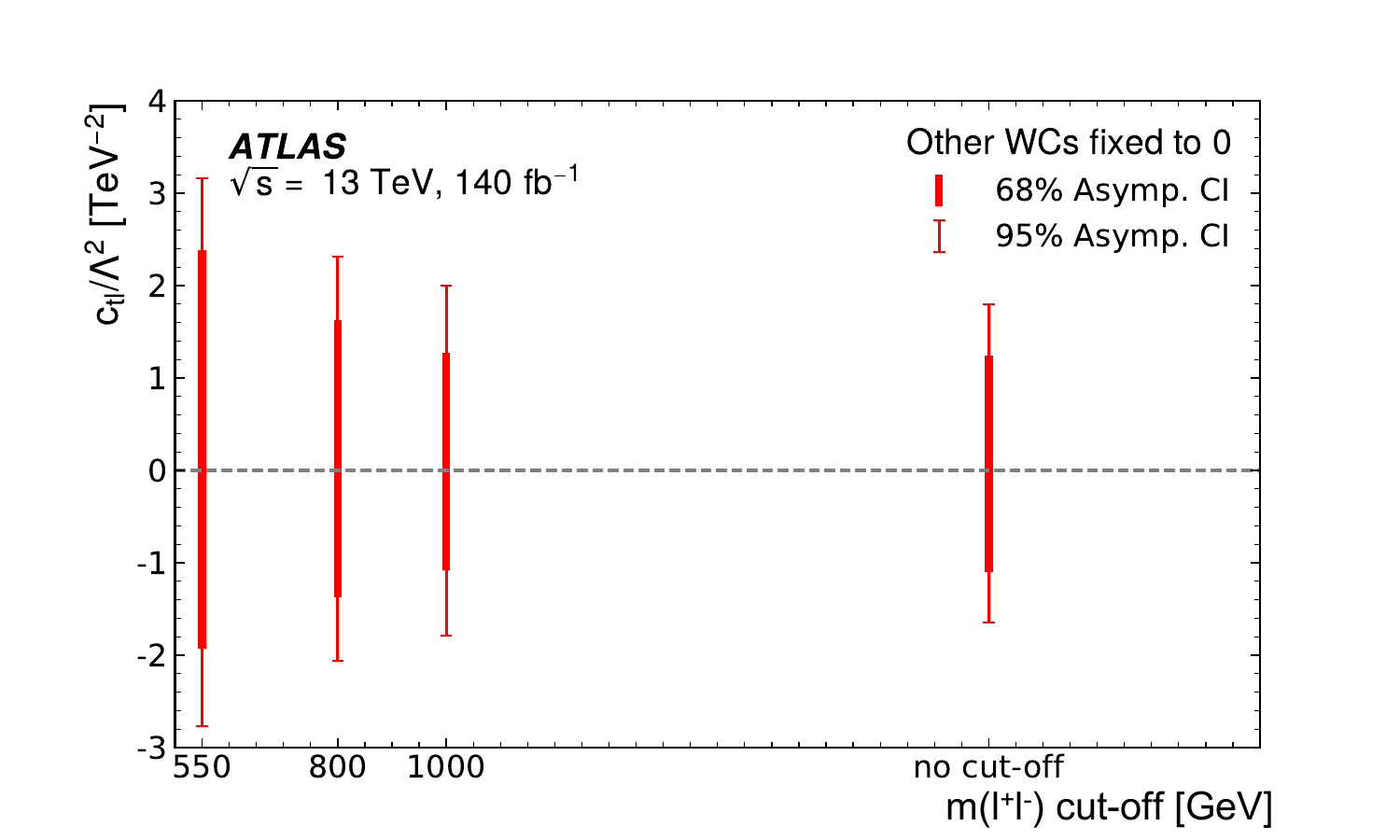}}
\subfloat[]{\includegraphics[width=0.48\linewidth]{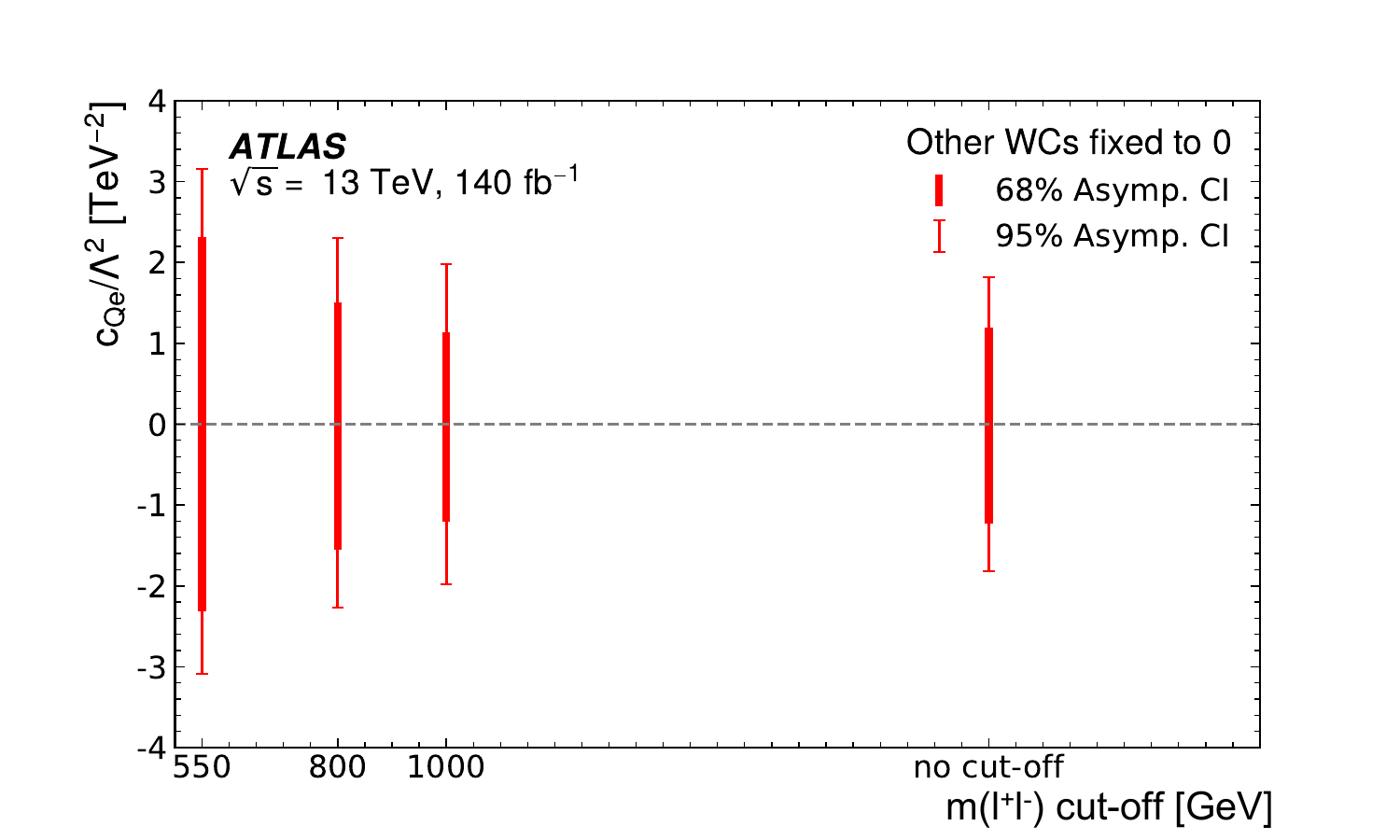}}
\caption{The confidence intervals for (a) $c_{tl}$ and (b) $c_{Qe}$ from fits to data for different values of the \mll upper cut-off. The results use the asymptotic assumption to calculate the intervals (Asymp.).}
\label{fig:Results:ctl_mllcutoff}
\end{figure}

\subsection{EFT flavour-separated fit results}

Table~\ref{tab:Res:EFT_flavsep} shows the results of the
flavour-separated EFT WC fits to data.  Due to the limited sensitivity
of the $\mathcal{O}(\Lambda^{-2})$ contribution alone results are considered only with full EFT $\mathcal{O}(\Lambda^{-4})$ effects taken into
account.
The sensitivity is diluted relative to
the equivalent flavour-inclusive results, as expected due to the
additional degrees of freedom in the fit and reduction in number of
sensitive bins per WC, but significant constraining power still
remains due to the good lepton flavour identification efficiency of
ATLAS.  Again, all WCs are compatible with the SM, as expected from the
good agreement between data and expectation for each flavour-split
signal region.  These results are also summarised in
Figure~\ref{fig:Results:EFT_summary_flav}.

\begin{table}[t!]
\caption{Fit results for the considered EFT flavour specific WCs. Results are shown for the 68\% and 96\% confidence intervals for the asymptotic assumption (Asymp.) and the coverage calculated from pseudodata (Pseud.). In all cases only one EFT operator is considered at a time but both of the corresponding flavour-dependent WCs are constrained at the same time.}
\begin{center}
\renewcommand{\arraystretch}{1.2}
\begin{tabular}{l c c c c c}
\hline
& \multicolumn{4}{c}{$\mathcal{O}(\Lambda^{-4}) [\TeV^{-2}]$} \\
& Best fit & 68\% Asymp. CI & 68\% Pseud. CI & 95\% Asymp. CI & 95\% Pseud. CI\\
\hline\hline
$c_{tl(11)}$ & $0.22$ &  $[-1.10,1.24]$  & $[-1.20,1.36]$ & $[-2.23,2.37]$ & $[-2.31,2.52]$\\
$c_{tl(22)}$ & $-0.81$ &  $[-1.57,1.79]$  & $[-1.64,1.85]$ & $[-2.25,2.49]$ &  $[-2.28,2.51]$ \\
\hline
$c_{te(11)}$ & $-0.24$ &  $[-1.19,1.21]$   & $[-1.44,1.55]$ & $[-2.32,2.47]$ & $[-2.48,3.10]$ \\
$c_{te(22)}$ & $-0.67$ &  $[-1.40,1.79]$   & $[-1.56,2.07]$ & $[-2.06,2.52]$ & $[-2.09,2.66]$\\
\hline
$c_{Qe(11)}$ & $-0.24$ &  $[-1.28,1.22]$   & $[-1.40,1.31]$ & $[-2.52,2.45]$ & $[-2.61,2.58]$\\
$c_{Qe(22)}$ & $-0.89$ &  $[-1.72,1.77]$   & $[-1.76,1.84]$ & $[-2.46,2.53]$ & $[-2.48,2.55]$\\
\hline
$c_{Ql(11)}^{-}$ & $0.48$ &  $[-1.00,1.49]$   & $[-1.30,2.09]$ & $[-2.16,2.81]$ & $[-2.30,3.43]$\\
$c_{Ql(22)}^{-}$ & $-0.66$ &  $[-1.46,1.99]$  & $[-1.65,2.44]$ & $[-2.17,2.78]$ & $[-2.45,3.10]$\\
\hline
$c_{leQt(11)}^{1}$ & $-0.22$ &  $[-1.57,1.58]$   & $[-1.67,1.70]$ & $[-3.13,3.13]$ & $[-3.12,3.12]$\\
$c_{leQt(22)}^{1}$ & $-1.08$ &  $[-2.21,2.20]$   & $[-2.27,2.26]$ & $[-3.22,3.22]$ & $[-3.21,3.21]$\\
\hline
$c_{leQt(11)}^{3}$  & $0.0$ &  $[-0.26,0.26]$   & $[-0.28,0.27]$ & $[-0.50,0.50]$     & $[-0.50,0.50]$\\
$c_{leQt(22)}^{3}$ & $-0.20$ &  $[-0.35,0.35]$   & $[-0.36,0.35]$ & $[-0.48,0.48]$ & $[-0.48,0.48]$\\
\hline
\end{tabular}
\end{center}

\label{tab:Res:EFT_flavsep}
\end{table}

\begin{figure}[t!]
\centering
\subfloat[]{\includegraphics[width=0.49\linewidth]{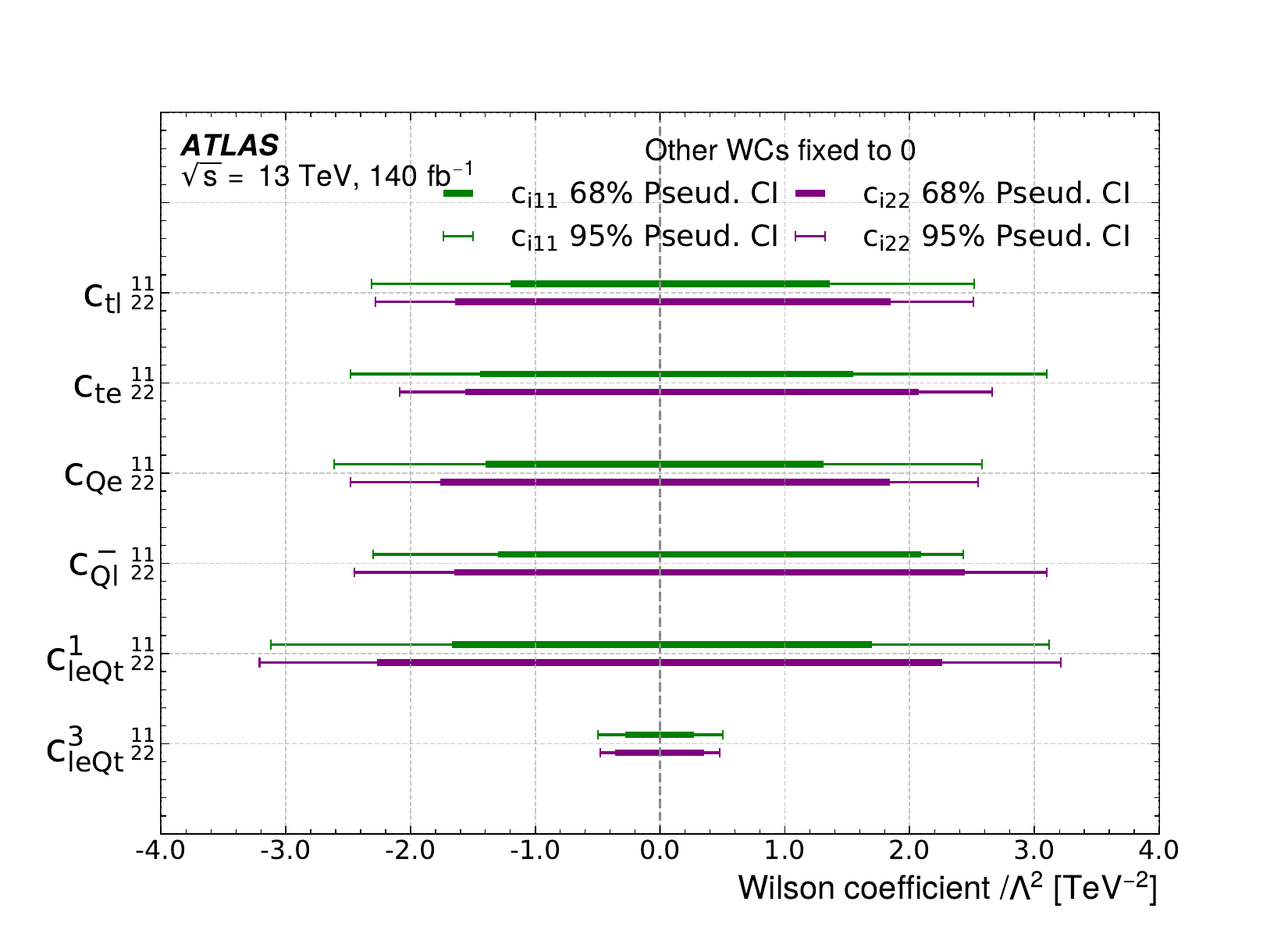}}
\subfloat[]{\includegraphics[width=0.49\linewidth]{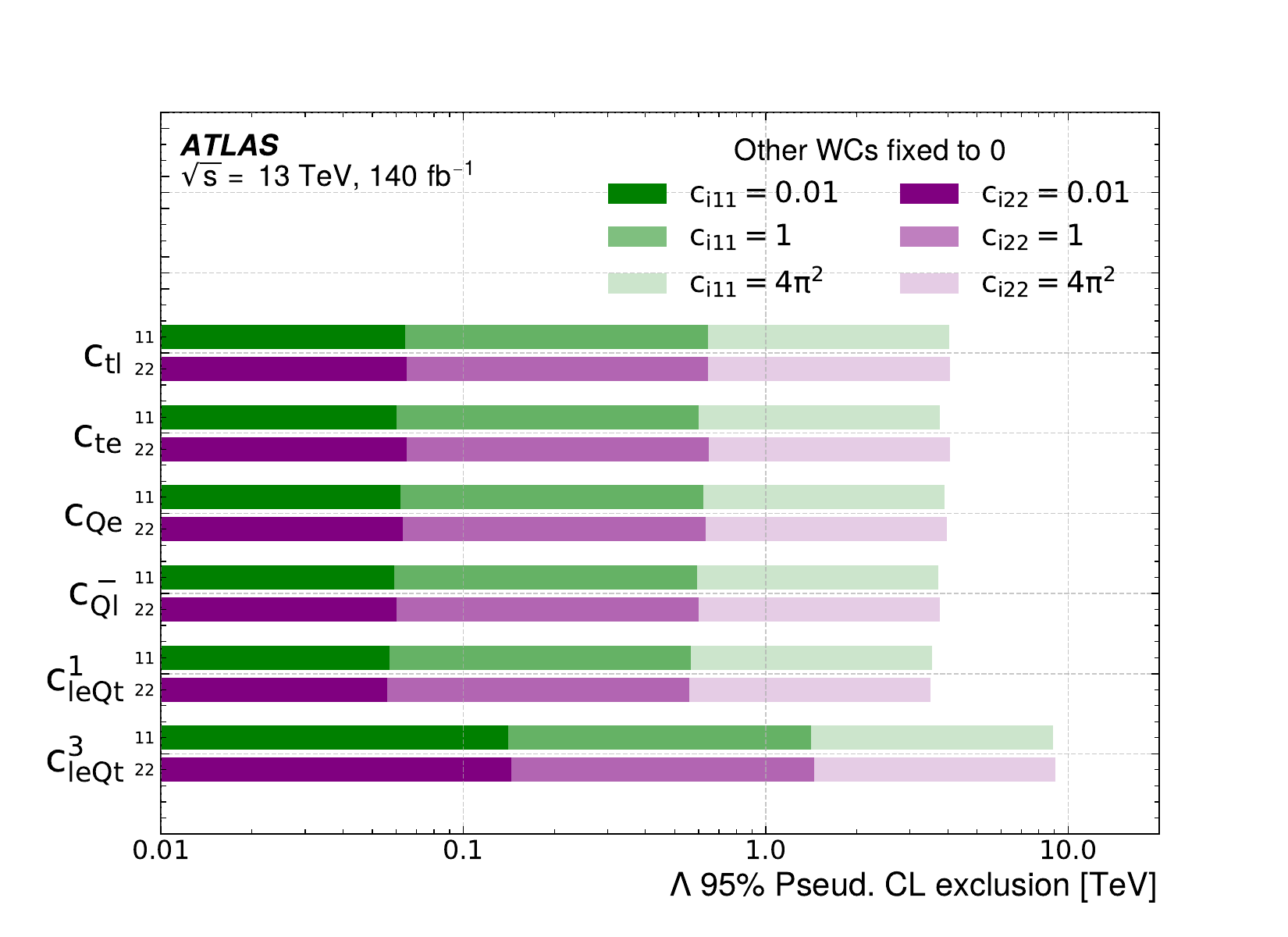}}\\
\subfloat[]{\includegraphics[width=0.49\linewidth]{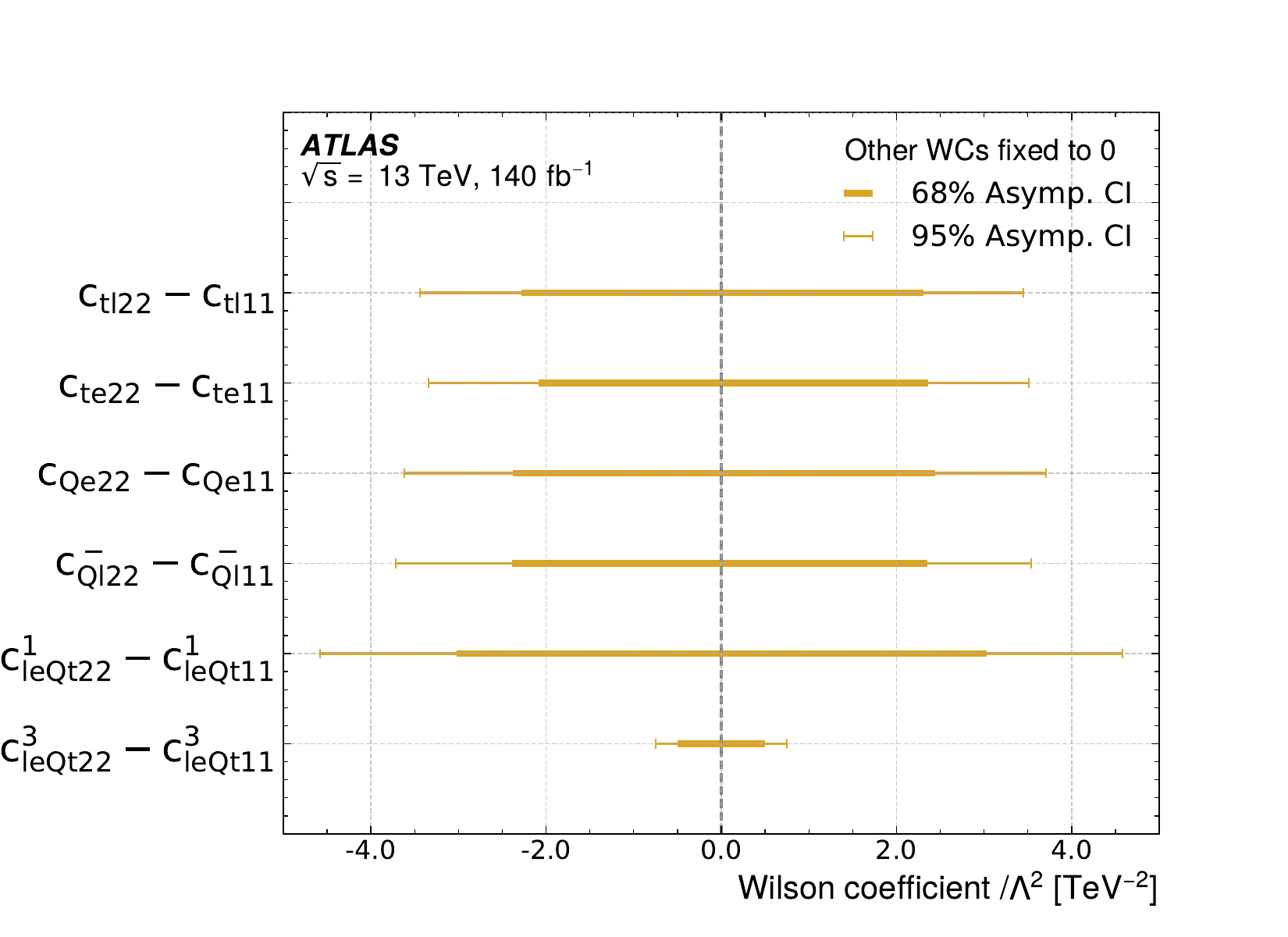}}
\caption{Confidence intervals on the WCs for a fit to data with only one WC constrained at a time for (a) the flavour-dependent WCs and (b) the corresponding 95\% exclusion limits on new physics scale $\Lambda$ for different assumptions on the value of the (dimensionless) coupling. Both of these results account for the coverage calculated from pseudodata (Pseud.). In addition (c) limits on the difference between flavour dependent WCs per operator are shown. These results use the asymptotic assumption to calculate the intervals (Asymp.). }
\label{fig:Results:EFT_summary_flav}
\end{figure}

\subsection{EFT lepton flavour universality violation sensitive fits}

Flavour-relative fits to data are also performed in the configuration
targetting LFU-violation in the EFT effects.  This is captured by the
difference between the flavour-split WCs for different operators,
$O_X$, defined as $\Delta(c_{[X](22)}-c_{[X](11)})$, that are probed
to search for EFT effects that apply preferentially on either electron
or muon couplings to the top quark. In these fits the difference
between flavours is a free parameter but apply relative to a
single-flavour reference WC. A convention is chosen to make the
electron flavour WC, $c_{[X](11)}$, the reference, so this is
additionally a free parameter in the
fit. Table~\ref{tab:Results:EFT_flavratio} shows the results of the
fits to data, in the case $\mathcal{O}(\Lambda^{-4})$ terms are
considered and only single operators, $O_X$, have
WCs included in the fit at a time (other WCs fixed to 0). These
results are also summarised in
Figure~\ref{fig:Results:EFT_summary_flav}.

There is potential for systematic uncertainties to cancel out in the
$\Delta(c_{[X](22)}-c_{[X](11)})$, but this analysis is now limited by
the available Run 2 data sample size, so it is observed that the
sensitivity to the flavour-split WC difference is lower than that of
the individual flavour-specific WCs.  Nonetheless, this marks the
first search for LFU-violation for EFT effects using this final state
and cancellation of systematic uncertainties is observed which will
become valuable with a larger data sample size. For example,
systematic uncertainties in backgrounds common to both
$\ttbar{}e^{+}e^{-}$ and $\ttbar{}\mu^{+}\mu^{-}$ SRs, such as the
\ttW and fake lepton backgrounds, are reduced while uncertainties relating
specifically one lepton flavour, such as those in the electron
isolation SFs, become of higher relative importance.

As with the other results, all
fitted WC values are in good agreement with the SM, both for the
flavour-split results and for the different WCs between flavours.  In
most cases the central value of the fitted
$\Delta(c_{[X](22)}-c_{[X](11)})$ value corresponds to the calculation based on the results in Table~\ref{tab:Res:EFT_flavsep},
however for $c_{tl}$ and $c_{leQt(11)}^{1}$ this is not the case. This is due to the
double-minimum structure as can be seen in
Figure~\ref{fig:Results:coverage}(a), where in this case an
alternative minimum is preferred. Overall, the confidence intervals
are the more meaningful quantities when interpreting the
flavour-relative results.

\begin{table}[t!]
\caption{Results of the fit to data of the difference between the flavour-specific WCs. The results use the asymptotic assumption to calculate the intervals (Asymp. CI).}
\begin{center}
\renewcommand{\arraystretch}{1.2}
\begin{tabular}{l c c c}
\hline
& \multicolumn{3}{c}{$\mathcal{O}(\Lambda^{-4}) [\TeV^{-2}]$}                                \\

& Best fit & 68\% Asymp. CI  & 95\% Asymp. CI\\
\hline    \hline
$c_{tl(22)} - c_{tl(11)}$ & $\;\;0.79$ &  $[-2.28,2.32]$ &  $[-3.44,3.45]$ \\
\hline
$c_{te(22)} - c_{te(11)}$ & $-0.43$ &  $[-2.10,2.36]$ &  $[-3.34,3.51]$\\
\hline
$c_{Qe(22)} - c_{Qe(11)}$ & $-0.66$ &  $[-2.39,2.43]$ &  $[-3.62,3.71]$\\
\hline
$c_{Ql(22)}^{-} - c_{Ql(11)}^{-}$ &  $-1.14$ &  $[-2.38,2.37]$&  $[-3.72,3.54]$\\
\hline
$c_{leQt(22)}^{1} - c_{leQt(11)}^{1}$ & $-1.32$ &  $[-3.03,3.03]$ &  $[-4.58,4.58]$\\
\hline
$c_{leQt(22)}^{3} - c_{leQt(11)}^{3}$  & $-0.20$ &  $[-0.50,0.50]$ &  $[-0.75,0.75]$\\
\hline
\end{tabular}
\end{center}

\label{tab:Results:EFT_flavratio}
\end{table}

\subsection{Discussion of multi-operator EFT fit sensitivity}
\label{sec:multiEFT}

Despite there being differences in behaviour as a function of
$m(\ell^{+}\ell^{-})$ for the different EFT operators of interest it is
challenging to simultaneously constrain all six operators of
interest. This is expected given that sensitivity comes from only a
limited number of high-mass bins.  The behaviour of the multi-operator
fits is non-trivial as correlations are hidden in the case that the WC
values are close to zero, as happens when calculating expected
sensitivity with Asimov fits. Therefore, many tests are
performed with and without EFT signal injection to show how the
expected sensitivity can decrease when a signal is present and the
fit is exposed to the possible correlations.
Figure~\ref{fig:Results:2D_vs_3D_inj} demonstrates the lack of an
abilitity to probe all operators at once by showing results in two
scenarios, one without any signal injection i.e. all WCs with a
nominal value of zero (Figure~\ref{fig:Results:2D_vs_3D_inj} top), and
another where a signal is injected with $c_{tl}=1.53$, chosen as it is
close to the 95\% confidence interval limit
(Figure~\ref{fig:Results:2D_vs_3D_inj} bottom). These results are
derived with only statistical uncertainties considered in the fits to
the pseudodata as this is the dominant uncertainty component and the
inclusion of full systematic uncertainties is very computationally
demanding.  The distributions considered in each case are the fitted
pseudodata (a), 1D likelihood scans (b), 2D
likelihood scan for a 2D fit (c) and 2D likelihood scan for
a 3D fit (d). These results show that although it can seem
that there is minimal correlation between WCs in a multi-operator
Asimov fit with no signal injection, this is misleading. There cannot be anticorrelations
between WCs with large quadratic component to accomodate low observed
signal, as changes to the WCs can only increase the signal yield, not decrease
it. However, if a sizeable signal were to be observed then the
correlation is much more strongly manifested as intermediate values of the
WCs between zero and the injected value can be absorbed by smaller
non-zero values of multiple WCs. This is illustrated in 1D in figure
Figure~\ref{fig:Results:2D_vs_3D_inj} (b-bottom) and in 2D
for Figure~\ref{fig:Results:2D_vs_3D_inj} (d-bottom). It is these results that motivate not including multi-operator
fits in the results even though there is sensitivity when the observed
EFT signal is small.

\begin{figure}[t!]
\centering
\includegraphics[width=0.24\linewidth]{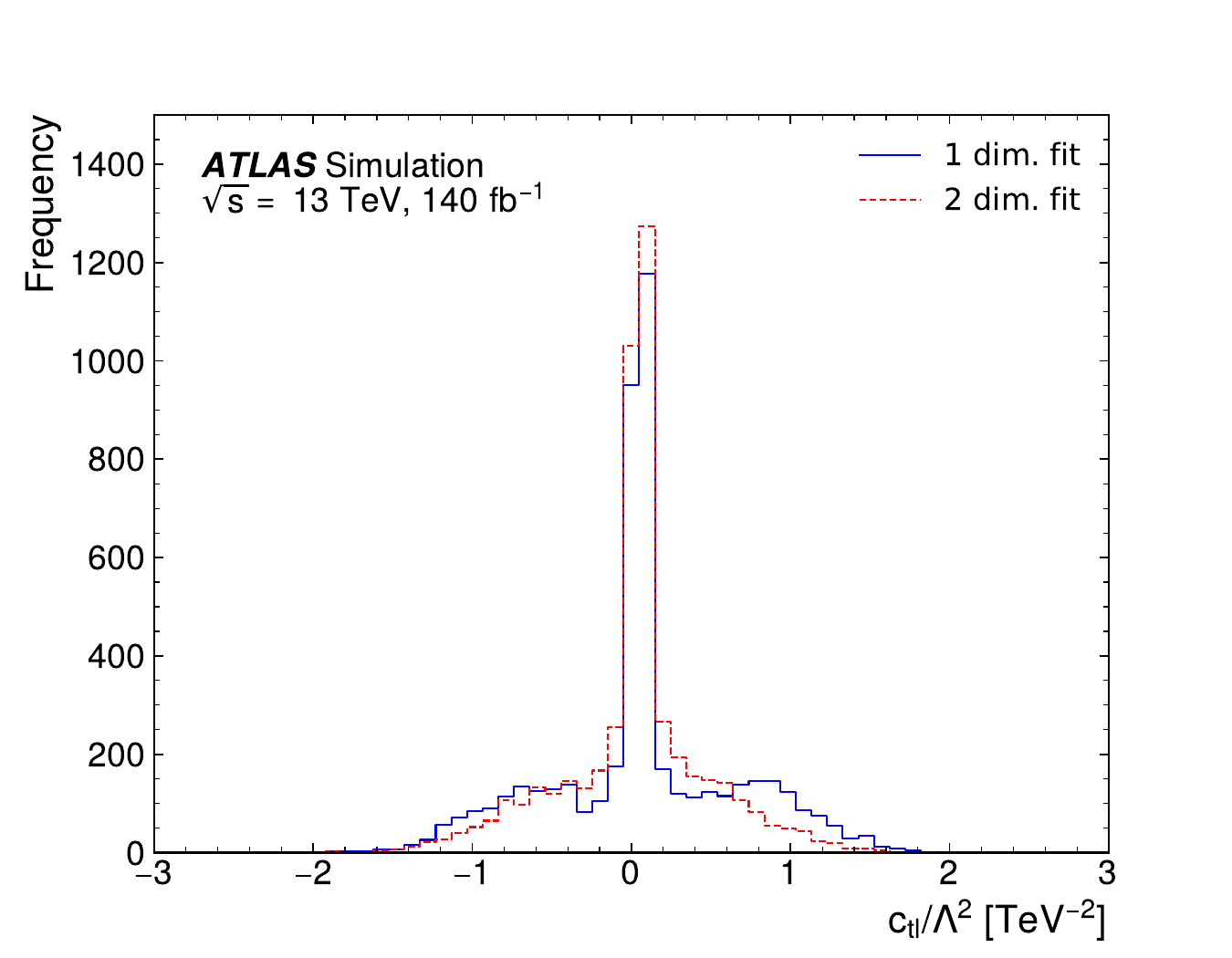}
\includegraphics[width=0.24\linewidth]{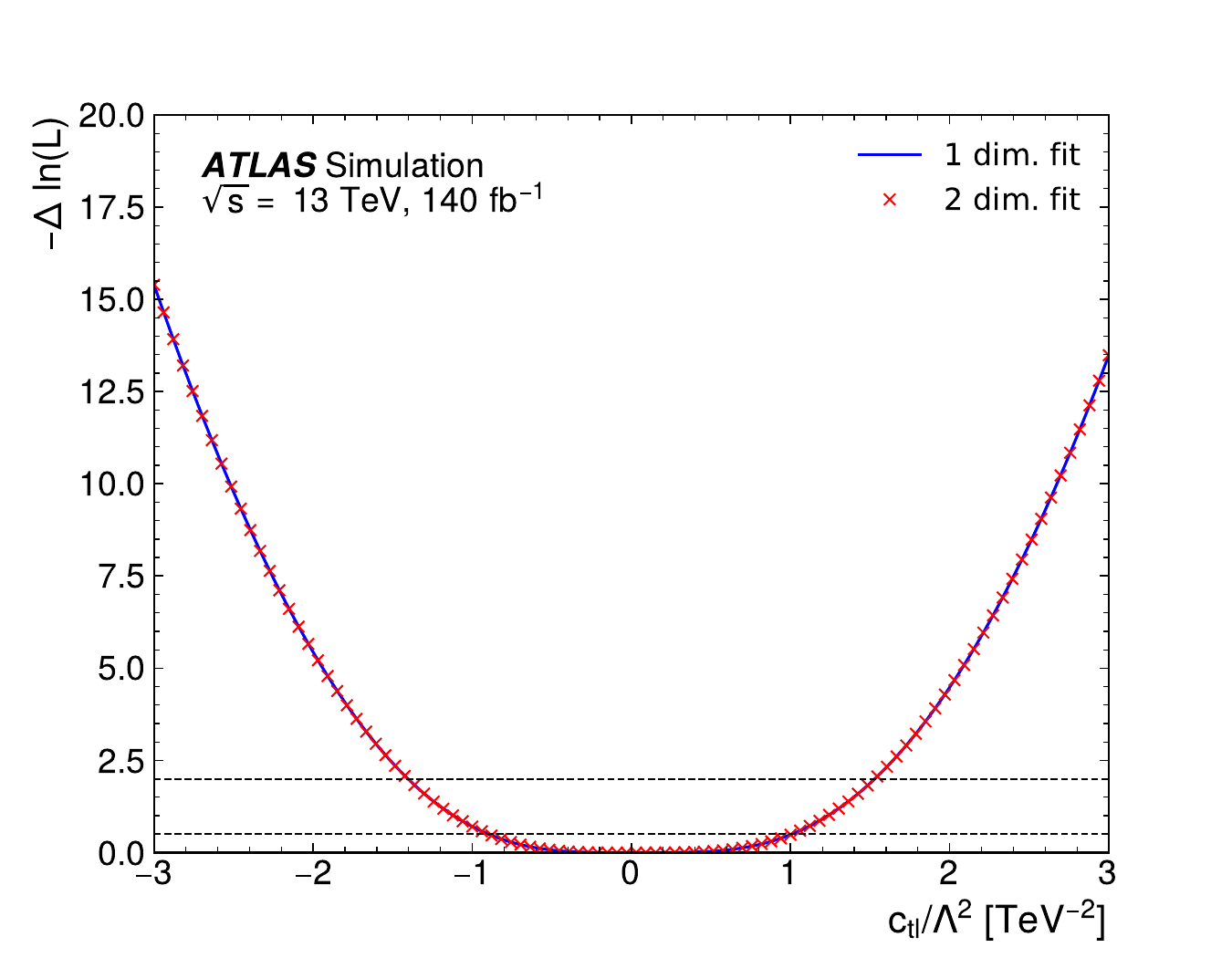}
\includegraphics[width=0.24\linewidth]{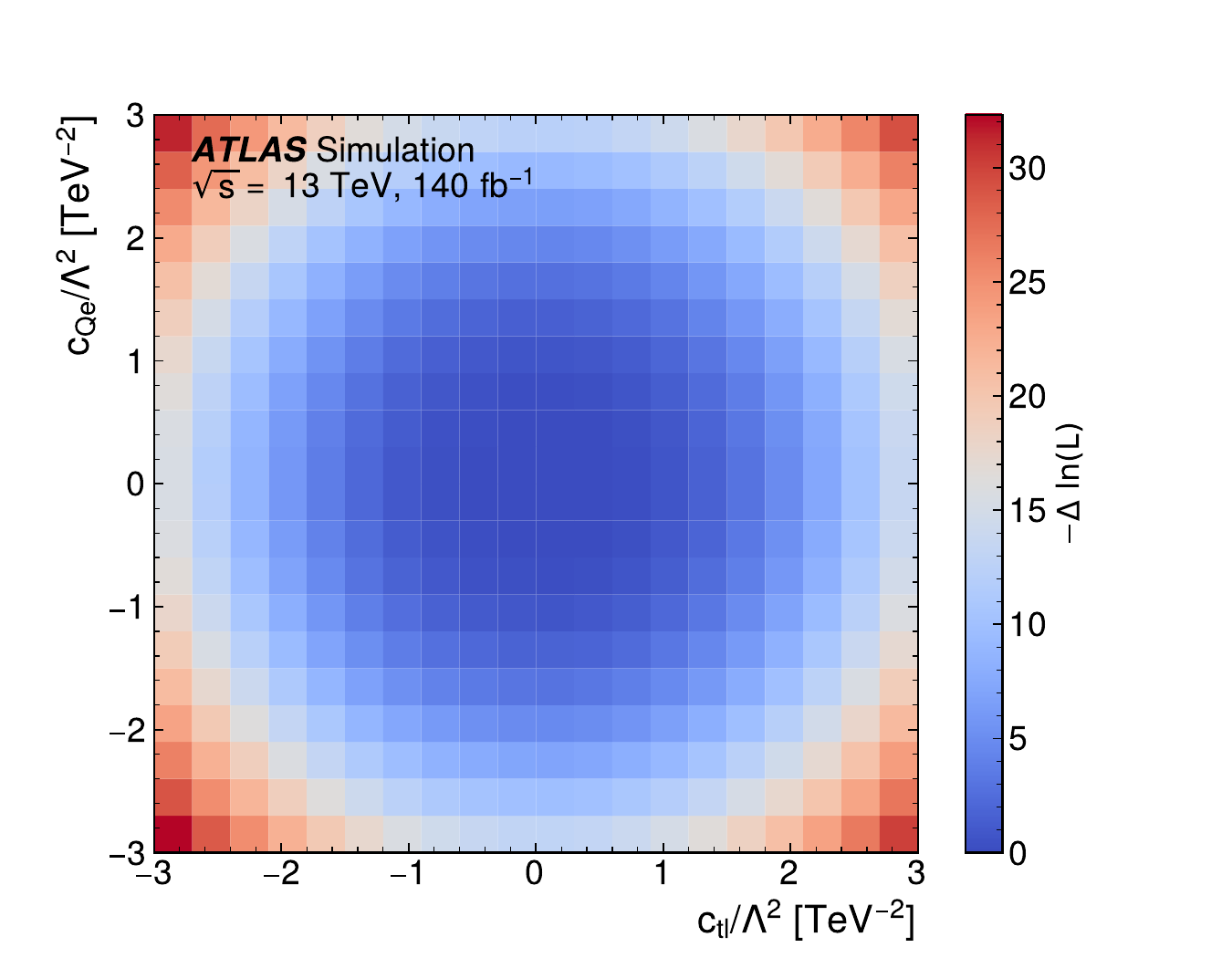}
\includegraphics[width=0.24\linewidth]{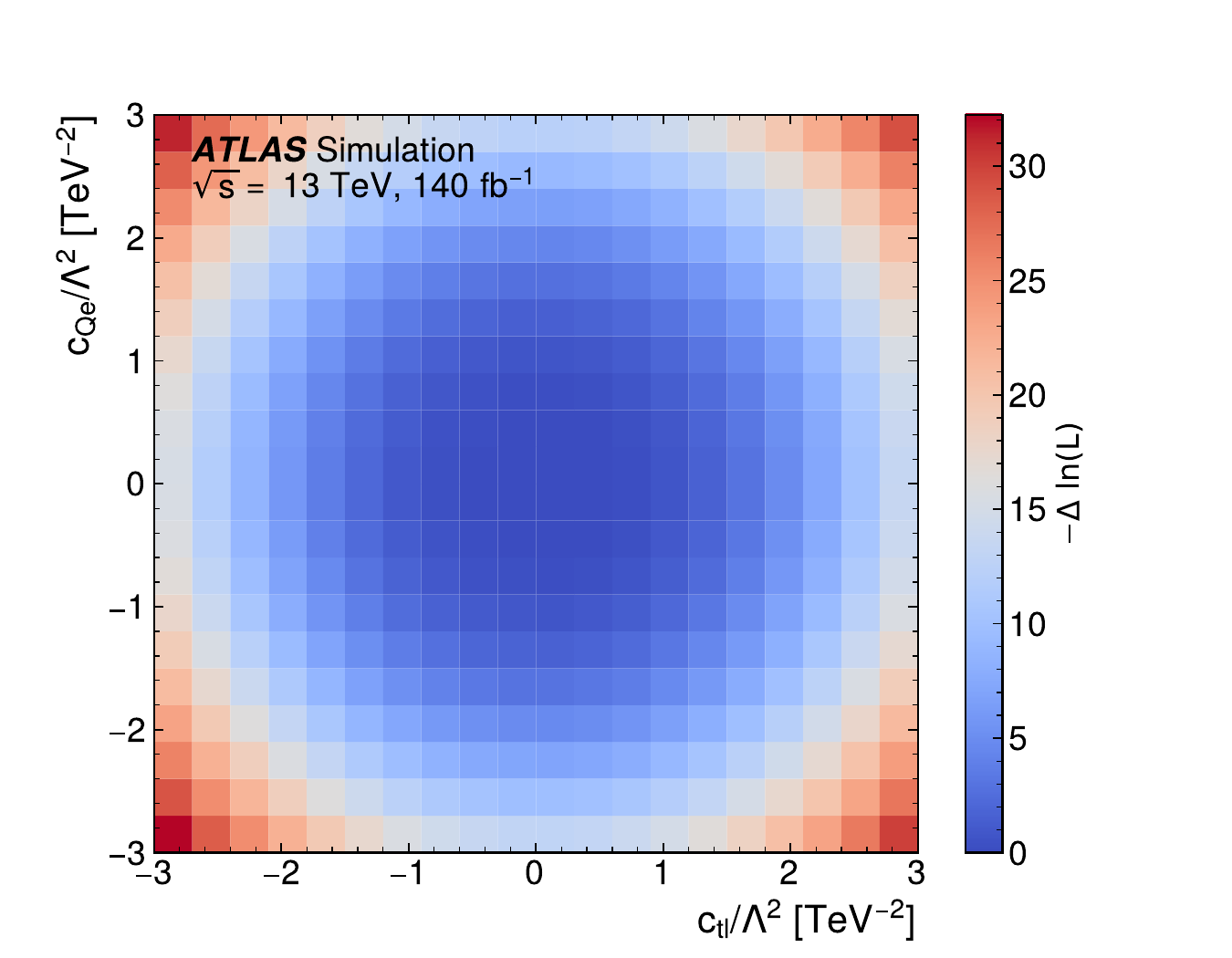} \\
\subfloat[]{\includegraphics[width=0.24\linewidth]{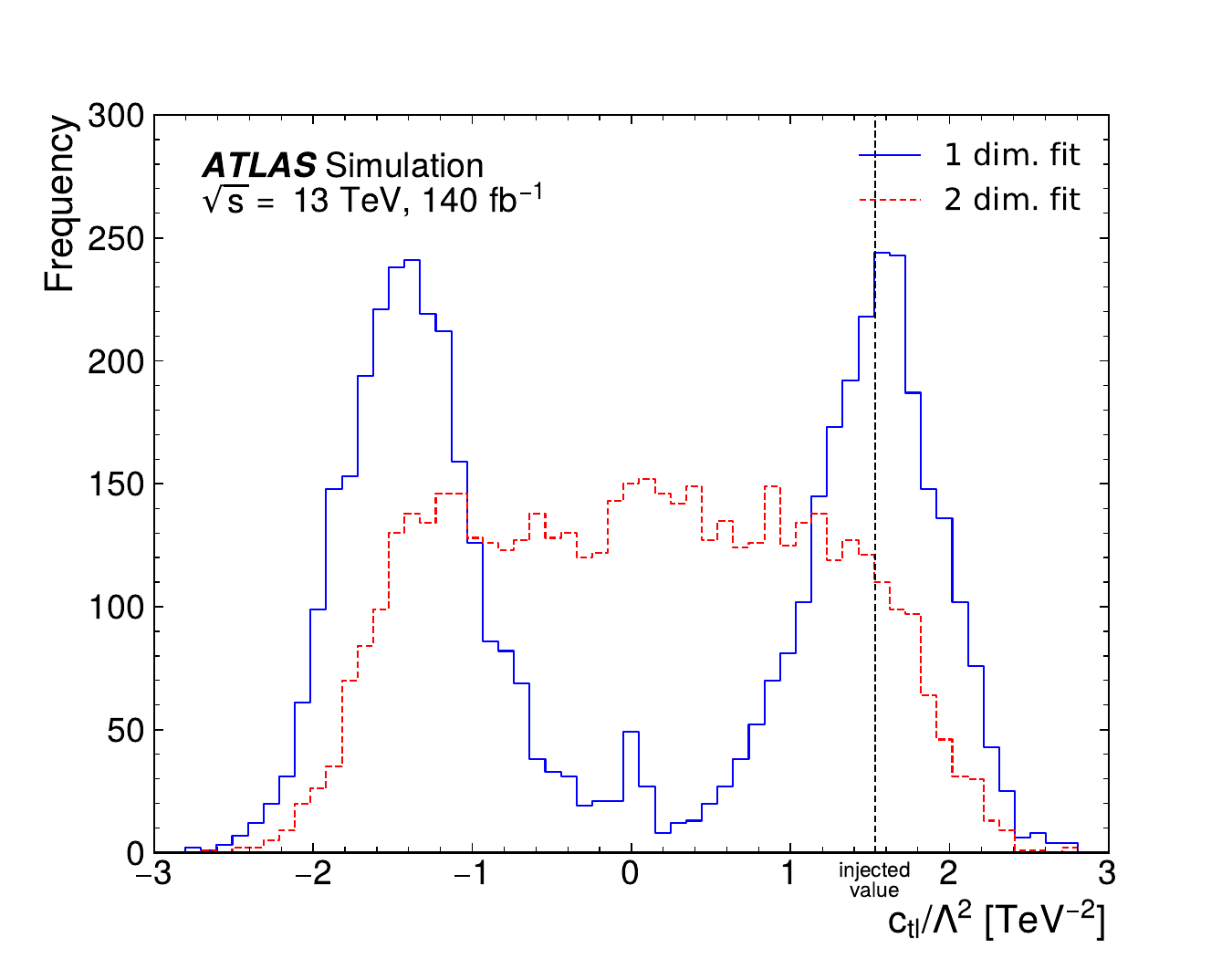}}
\subfloat[]{\includegraphics[width=0.24\linewidth]{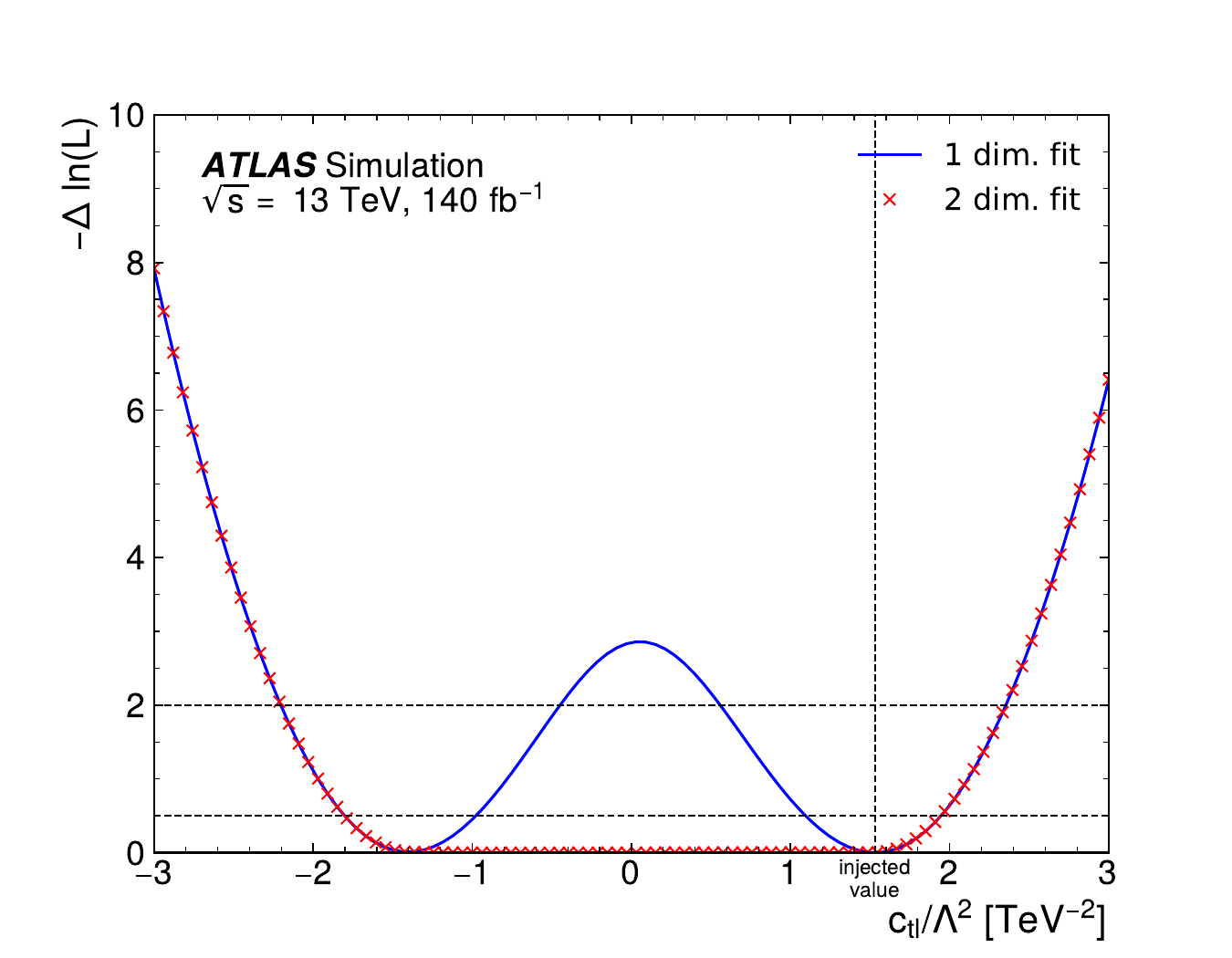}}
\subfloat[]{\includegraphics[width=0.24\linewidth]{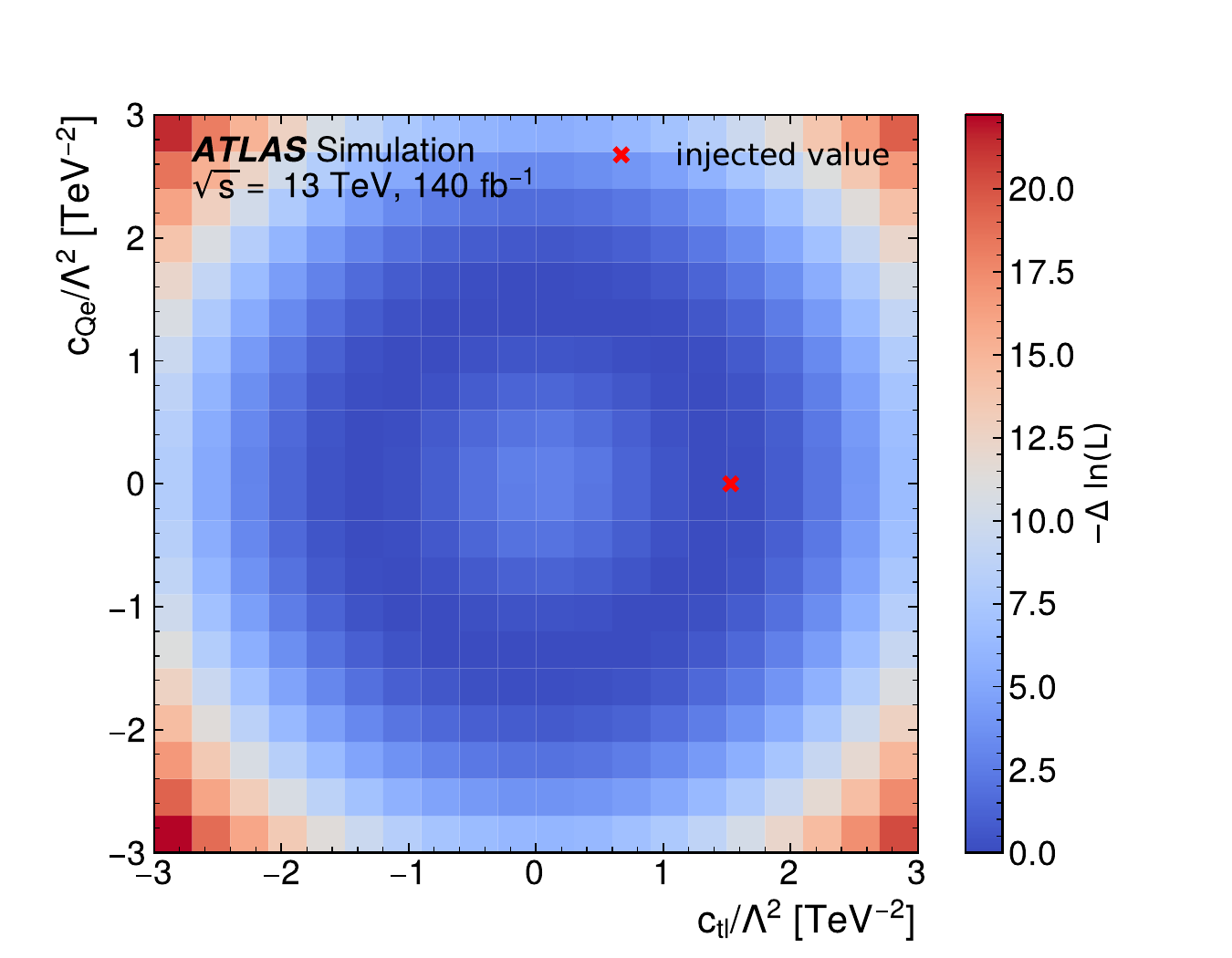}}
\subfloat[]{\includegraphics[width=0.24\linewidth]{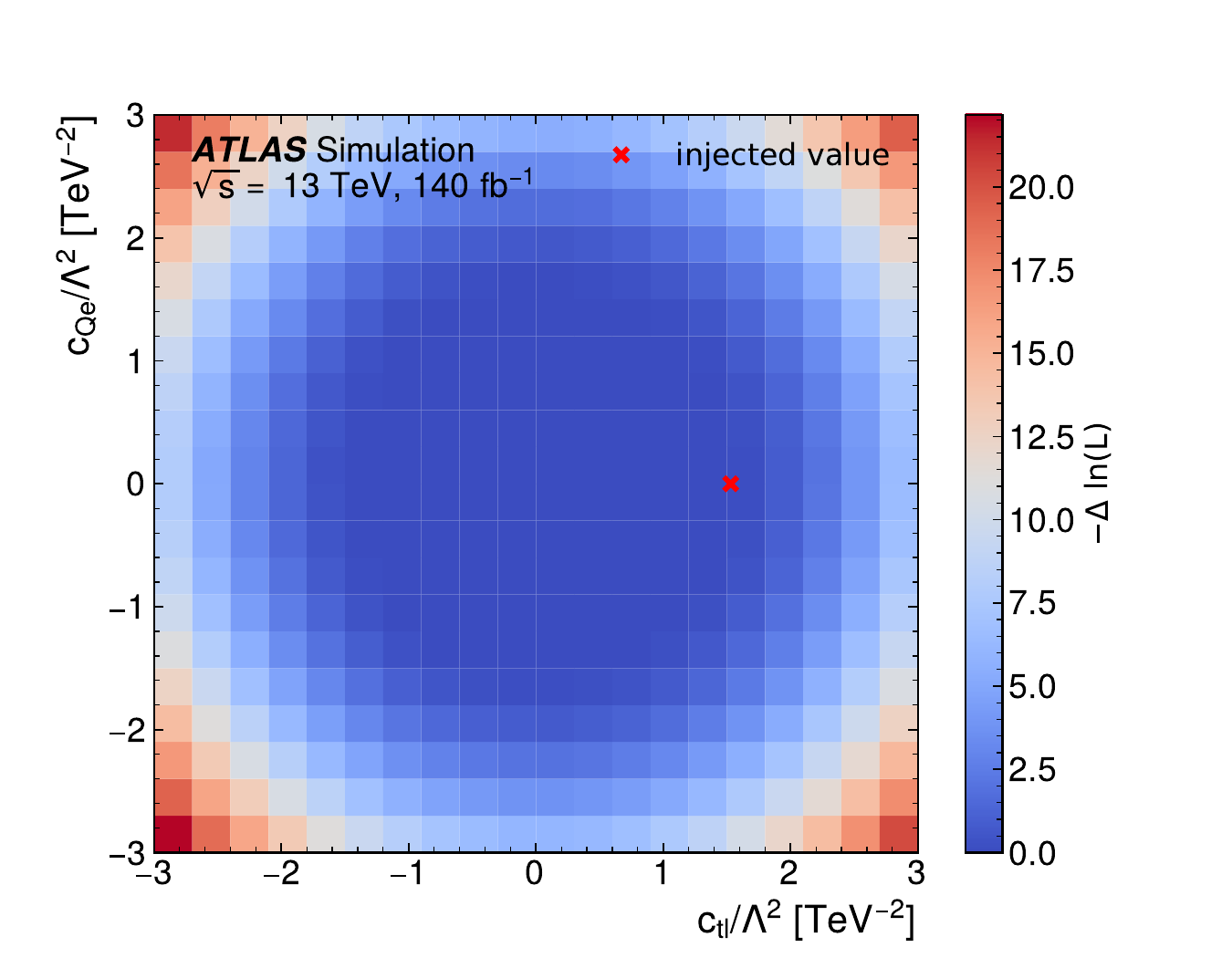}}
\caption{Studies on the sensitivity to multiple WCs at a time. Results are shown for fits considering only the statistical uncertainties for the SM expectation, for (top) all WCs$=0$, and (bottom) for an injected value of $c_{tl}=1.53$, close to the 95\% confidence interval limit. The distributions considered in each case are (a) the fitted pseudodata, (b) the 1D likelihood scans, (c) the 2D likelihood scan for a 2D fit and (d) the 2D likelihood scan for a 3D fit. }
\label{fig:Results:2D_vs_3D_inj}
\end{figure}

%
%
%
%
%
%
%
%
%
%
%
%
%
%
%
%
%
%
%
%
%
%
%
%
%
%
%
%
%
%
%
%
%
%


%

%
%
%

%
\section{Conclusion}
\label{sec:conclusion}

This paper presents a search for new physics in the \ttll final state.
This process is highly interesting as it provides a probe of the \ttll
interaction vertex, which offers unique sensitivity to four-fermion
EFT operators.  To exploit this sensitivity the measurement is focused
on the region of high invariant mass of $\ell^+\ell^-$ pair that has
not previously been probed directly by ATLAS.  Data from
proton--proton collisions corresponding to an integrated luminosity of
140~$\mathrm{fb^{-1}}$ at $\sqrt{s}=13$~\TeV are used.

The three-lepton final state is targeted due to the high \ttll
purity. Important backgrounds arise through the \ttW process and fake
leptons from \ttbar events, these are controlled using dedicated
control regions.

Profile likelihood fits to the data are used for several interpretations. This includes measurement of the signal strength of SM \ttll for $m(\ell^+\ell^-)>m_Z+10$~\GeV, computation of cross-section upper limits for a BSM signature in the region of $m(\ell^+\ell^-)>550$~\GeV, and extraction of EFT
coefficients in various combinations. These include flavour-inclusive and, for the first time,
flavour-split results, with configurations that test for lepton
flavour universality-violating signals.

The best-fit values are compatible with the SM and limits on the
relevant EFT WCs are set. These limits improve upon previous
constraints on these operators from the CMS Collaboration~\cite{CMS-TOP-22-006}.

The search for lepton flavour universality-violating EFT effects is
still limited by the available Run 2 statistical precision, but this result
demonstrates that cancellation of systematic uncertainties is possible
and with more data this method could provide higher-precision tests
of the SM. Again, no deviations from the SM or signs of lepton flavour universality violation are observed.

\section*{Acknowledgements}
%

%
%

%
%

We thank CERN for the very successful operation of the LHC and its injectors, as well as the support staff at
CERN and at our institutions worldwide without whom ATLAS could not be operated efficiently.

The crucial computing support from all WLCG partners is acknowledged gratefully, in particular from CERN, the ATLAS Tier-1 facilities at TRIUMF/SFU (Canada), NDGF (Denmark, Norway, Sweden), CC-IN2P3 (France), KIT/GridKA (Germany), INFN-CNAF (Italy), NL-T1 (Netherlands), PIC (Spain), RAL (UK) and BNL (USA), the Tier-2 facilities worldwide and large non-WLCG resource providers. Major contributors of computing resources are listed in Ref.~\cite{ATL-SOFT-PUB-2025-001}.

We gratefully acknowledge the support of ANPCyT, Argentina; YerPhI, Armenia; ARC, Australia; BMWFW and FWF, Austria; ANAS, Azerbaijan; CNPq and FAPESP, Brazil; NSERC, NRC and CFI, Canada; CERN; ANID, Chile; CAS, MOST and NSFC, China; Minciencias, Colombia; MEYS CR, Czech Republic; DNRF and DNSRC, Denmark; IN2P3-CNRS and CEA-DRF/IRFU, France; SRNSFG, Georgia; BMFTR, HGF and MPG, Germany; GSRI, Greece; RGC and Hong Kong SAR, China; ICHEP and Academy of Sciences and Humanities, Israel; INFN, Italy; MEXT and JSPS, Japan; CNRST, Morocco; NWO, Netherlands; RCN, Norway; MNiSW, Poland; FCT, Portugal; MNE/IFA, Romania; MSTDI, Serbia; MSSR, Slovakia; ARIS and MVZI, Slovenia; DSI/NRF, South Africa; MICIU/AEI, Spain; SRC and Wallenberg Foundation, Sweden; SERI, SNSF and Cantons of Bern and Geneva, Switzerland; NSTC, Taipei; TENMAK, T\"urkiye; STFC/UKRI, United Kingdom; DOE and NSF, United States of America.

Individual groups and members have received support from BCKDF, CANARIE, CRC and DRAC, Canada; CERN-CZ, FORTE and PRIMUS, Czech Republic; COST, ERC, ERDF, Horizon 2020, ICSC-NextGenerationEU and Marie Sk{\l}odowska-Curie Actions, European Union; Investissements d'Avenir Labex, Investissements d'Avenir Idex and ANR, France; DFG and AvH Foundation, Germany; Herakleitos, Thales and Aristeia programmes co-financed by EU-ESF and the Greek NSRF, Greece; BSF-NSF and MINERVA, Israel; NCN and NAWA, Poland; La Caixa Banking Foundation, CERCA Programme Generalitat de Catalunya and PROMETEO and GenT Programmes Generalitat Valenciana, Spain; G\"{o}ran Gustafssons Stiftelse, Sweden; The Royal Society and Leverhulme Trust, United Kingdom.

In addition, individual members wish to acknowledge support from Armenia: Yerevan Physics Institute (FAPERJ); CERN: European Organization for Nuclear Research (CERN DOCT); Chile: Agencia Nacional de Investigaci\'on y Desarrollo (FONDECYT 1230812, FONDECYT 1240864, Fondecyt 3240661); China: Chinese Ministry of Science and Technology (MOST-2023YFA1605700, MOST-2023YFA1609300), National Natural Science Foundation of China (NSFC - 12175119, NSFC 12275265); Czech Republic: Czech Science Foundation (GACR - 24-11373S), Ministry of Education Youth and Sports (ERC-CZ-LL2327, FORTE CZ.02.01.01/00/22\_008/0004632), PRIMUS Research Programme (PRIMUS/21/SCI/017); EU: H2020 European Research Council (ERC - 101002463); European Union: European Research Council (BARD No. 101116429, ERC - 948254, ERC 101089007), European Regional Development Fund (SMASH COFUND 101081355, SLO ERDF), Horizon 2020 Framework Programme (MUCCA - CHIST-ERA-19-XAI-00), European Union, Future Artificial Intelligence Research (FAIR-NextGenerationEU PE00000013), Horizon 2020 (EuroHPC - EHPC-DEV-2024D11-051), Italian Center for High Performance Computing, Big Data and Quantum Computing (ICSC, NextGenerationEU); France: Agence Nationale de la Recherche (ANR-21-CE31-0022, ANR-22-EDIR-0002); Germany: Baden-Württemberg Stiftung (BW Stiftung-Postdoc Eliteprogramme), Deutsche Forschungsgemeinschaft (DFG - 469666862, DFG - CR 312/5-2); China: Research Grants Council (GRF); Italy: Istituto Nazionale di Fisica Nucleare (ICSC, NextGenerationEU), Ministero dell'Università e della Ricerca (NextGenEU 153D23001490006 M4C2.1.1, NextGenEU I53D23000820006 M4C2.1.1, NextGenEU I53D23001490006 M4C2.1.1); Japan: Japan Society for the Promotion of Science (JSPS KAKENHI JP22H01227, JSPS KAKENHI JP22H04944, JSPS KAKENHI JP22KK0227, JSPS KAKENHI JP23KK0245, JSPS KAKENHI JP24K23939); Norway: Research Council of Norway (RCN-314472); Poland: Ministry of Science and Higher Education (IDUB AGH, POB8, D4 no 9722), Polish National Science Centre (NCN 2021/42/E/ST2/00350, NCN OPUS 2023/51/B/ST2/02507, NCN OPUS nr 2022/47/B/ST2/03059, NCN UMO-2019/34/E/ST2/00393, UMO-2022/47/O/ST2/00148, UMO-2023/49/B/ST2/04085, UMO-2023/51/B/ST2/00920, UMO-2024/53/N/ST2/00869); Portugal: Foundation for Science and Technology (FCT); Spain: Generalitat Valenciana (Artemisa, FEDER, IDIFEDER/2018/048), Ministry of Science and Innovation (MCIN \& NextGenEU PCI2022-135018-2, MICIN \& FEDER PID2021-125273NB, RYC2019-028510-I, RYC2020-030254-I, RYC2021-031273-I, RYC2022-038164-I); Sweden: Carl Trygger Foundation (Carl Trygger Foundation CTS 22:2312), Swedish Research Council (Swedish Research Council 2023-04654, VR 2021-03651, VR 2022-03845, VR 2022-04683, VR 2023-03403, VR 2024-05451), Knut and Alice Wallenberg Foundation (KAW 2018.0458, KAW 2022.0358, KAW 2023.0366); Switzerland: Swiss National Science Foundation (SNSF - PCEFP2\_194658); United Kingdom: Leverhulme Trust (Leverhulme Trust RPG-2020-004), Royal Society (NIF-R1-231091); United States of America: U.S. Department of Energy (ECA DE-AC02-76SF00515), Neubauer Family Foundation.

%
%


%
%

%
%
%
%
%
%
%

%
\clearpage
\printbibliography

@article{CMS-TOP-22-006,
%      author        = "{CMS Collaboration}",
%      collaboration = "CMS",
%      title         = "{Search for physics beyond the standard model in top quark
%                       production with additional leptons in the context of
%                       effective field theory}",
%      archivePrefix = "arXiv",
%      eprint        = "2307.15761",
%      reportNumber  = "CMS-TOP-22-006, CERN-EP-2023-124, CMS-TOP-22-006-003",
%      journal       = "JHEP",
%      volume        = "2312",
%      pages         = "068",
%      year          = "2023",
%      url           = "https://cds.cern.ch/record/2866292",
%      note          = "All the figures and tables can be found
%                       at
%http://cms-results.web.cern.ch/cms-results/public-results/publications/TOP-22-006
%(CMS
%                       Public Pages)",
%      doi           = "10.1007/JHEP12(2023)068",
%}

@article{Aguilar-Saavedra:2009xmz,
    author = "Aguilar-Saavedra, J. A.",
    title = "{Identifying top partners at LHC}",
    eprint = "0907.3155",
    archivePrefix = "arXiv",
    primaryClass = "hep-ph",
    doi = "10.1088/1126-6708/2009/11/030",
    journal = "JHEP",
    volume = "11",
    pages = "030",
    year = "2009"
}

@article{Aguilar-Saavedra:2013qpa,
    author = "Aguilar-Saavedra, J. A. and Benbrik, R. and Heinemeyer, S. and P\'erez-Victoria, M.",
    title = "{Handbook of vectorlike quarks: Mixing and single production}",
    eprint = "1306.0572",
    archivePrefix = "arXiv",
    primaryClass = "hep-ph",
    doi = "10.1103/PhysRevD.88.094010",
    journal = "Phys. Rev. D",
    volume = "88",
    number = "9",
    pages = "094010",
    year = "2013"
}

@article{Perelstein:2005ka,
    author = "Perelstein, Maxim",
    title = "{Little Higgs models and their phenomenology}",
    eprint = "hep-ph/0512128",
    archivePrefix = "arXiv",
    doi = "10.1016/j.ppnp.2006.04.001",
    journal = "Prog. Part. Nucl. Phys.",
    volume = "58",
    pages = "247--291",
    year = "2007"
}

@article{Chivukula:1992ap,
    author = "Chivukula, R. Sekhar and Selipsky, Stephen B. and Simmons, Elizabeth H.",
    title = "{Nonoblique effects in the $\mathrm{Z}b\overline{b}$ vertex from extended technicolor dynamics}",
    eprint = "hep-ph/9204214",
    archivePrefix = "arXiv",
    reportNumber = "BUHEP-92-12, HUTP-92-A017",
    doi = "10.1103/PhysRevLett.69.575",
    journal = "Phys. Rev. Lett.",
    volume = "69",
    pages = "575--577",
    year = "1992"
}

@article{Chivukula:1994mn,
    author = "Chivukula, R. S. and Simmons, Elizabeth H. and Terning, J.",
    title = "{A heavy top quark and the $\mathrm{Z}b\overline{b}$ vertex in non-commuting extended technicolor}",
    eprint = "hep-ph/9404209",
    archivePrefix = "arXiv",
    reportNumber = "BUHEP-94-8",
    doi = "10.1016/0370-2693(94)91068-5",
    journal = "Phys. Lett. B",
    volume = "331",
    pages = "383--389",
    year = "1994"
}

@article{PhysRevD.52.5374,
  title = {Extended technicolor contribution to the Zbb vertex},
  author = {Hagiwara, Kaoru and Kitazawa, Noriaki},
  eprint = "hep-ph/9504332",
  archivePrefix = "arXiv",
  journal = {Phys. Rev. D},
  volume = {52},
  pages = {5374--5377},
  numpages = {0},
  year = {1995},
  month = {Nov},
  publisher = {American Physical Society},
  doi = {10.1103/PhysRevD.52.5374},
}

@article{PhysRevD.55.5848,
  title = {Noncommuting ETC corrections to the $\mathrm{Z}t\overline{t}$ vertex},
  author = {Mahanta, Uma},
  eprint = "hep-ph/9611289",
  archivePrefix = "arXiv",
  journal = {Phys. Rev. D},
  volume = {55},
  pages = {5848--5850},
  numpages = {0},
  year = {1997},
  month = {May},
  publisher = {American Physical Society},
  doi = {10.1103/PhysRevD.55.5848},
}

@article{PhysRevD.56.402,
  title = {Probing noncommuting extended technicolor effects by ${e}^{+}{e}^{\ensuremath{-}}\ensuremath{\rightarrow}t\overline{t}$ at the Next Linear Collider},
  author = {Mahanta, Uma},
  journal = {Phys. Rev. D},
  volume = {56},
  pages = {402--406},
  numpages = {0},
  year = {1997},
  month = {Jul},
  publisher = {American Physical Society},
  doi = {10.1103/PhysRevD.56.402},
}

@article{EFT:Bylund,
     author    = "Bylund, Olga Bessidskaia and Maltoni, Fabio and Tsinikos, Ioannis and Vryonidou, Eleni and Zhang, Cen",
     title     = "{Probing top quark neutral couplings in the Standard Model Effective Field Theory at NLO in QCD}",
     journal   = "JHEP",
     volume    = "05",
     year      = "2016",
     pages     = "052",
     eprint    = "1601.08193",
     archivePrefix = "arXiv",
     primaryClass  =  "hep-ph",
     doi       = "10.1007/JHEP05(2016)052",
     SLACcitation  = "%%CITATION = ARXIV:1601.08193;%%"
}

@article{Bissmann:2020mfi,
    author = {Bi\ss{}mann, Stefan and Grunwald, Cornelius and Hiller, Gudrun and Kr\"oninger, Kevin},
    title = "{Top and Beauty synergies in SMEFT-fits at present and future colliders}",
    eprint = "2012.10456",
    archivePrefix = "arXiv",
    primaryClass = "hep-ph",
    reportNumber = "DO-TH 20/14",
    doi = "10.1007/JHEP06(2021)010",
    journal = "JHEP",
    volume = "06",
    pages = "010",
    year = "2021"
}

@article{Grunwald:2023nli,
    author = {Grunwald, Cornelius and Hiller, Gudrun and Kr\"oninger, Kevin and Nollen, Lara},
    title = "{More synergies from beauty, top, Z and Drell-Yan measurements in SMEFT}",
    eprint = "2304.12837",
    archivePrefix = "arXiv",
    primaryClass = "hep-ph",
    reportNumber = "DO-TH 23/03",
    doi = "10.1007/JHEP11(2023)110",
    journal = "JHEP",
    volume = "11",
    pages = "110",
    year = "2023"
}

@article{Bartocci:2023nvp,
    author = {Bartocci, Riccardo and Biek\"otter, Anke and Hurth, Tobias},
    title = "{A global analysis of the SMEFT under the minimal MFV assumption}",
    eprint = "2311.04963",
    archivePrefix = "arXiv",
    primaryClass = "hep-ph",
    reportNumber = "MITP/23-063",
    doi = "10.1007/JHEP05(2024)074",
    journal = "JHEP",
    volume = "05",
    pages = "074",
    year = "2024"
}

@article{Allwicher:2023shc,
    author = "Allwicher, Lukas and Cornella, Claudia and Isidori, Gino and Stefanek, Ben A.",
    title = "{New physics in the third generation. A comprehensive SMEFT analysis and future prospects}",
    eprint = "2311.00020",
    archivePrefix = "arXiv",
    primaryClass = "hep-ph",
    reportNumber = "ZU-TH 71/23, MITP-23-060, KCL-PH-TH/2023-59",
    doi = "10.1007/JHEP03(2024)049",
    journal = "JHEP",
    volume = "03",
    pages = "049",
    year = "2024"
}

@article{Fuentes-Martin:2020lea,
    author = "Fuentes-Martin, Javier and Greljo, Admir and Camalich, Jorge Martin and Ruiz-Alvarez, Jos\'e David",
    title = "{Charm physics confronts high-p$_{T}$ lepton tails}",
    eprint = "2003.12421",
    archivePrefix = "arXiv",
    primaryClass = "hep-ph",
    reportNumber = "CERN-TH-2020-047, ZU-TH 07/20",
    doi = "10.1007/JHEP11(2020)080",
    journal = "JHEP",
    volume = "11",
    pages = "080",
    year = "2020"
}

@article{BaBar:2012obs,
%    author = "Lees, J. P. and others",
    author = "{BaBar Collaboration}",
    collaboration = "BaBar",
    title = "{Evidence for an excess of $\bar{B} \to D^{(*)} \tau^-\bar{\nu}_\tau$ decays}",
    eprint = "1205.5442",
    archivePrefix = "arXiv",
    primaryClass = "hep-ex",
    reportNumber = "BABAR-PUB-12-012, SLAC-PUB-15028",
    doi = "10.1103/PhysRevLett.109.101802",
    journal = "Phys. Rev. Lett.",
    volume = "109",
    pages = "101802",
    year = "2012"
}

@article{BaBar:2013mob,
%    author = "Lees, J. P. and others",
    author = "{BaBar Collaboration}",
    collaboration = "BaBar",
    title = "{Measurement of an Excess of $\bar{B} \to D^{(*)}\tau^- \bar{\nu}_\tau$ Decays and Implications for Charged Higgs Bosons}",
    eprint = "1303.0571",
    archivePrefix = "arXiv",
    primaryClass = "hep-ex",
    reportNumber = "BABAR-PUB-13-001, SLAC-PUB-15381",
    doi = "10.1103/PhysRevD.88.072012",
    journal = "Phys. Rev. D",
    volume = "88",
    number = "7",
    pages = "072012",
    year = "2013"
}

@article{Belle:2015qfa,
%    author = "Huschle, M. and others",
    author = "{Belle Collaboration}",
    collaboration = "Belle",
    title = "{Measurement of the branching ratio of $\bar{B} \to D^{(\ast)} \tau^- \bar{\nu}_\tau$ relative to $\bar{B} \to D^{(\ast)} \ell^- \bar{\nu}_\ell$ decays with hadronic tagging at Belle}",
    eprint = "1507.03233",
    archivePrefix = "arXiv",
    primaryClass = "hep-ex",
    reportNumber = "KEK-REPORT-2015-18",
    doi = "10.1103/PhysRevD.92.072014",
    journal = "Phys. Rev. D",
    volume = "92",
    number = "7",
    pages = "072014",
    year = "2015"
}

@article{Belle:2016dyj,
%    author = "Hirose, S. and others",
    author = "{Belle Collaboration}",
    collaboration = "Belle",
    title = "{Measurement of the $\tau$ lepton polarization and $R(D^*)$ in the decay $\bar{B} \to D^* \tau^- \bar{\nu}_\tau$}",
    eprint = "1612.00529",
    archivePrefix = "arXiv",
    primaryClass = "hep-ex",
    reportNumber = "KEK-PREPRINT-2016-53, BELLE-PREPRINT-2016-14",
    doi = "10.1103/PhysRevLett.118.211801",
    journal = "Phys. Rev. Lett.",
    volume = "118",
    number = "21",
    pages = "211801",
    year = "2017"
}

@article{Belle:2017ilt,
%    author = "Hirose, S. and others",
    author = "{Belle Collaboration}",
    collaboration = "Belle",
    title = "{Measurement of the $\tau$ lepton polarization and $R(D^*)$ in the decay $\bar{B} \rightarrow D^* \tau^- \bar{\nu}_\tau$ with one-prong hadronic $\tau$ decays at Belle}",
    eprint = "1709.00129",
    archivePrefix = "arXiv",
    primaryClass = "hep-ex",
    reportNumber = "KEK-PREPRINT-2017-26, BELLE-PREPRINT-2017-18",
    doi = "10.1103/PhysRevD.97.012004",
    journal = "Phys. Rev. D",
    volume = "97",
    number = "1",
    pages = "012004",
    year = "2018"
}

@article{Belle:2019rba,
%    author = "Caria, G. and others",
    author = "{Belle Collaboration}",
    collaboration = "Belle",
    title = "{Measurement of $\mathcal{R}(D)$ and $\mathcal{R}(D^*)$ with a semileptonic tagging method}",
    eprint = "1910.05864",
    archivePrefix = "arXiv",
    primaryClass = "hep-ex",
    reportNumber = "Belle-2019-18, KEK-2019-40",
    doi = "10.1103/PhysRevLett.124.161803",
    journal = "Phys. Rev. Lett.",
    volume = "124",
    number = "16",
    pages = "161803",
    year = "2020"
}

@article{LHCb:2023zxo,
    collaboration = "LHCb",
    author = "{LHCb Collaboration}",
    title = "{Measurement of the ratios of branching fractions $\mathcal{R}(D^{*})$ and $\mathcal{R}(D^{0})$}",
    eprint = "2302.02886",
    archivePrefix = "arXiv",
    primaryClass = "hep-ex",
    reportNumber = "LHCb-PAPER-2022-039, CERN-EP-2022-284",
    doi = "10.1103/PhysRevLett.131.111802",
    journal = "Phys. Rev. Lett.",
    volume = "131",
    pages = "111802",
    year = "2023"
}

@article{LHCb:2023uiv,
%    author = "Aaij, Roel and others",
    author = "{LHCb Collaboration}",
    collaboration = "LHCb",
    title = "{Test of lepton flavor universality using $B^{0}\rightarrow D^{*-}\tau^{+}\nu_{\tau}$ decays with hadronic \ensuremath{\tau} channels}",
    eprint = "2305.01463",
    archivePrefix = "arXiv",
    primaryClass = "hep-ex",
    reportNumber = "LHCb-PAPER-2022-052, CERN-EP-2023-062",
    doi = "10.1103/PhysRevD.108.012018",
    journal = "Phys. Rev. D",
    volume = "108",
    number = "1",
    pages = "012018",
    year = "2023"
}

@article{Afik:2021xmi,
    author = "Afik, Yoav and Bar-Shalom, Shaouly and Pal, Kuntal and Soni, Amarjit and Wudka, Jose",
    title = "{Multi-lepton probes of new physics and lepton-universality in top-quark interactions}",
    eprint = "2111.13711",
    archivePrefix = "arXiv",
    primaryClass = "hep-ph",
    doi = "10.1016/j.nuclphysb.2022.115849",
    journal = "Nucl. Phys. B",
    volume = "980",
    pages = "115849",
    year = "2022"
}

@article{Atkinson:2024hqp,
    author = "Atkinson, Oliver and Englert, Christoph and Kirk, Matthew and Tetlalmatzi-Xolocotzi, Gilberto",
    title = "{Collider-flavour complementarity from the bottom to the top}",
    eprint = "2411.00940",
    archivePrefix = "arXiv",
    primaryClass = "hep-ph",
    reportNumber = "IPPP/24/72, SI-HEP-2024-24, P3H-24-086",
    doi = "10.1140/epjc/s10052-024-13739-w",
    journal = "Eur. Phys. J. C",
    volume = "85",
    number = "3",
    pages = "258",
    year = "2025"
}

@article{LUCID2,
  author  = {G. Avoni and others},
  title   = {The new LUCID-2 detector for luminosity measurement and monitoring in ATLAS},
  journal = "JINST",
  volume  = "13",
  number  = "07",
  pages   = "P07017",
  doi     = "10.1088/1748-0221/13/07/P07017",
  year    = "2018"
}

@article{Frederix:2021agh,
    author = "Frederix, Rikkert and Tsinikos, Ioannis",
    title = "{On improving NLO merging for $ \mathrm{t}\overline{\mathrm{t}}\mathrm{W} $ production}",
    eprint = "2108.07826",
    archivePrefix = "arXiv",
    primaryClass = "hep-ph",
    reportNumber = "LU-TP 21-34",
    doi = "10.1007/JHEP11(2021)029",
    journal = "JHEP",
    volume = "11",
    pages = "029",
    year = "2021"
}

@Article{Demartin2017,
     author    = "Demartin, Federico and Maier, Benedikt and Maltoni, Fabio and Mawatari, Kentarou and Zaro, Marco",
     title     = "tWH associated production at the LHC",
     journal   = "Eur. Phys. J. C",
     year      = "2017",
     month     = "1",
     day       = "18",
     volume    = "77",
     number    = "1",
     pages     = "34",
     archivePrefix = "arXiv",
     primaryClass  =  "hep-ph",
     eprint        = "1607.05862",
     doi           = "10.1140/epjc/s10052-017-4601-7",
}

@article{Brivio:2020onw,
    author = "Brivio, Ilaria",
    title = "{SMEFTsim 3.0 \textemdash{} a practical guide}",
    eprint = "2012.11343",
    archivePrefix = "arXiv",
    primaryClass = "hep-ph",
    doi = "10.1007/JHEP04(2021)073",
    journal = "JHEP",
    volume = "04",
    pages = "073",
    year = "2021"
}

@article{Brivio:2021yjb,
    author = "Brivio, Ilaria and Dawson, Sally and de Blas, Jorge and Durieux, Gauthier and Savard, Pierre and Denner, Ansgar and Freitas, Ayres and Hays, Chris and Pecjak, Ben and Vicini, Alessandro",
    title = "{Electroweak input parameters}",
    eprint = "2111.12515",
    archivePrefix = "arXiv",
    primaryClass = "hep-ph",
    reportNumber = "CERN-LPCC-2021-002",
    month = "11",
    year = "2021"
}

@article{Buccioni:2019sur,
    author = {Buccioni, Federico and Lang, Jean-Nicolas and Lindert, Jonas M. and Maierh\"ofer, Philipp and Pozzorini, Stefano and Zhang, Hantian and Zoller, Max F.},
    title = "{OpenLoops 2}",
    eprint = "1907.13071",
    archivePrefix = "arXiv",
    primaryClass = "hep-ph",
    reportNumber = "IPPP/19/62, FR-PHENO-2019-12, PSI-PR-19-15, ZU-TH 37/19",
    doi = "10.1140/epjc/s10052-019-7306-2",
    journal = "Eur. Phys. J. C",
    volume = "79",
    number = "10",
    pages = "866",
    year = "2019"
}

@article{Mattelaer:2016gcx,
    author = "Mattelaer, Olivier",
    title = "{On the maximal use of Monte Carlo samples: re-weighting events at NLO accuracy}",
    eprint = "1607.00763",
    archivePrefix = "arXiv",
    primaryClass = "hep-ph",
    reportNumber = "IPPP-16-55, MCNET-16-25",
    doi = "10.1140/epjc/s10052-016-4533-7",
    journal = "Eur. Phys. J. C",
    volume = "76",
    number = "12",
    pages = "674",
    year = "2016"
}

@article{NNPDF:2017mvq,
    author = "{NNPDF Collaboration} and Ball, Richard D. and others",
    collaboration = "NNPDF",
    title = "{Parton distributions from high-precision collider data}",
    eprint = "1706.00428",
    archivePrefix = "arXiv",
    primaryClass = "hep-ph",
    reportNumber = "EDINBURGH-2017-08, NIKHEF-2017-006, OUTP-17-04P, TIF-UNIMI-2017-3, CAVENDISH-HEP-17-06, CERN-TH-2017-077, Edinburgh 2017/08,
  Nikhef/2017-006, OUTP-17-04P,TIF-UNIMI-2017-3",
    doi = "10.1140/epjc/s10052-017-5199-5",
    journal = "Eur. Phys. J. C",
    volume = "77",
    number = "10",
    pages = "663",
    year = "2017"
}

@article{Cacciari:2008gn,
      author         = "Cacciari, Matteo and Salam, Gavin P. and Soyez, Gregory",
      title          = "{The Catchment Area of Jets}",
      journal        = "JHEP",
      volume         = "04",
      year           = "2008",
      pages          = "005",
      doi            = "10.1088/1126-6708/2008/04/005",
      eprint         = "0802.1188",
      archivePrefix  = "arXiv",
      primaryClass   = "hep-ph",
      reportNumber   = "LPTHE-07-02",
      SLACcitation   = "%%CITATION = ARXIV:0802.1188;%%"                                                                                                                                                          
}

@article{Brivio:2022pyi,
    author = "Brivio, Ilaria and others",
    title = "{Truncation, validity, uncertainties}",
    eprint = "2201.04974",
    archivePrefix = "arXiv",
    primaryClass = "hep-ph",
    reportNumber = "CERN-LHCEFTWG-2021-002, CERN-LPCC-2022-01",
    month = "1",
    year = "2022"
}

@article{Cranmer:2012sba,
    author = "{ROOT Collaboration}",
    title = "{HistFactory: A tool for creating statistical models for use with RooFit and RooStats}",
%    reportNumber = "CERN-OPEN-2012-016",
    journal = "CERN-OPEN-2012-016",
%    url="https://cds.cern.ch/record/1456844",
    year = "2012",
    doi = "10.17181/CERN-OPEN-2012-016",
}

@article{Bernlochner:2022oiw,
    author = "Urs Bernlochner, Florian and Fry, Daniel C. and Burns Menary, Stephen and Persson, Eric",
    title = "{Cover your bases: asymptotic distributions of the profile likelihood ratio when constraining effective field theories in high-energy physics}",
    eprint = "2207.01350",
    archivePrefix = "arXiv",
    primaryClass = "physics.data-an",
    doi = "10.21468/SciPostPhysCore.6.1.013",
    journal = "SciPost Phys. Core",
    volume = "6",
    pages = "013",
    year = "2023"
}

@article{Wilks:1938dza,
    author = "Wilks, S. S.",
    title = "{The Large-Sample Distribution of the Likelihood Ratio for Testing Composite Hypotheses}",
    doi = "10.1214/aoms/1177732360",
    journal = "Annals Math. Statist.",
    volume = "9",
    number = "1",
    pages = "60--62",
    year = "1938"
}

@Article{Evans:2008zzb,
      author         = "Evans, Lyndon and Bryant, Philip",
      title          = "{LHC Machine}",
      journal        = "JINST",
      volume         = "3",
      pages          = "S08001",
      doi            = "10.1088/1748-0221/3/08/S08001",
      year           = "2008",
      SLACcitation   = "%%CITATION = JINST,3,S08001;%%",
}

@Article{Cacciari:2008gp,
     author    = "Cacciari, Matteo and Salam, Gavin P. and Soyez, Gregory",
     title     = "{The anti-\(k_{t}\) jet clustering algorithm}",
     journal   = "JHEP",
     volume    = "04",
     year      = "2008",
     pages     = "063",
     eprint    = "0802.1189",
     archivePrefix = "arXiv",
     primaryClass  =  "hep-ph",
     doi       = "10.1088/1126-6708/2008/04/063",
     SLACcitation  = "%%CITATION = 0802.1189;%%"
}

@Article{Fastjet,
      author         = "Cacciari, Matteo and Salam, Gavin P. and Soyez, Gregory",
      title          = "{FastJet user manual}",
      journal        = "Eur. Phys. J. C",
      volume         = "72",
      year           = "2012",
      pages          = "1896",
      doi            = "10.1140/epjc/s10052-012-1896-2",
      eprint         = "1111.6097",
      archivePrefix  = "arXiv",
      primaryClass   = "hep-ph",
      reportNumber   = "CERN-PH-TH-2011-297",
      SLACcitation   = "%%CITATION = ARXIV:1111.6097;%%"
}

@Article{Butterworth:2015oua,
      author         = "Butterworth, Jon and others",
      title          = "{PDF4LHC recommendations for LHC Run II}",
      journal        = "J. Phys. G",
      volume         = "43",
      year           = "2016",
      pages          = "023001",
      doi            = "10.1088/0954-3899/43/2/023001",
      eprint         = "1510.03865",
      archivePrefix  = "arXiv",
      primaryClass   = "hep-ph",
      reportNumber   = "OUTP-15-17P, SMU-HEP-15-12, TIF-UNIMI-2015-14,
                        LCTS-2015-27, CERN-PH-TH-2015-249",
      SLACcitation   = "%%CITATION = ARXIV:1510.03865;%%"
}

@article{Ball:2014uwa,
      author         = "{NNPDF Collaboration} and Ball, Richard D. and others",
      title          = "{Parton distributions for the LHC run II}",
      journal        = "JHEP",
      volume         = "04",
      year           = "2015",
      pages          = "040",
      doi            = "10.1007/JHEP04(2015)040",
      eprint         = "1410.8849",
      archivePrefix  = "arXiv",
      primaryClass   = "hep-ph",
      reportNumber   = "EDINBURGH-2014-15, IFUM-1034-FT, CERN-PH-TH-2013-253,
                        OUTP-14-11P, CAVENDISH-HEP-14-11",
      SLACcitation   = "%%CITATION = ARXIV:1410.8849;%%"
}

@Article{Sjostrand:2007gs,
	author    = "Sj{\"o}strand, T. and Mrenna, S. and Skands, P.",
	title     = "{A brief introduction to PYTHIA 8.1}",
	journal   = "Comput. Phys. Commun.",
	volume    = "178",
	year      = "2008",
	pages     = "852-867",
	eprint    = "0710.3820",
	archivePrefix = "arXiv",
	primaryClass  =  "hep-ph",
	doi       = "10.1016/j.cpc.2008.01.036",
	SLACcitation  = "%%CITATION = 0710.3820;%%"
}

@article{Sjostrand:2014zea,
      author         = "Sj{\"o}strand, Torbj{\"o}rn and Ask, Stefan and Christiansen,
                        Jesper R. and Corke, Richard and Desai, Nishita and Ilten,
                        Philip and Mrenna, Stephen and Prestel, Stefan and
                        Rasmussen, Christine O. and Skands, Peter Z.",
      title          = "{An introduction to PYTHIA 8.2}",
      journal        = "Comput. Phys. Commun.",
      volume         = "191",
      year           = "2015",
      pages          = "159",
      doi            = "10.1016/j.cpc.2015.01.024",
      eprint         = "1410.3012",
      archivePrefix  = "arXiv",
      primaryClass   = "hep-ph",
      reportNumber   = "LU-TP-14-36, MCNET-14-22, CERN-PH-TH-2014-190,
                        FERMILAB-PUB-14-316-CD, DESY-14-178, SLAC-PUB-16122,
                        --FERMILAB-PUB-14-316-CD",
      SLACcitation   = "%%CITATION = ARXIV:1410.3012;%%"
}

@Article{Lange:2001uf,
      author         = "Lange, D. J.",
      title          = "{The EvtGen particle decay simulation package}",
      booktitle      = "{Proceedings, 7th International Conference on B physics
                        at hadron machines (BEAUTY 2000)}",
      journal        = "Nucl. Instrum. Meth. A",
      volume         = "462",
      year           = "2001",
      pages          = "152",
      doi            = "10.1016/S0168-9002(01)00089-4",
      SLACcitation   = "%%CITATION = NUIMA,A462,152;%%"
}

@Article{Alwall:2014hca,
      author         = "Alwall, J. and Frederix, R. and Frixione, S. and Hirschi,
                        V. and Maltoni, F. and Mattelaer, O. and Shao, H. -S. and
                        Stelzer, T. and Torrielli, P. and Zaro, M.",
      title          = "{The automated computation of tree-level and
                        next-to-leading order differential cross sections, and
                        their matching to parton shower simulations}",
      journal        = "JHEP",
      volume         = "07",
      year           = "2014",
      pages          = "079",
      doi            = "10.1007/JHEP07(2014)079",
      eprint         = "1405.0301",
      archivePrefix  = "arXiv",
      primaryClass   = "hep-ph",
      reportNumber   = "CERN-PH-TH-2014-064, CP3-14-18, LPN14-066, MCNET-14-09,
                        ZU-TH-14-14",
      SLACcitation   = "%%CITATION = ARXIV:1405.0301;%%"
}

@article{Frederix:2012ps,
      author         = "Frederix, Rikkert and Frixione, Stefano",
      title          = "{Merging meets matching in MC@NLO}",
      journal        = "JHEP",
      volume         = "12",
      year           = "2012",
      pages          = "061",
      doi            = "10.1007/JHEP12(2012)061",
      eprint         = "1209.6215",
      archivePrefix  = "arXiv",
      primaryClass   = "hep-ph",
      reportNumber   = "CERN-PH-TH-2012-247, ZU-TH-21-12",
      SLACcitation   = "%%CITATION = ARXIV:1209.6215;%%"
}

@article{Artoisenet:2012st,
      author         = "Artoisenet, Pierre and Frederix, Rikkert and Mattelaer,
                        Olivier and Rietkerk, Robbert",
      title          = "{Automatic spin-entangled decays of heavy resonances in
                        Monte Carlo simulations}",
      journal        = "JHEP",
      volume         = "03",
      pages          = "015",
      doi            = "10.1007/JHEP03(2013)015",
      year           = "2013",
      eprint         = "1212.3460",
      archivePrefix  = "arXiv",
      primaryClass   = "hep-ph",
      reportNumber   = "NIKHEF-2012-021, CERN-PH-TH-2012-329",
      SLACcitation   = "%%CITATION = ARXIV:1212.3460;%%",
}

@article{Frixione:2007zp,
      author         = "Frixione, Stefano and Laenen, Eric and Motylinski,
                        Patrick and Webber, Bryan R.",
      title          = "{Angular correlations of lepton pairs from vector boson
                        and top quark decays in Monte Carlo simulations}",
      journal        = "JHEP",
      volume         = "04",
      year           = "2007",
      pages          = "081",
      doi            = "10.1088/1126-6708/2007/04/081",
      eprint         = "hep-ph/0702198",
      archivePrefix  = "arXiv",
      OPTprimaryClass   = "HEP-PH",
      reportNumber   = "CAVENDISH-HEP-07-01, GEF-TH-09-2007, ITP-UU-07-10,
                        NIKHEF-2007-004",
      SLACcitation   = "%%CITATION = HEP-PH/0702198;%%"
}

@Article{Bahr:2008pv,
      author         = "B{\"a}hr, M. and others",
      title          = "{Herwig++ physics and manual}",
      journal        = "Eur. Phys. J. C",
      volume         = "58",
      year           = "2008",
      pages          = "639",
      doi            = "10.1140/epjc/s10052-008-0798-9",
      eprint         = "0803.0883",
      archivePrefix  = "arXiv",
      primaryClass   = "hep-ph",
      reportNumber   = "CERN-PH-TH-2008-038, CAVENDISH-HEP-08-03, KA-TP-05-2008,
                        DCPT-08-22, IPPP-08-11, CP3-08-05",
      SLACcitation   = "%%CITATION = ARXIV:0803.0883;%%"
}

@Article{Bellm:2015jjp,
      author         = "Bellm, Johannes and others",
      title          = "{Herwig 7.0/Herwig++ 3.0 release note}",
      journal        = "Eur. Phys. J. C",
      volume         = "76",
      year           = "2016",
      number         = "4",
      pages          = "196",
      doi            = "10.1140/epjc/s10052-016-4018-8",
      eprint         = "1512.01178",
      archivePrefix  = "arXiv",
      primaryClass   = "hep-ph",
      reportNumber   = "CERN-PH-TH-2015-289, MAN-HEP-2015-15, IFJPAN-IV-2015-13,
                        HERWIG-2015-01, KA-TP-18-2015, DCPT-15-142, MCNET-15-28,
                        IPPP-15-71, --HERWIG-2015-01",
      SLACcitation   = "%%CITATION = ARXIV:1512.01178;%%"
}

@Article{Alioli:2010xd,
      author         = "Alioli, Simone and Nason, Paolo and Oleari, Carlo and Re,
                        Emanuele",
      title          = "{A general framework for implementing NLO calculations in
                        shower Monte Carlo programs: the POWHEG BOX}",
      journal        = "JHEP",
      volume         = "06",
      pages          = "043",
      doi            = "10.1007/JHEP06(2010)043",
      year           = "2010",
      eprint         = "1002.2581",
      archivePrefix  = "arXiv",
      primaryClass   = "hep-ph",
}

@article{deFlorian:2016spz,
      author         = "de Florian, D. and others",
      title          = "{Handbook of LHC Higgs Cross Sections: 4. Deciphering the Nature of the Higgs Sector}",
      collaboration  = "LHC Higgs Cross Section Working Group",
      doi            = "10.23731/CYRM-2017-002",
      year           = "2017",
      eprint         = "1610.07922",
      archivePrefix  = "arXiv",
      primaryClass   = "hep-ph",
      reportNumber   = "FERMILAB-FN-1025-T, CERN-2017-002-M",
      SLACcitation   = "%%CITATION = ARXIV:1610.07922;%%"
}

@Article{Gleisberg:2008ta,
      author         = "Gleisberg, T. and H{\"o}che, S. and Krauss, F. and Sch{\"o}nherr, M. and Schumann, S. and Siegert, F. and Winter J.",
      title          = "{Event generation with SHERPA 1.1}",
      journal        = "JHEP",
      volume         = "02",
      pages          = "007",
      doi            = "10.1088/1126-6708/2009/02/007",
      year           = "2009",
      eprint         = "0811.4622",
      archivePrefix  = "arXiv",
      primaryClass   = "hep-ph",
}

@Article{Hoeche:2009rj,
      author         = "H{\"o}che, Stefan and Krauss, Frank and Schumann, Steffen and Siegert, Frank",
      title          = "{QCD matrix elements and truncated showers}",
      journal        = "JHEP",
      volume         = "05",
      pages          = "053",
      doi            = "10.1088/1126-6708/2009/05/053",
      year           = "2009",
      eprint         = "0903.1219",
      archivePrefix  = "arXiv",
      primaryClass   = "hep-ph",
}

@Article{Gleisberg:2008fv,
      author         = "Gleisberg, Tanju and H{\"o}che, Stefan",
      title          = "{Comix, a new matrix element generator}",
      journal        = "JHEP",
      volume         = "12",
      pages          = "039",
      doi            = "10.1088/1126-6708/2008/12/039",
      year           = "2008",
      eprint         = "0808.3674",
      archivePrefix  = "arXiv",
      primaryClass   = "hep-ph",
}

@Article{Schumann:2007mg,
      author         = "Schumann, Steffen and Krauss, Frank",
      title          = "{A parton shower algorithm based on Catani--Seymour dipole factorisation}",
      journal        = "JHEP",
      volume         = "03",
      pages          = "038",
      doi            = "10.1088/1126-6708/2008/03/038",
      year           = "2008",
      eprint         = "0709.1027",
      archivePrefix  = "arXiv",
      primaryClass   = "hep-ph",
}

@article{Hoeche:2012yf,
      author         = "H{\"o}che, Stefan and Krauss, Frank and Sch{\"o}nherr, Marek and
                        Siegert, Frank",
      title          = "{QCD matrix elements + parton showers. The NLO case}",
      journal        = "JHEP",
      volume         = "04",
      year           = "2013",
      pages          = "027",
      doi            = "10.1007/JHEP04(2013)027",
      eprint         = "1207.5030",
      archivePrefix  = "arXiv",
      primaryClass   = "hep-ph",
      reportNumber   = "SLAC-PUB-15191, IPPP-12-52, DCPT-12-104, LPN12-081,
                        FR-PHENO-2012-017, MCNET-12-09, --FR-PHENO-2012-017",
      SLACcitation   = "%%CITATION = ARXIV:1207.5030;%%"
}

@article{Bothmann:2019yzt,
      author = "Bothmann, Enrico and others",
      title = "{Event generation with Sherpa 2.2}",
      journal = "SciPost Phys.",
      volume = "7",
      year = "2019",
      number = "3",
      pages = "034",
      doi = "10.21468/SciPostPhys.7.3.034",
      reportNumber = "FERMILAB-PUB-19-218-T, SLAC-PUB-17433, IPPP/19/42, MCNET-19-11",
      eprint = "1905.09127",
      archivePrefix = "arXiv",
      primaryClass = "hep-ph",
}

@article{Hoeche:2011fd,
      author         = "H{\"o}che, Stefan and Krauss, Frank and Sch{\"o}nherr, Marek and
                        Siegert, Frank",
      title          = "{A critical appraisal of NLO+PS matching methods}",
      journal        = "JHEP",
      volume         = "09",
      year           = "2012",
      pages          = "049",
      doi            = "10.1007/JHEP09(2012)049",
      eprint         = "1111.1220",
      archivePrefix  = "arXiv",
      primaryClass   = "hep-ph",
      reportNumber   = "SLAC-PUB-14661, IPPP-11-67, DCPT-11-134, LPN11-58,
                        FR-PHENO-2011-019, MCNET-11-24",
      SLACcitation   = "%%CITATION = ARXIV:1111.1220;%%"
}

@article{Catani:2001cc,
      author         = "Catani, S. and Krauss, F. and Webber, B. R. and Kuhn, R.",
      title          = "{QCD Matrix Elements + Parton Showers}",
      journal        = "JHEP",
      volume         = "11",
      year           = "2001",
      pages          = "063",
      doi            = "10.1088/1126-6708/2001/11/063",
      eprint         = "hep-ph/0109231",
      archivePrefix  = "arXiv",
      reportNumber   = "CERN-TH-2000-367, CAVENDISH-HEP-00-03",
      SLACcitation   = "%%CITATION = HEP-PH/0109231;%%"
}

@article{Cascioli:2011va,
      author         = "Cascioli, Fabio and Maierh{\"o}fer, Philipp and Pozzorini, Stefano",
      title          = "{Scattering Amplitudes with Open Loops}",
      journal        = "Phys. Rev. Lett.",
      volume         = "108",
      year           = "2012",
      pages          = "111601",
      doi            = "10.1103/PhysRevLett.108.111601",
      eprint         = "1111.5206",
      archivePrefix  = "arXiv",
      primaryClass   = "hep-ph",
      reportNumber   = "ZU-TH-23-11, LPN11-66",
      SLACcitation   = "%%CITATION = ARXIV:1111.5206;%%"
}

@article{Denner:2016kdg,
      author         = "Denner, Ansgar and Dittmaier, Stefan and Hofer, Lars",
      title          = "{\textsc{Collier}: A fortran-based complex one-loop library in
                        extended regularizations}",
      journal        = "Comput. Phys. Commun.",
      volume         = "212",
      year           = "2017",
      pages          = "220-238",
      doi            = "10.1016/j.cpc.2016.10.013",
      eprint         = "1604.06792",
      archivePrefix  = "arXiv",
      primaryClass   = "hep-ph",
      reportNumber   = "FR-PHENO-2016-003, ICCUB-16-016",
      SLACcitation   = "%%CITATION = ARXIV:1604.06792;%%"
}

@article{Agostinelli:2002hh,
      author         = "Agostinelli, S. and others",
      title          = "{\textsc{Geant4} -- a simulation toolkit}",
      journal        = "Nucl. Instrum. Meth. A",
      volume         = "506",
      year           = "2003",
      pages          = "250",
      doi            = "10.1016/S0168-9002(03)01368-8",
      reportNumber   = "SLAC-PUB-9350, FERMILAB-PUB-03-339",
      SLACcitation   = "%%CITATION = NUIMA,A506,250;%%"
}

@Article{Cowan:2010js,
   Author = {Cowan, Glen and Cranmer, Kyle and Gross, Eilam and Vitells, Ofer},
   Title = {{Asymptotic formulae for likelihood-based tests of new physics}},
   Journal = {Eur. Phys. J. C},
   Volume = {71},
   Year = {2011},
   Pages = {1554},
   doi = "10.1140/epjc/s10052-011-1554-0",
   Eprint = "1007.1727",
   Archiveprefix = {arXiv},
   Primaryclass = {physics.data-an},
   related = "Cowan:2010js-err",
   relatedstring  = "Erratum:",
}

@article{Barlow:1993dm,
    author = "Barlow, Roger and Beeston, Christine",
    title = "{Fitting using finite Monte Carlo samples}",
    reportNumber = "MAN-HEP-93-1",
    doi = "10.1016/0010-4655(93)90005-W",
    journal = "Comput. Phys. Commun.",
    volume = "77",
    pages = "219",
    year = "1993"
}

@Article{PERF-2007-01,
    author         = "{ATLAS Collaboration}",
    title          = "{The ATLAS Experiment at the CERN Large Hadron Collider}",
    journal        = "JINST",
    volume         = "3",
    year           = "2008",
    pages          = "S08003",
    doi            = "10.1088/1748-0221/3/08/S08003",
    primaryClass   = "hep-ex",
}

@Article{SOFT-2010-01,
    author         = "{ATLAS Collaboration}",
    title          = "{The ATLAS Simulation Infrastructure}",
    journal        = "Eur. Phys. J. C",
    volume         = "70",
    year           = "2010",
    pages          = "823",
    doi            = "10.1140/epjc/s10052-010-1429-9",
    eprint         = "1005.4568",
    archivePrefix  = "arXiv",
    primaryClass   = "physics.ins-det",
}

@Article{PERF-2015-09,
    author         = "{ATLAS Collaboration}",
    title          = "{Jet reconstruction and performance using particle flow with the ATLAS Detector}",
    journal        = "Eur. Phys. J. C",
    volume         = "77",
    year           = "2017",
    pages          = "466",
    doi            = "10.1140/epjc/s10052-017-5031-2",
    reportNumber   = "CERN-EP-2017-024",
    eprint         = "1703.10485",
    archivePrefix  = "arXiv",
    primaryClass   = "hep-ex",
}

@Article{TRIG-2016-01,
    author         = "{ATLAS Collaboration}",
    title          = "{Performance of the ATLAS trigger system in 2015}",
    journal        = "Eur. Phys. J. C",
    volume         = "77",
    year           = "2017",
    pages          = "317",
    doi            = "10.1140/epjc/s10052-017-4852-3",
    reportNumber   = "CERN-EP-2016-241",
    eprint         = "1611.09661",
    archivePrefix  = "arXiv",
    primaryClass   = "hep-ex",
}

@Article{HIGG-2017-02,
    author         = "{ATLAS Collaboration}",
    title          = "{Evidence for the associated production of the Higgs boson and a top quark pair with the ATLAS detector}",
    journal        = "Phys. Rev. D",
    volume         = "97",
    year           = "2018",
    pages          = "072003",
    doi            = "10.1103/PhysRevD.97.072003",
    reportNumber   = "CERN-EP-2017-281",
    eprint         = "1712.08891",
    archivePrefix  = "arXiv",
    primaryClass   = "hep-ex",
}

@Article{DAPR-2018-01,
    author         = "{ATLAS Collaboration}",
    title          = "{ATLAS data quality operations and performance for 2015--2018 data-taking}",
    journal        = "JINST",
    volume         = "15",
    year           = "2020",
    pages          = "P04003",
    doi            = "10.1088/1748-0221/15/04/P04003",
    reportNumber   = "CERN-EP-2019-207",
    eprint         = "1911.04632",
    archivePrefix  = "arXiv",
    primaryClass   = "physics.ins-det",
}

@Article{EGAM-2018-01,
    author         = "{ATLAS Collaboration}",
    title          = "{Electron and photon performance measurements with the ATLAS detector using the 2015--2017 LHC proton--proton collision data}",
    journal        = "JINST",
    volume         = "14",
    year           = "2019",
    pages          = "P12006",
    doi            = "10.1088/1748-0221/14/12/P12006",
    reportNumber   = "CERN-EP-2019-145",
    eprint         = "1908.00005",
    archivePrefix  = "arXiv",
    primaryClass   = "hep-ex",
}

@Article{FTAG-2018-01,
    author         = "{ATLAS Collaboration}",
    title          = "{ATLAS \(b\)-jet identification performance and efficiency measurement with \(t\bar{t}\) events in \(pp\) collisions at \(\sqrt{s} = 13\,\text{TeV}\)}",
    journal        = "Eur. Phys. J. C",
    volume         = "79",
    year           = "2019",
    pages          = "970",
    doi            = "10.1140/epjc/s10052-019-7450-8",
    reportNumber   = "CERN-EP-2019-132",
    eprint         = "1907.05120",
    archivePrefix  = "arXiv",
    primaryClass   = "hep-ex",
}

@Article{JETM-2018-05,
    author         = "{ATLAS Collaboration}",
    title          = "{Jet energy scale and resolution measured in proton--proton collisions at \(\sqrt{s} = 13\,\text{TeV}\) with the ATLAS detector}",
    journal        = "Eur. Phys. J. C",
    volume         = "81",
    year           = "2021",
    pages          = "689",
    doi            = "10.1140/epjc/s10052-021-09402-3",
    reportNumber   = "CERN-EP-2020-083",
    eprint         = "2007.02645",
    archivePrefix  = "arXiv",
    primaryClass   = "hep-ex",
}

@Article{MUON-2018-03,
    author         = "{ATLAS Collaboration}",
    title          = "{Muon reconstruction and identification efficiency in ATLAS using the full Run~2 \(pp\) collision data set at \(\sqrt{s} = 13\,\text{TeV}\)}",
    journal        = "Eur. Phys. J. C",
    volume         = "81",
    year           = "2021",
    pages          = "578",
    doi            = "10.1140/epjc/s10052-021-09233-2",
    reportNumber   = "CERN-EP-2020-199",
    eprint         = "2012.00578",
    archivePrefix  = "arXiv",
    primaryClass   = "hep-ex",
}

@Article{PIX-2018-001,
    author         = "Abbott, B. and others",
    title          = "{Production and integration of the ATLAS Insertable B-Layer}",
    journal        = "JINST",
    volume         = "13",
    year           = "2018",
    pages          = "T05008",
    doi            = "10.1088/1748-0221/13/05/T05008",
    eprint         = "1803.00844",
    archivePrefix  = "arXiv",
    primaryClass   = "physics.ins-det",
}

@Article{TOPQ-2018-01,
    author         = "{ATLAS Collaboration}",
    title          = "{Observation of the associated production of a top quark and a \(Z\) boson in \(pp\) collisions at \(\sqrt{s} = 13\,\text{TeV}\) with the ATLAS detector}",
    journal        = "JHEP",
    volume         = "07",
    year           = "2020",
    pages          = "124",
    doi            = "10.1007/JHEP07(2020)124",
    reportNumber   = "CERN-EP-2019-273",
    eprint         = "2002.07546",
    archivePrefix  = "arXiv",
    primaryClass   = "hep-ex",
}

@Article{TRIG-2018-01,
    author         = "{ATLAS Collaboration}",
    title          = "{Performance of the ATLAS muon triggers in Run~2}",
    journal        = "JINST",
    volume         = "15",
    year           = "2020",
    pages          = "P09015",
    doi            = "10.1088/1748-0221/15/09/p09015",
    reportNumber   = "CERN-EP-2020-031",
    eprint         = "2004.13447",
    archivePrefix  = "arXiv",
    primaryClass   = "physics.ins-det",
}

@Article{TRIG-2018-05,
    author         = "{ATLAS Collaboration}",
    title          = "{Performance of electron and photon triggers in ATLAS during LHC Run~2}",
    journal        = "Eur. Phys. J. C",
    volume         = "80",
    year           = "2020",
    pages          = "47",
    doi            = "10.1140/epjc/s10052-019-7500-2",
    reportNumber   = "CERN-EP-2019-169",
    eprint         = "1909.00761",
    archivePrefix  = "arXiv",
    primaryClass   = "hep-ex",
}

@Article{TOPQ-2019-30,
    author         = "{ATLAS Collaboration}",
    title          = "{Measurement of the total and differential cross-sections of \(t\bar{t}W\) production in \(pp\) collisions at \(\sqrt{s}=13\,\text{TeV}\) with the ATLAS detector}",
    journal        = "JHEP",
    volume         = "05",
    year           = "2024",
    pages          = "131",
    doi            = "10.1007/JHEP05(2024)131",
    reportNumber   = "CERN-EP-2023-260",
    eprint         = "2401.05299",
    archivePrefix  = "arXiv",
    primaryClass   = "hep-ex",
}

@Article{TRIG-2019-03,
    author         = "{ATLAS Collaboration}",
    title          = "{The ATLAS inner detector trigger performance in \(pp\) collisions at \(13\,\text{TeV}\) during LHC Run~2}",
    journal        = "Eur. Phys. J. C",
    volume         = "82",
    year           = "2022",
    pages          = "206",
    doi            = "10.1140/epjc/s10052-021-09920-0",
    reportNumber   = "CERN-EP-2021-076",
    eprint         = "2107.02485",
    archivePrefix  = "arXiv",
    primaryClass   = "hep-ex",
}

@Article{JETM-2020-03,
    author         = "{ATLAS Collaboration}",
    title          = "{The performance of missing transverse momentum reconstruction and its significance with the ATLAS detector using \(140\,\text{fb}^{-1}\) of \(\sqrt{s} = 13\,\text{TeV}\) \(pp\) collisions}",
    journal        = "Eur. Phys. J. C",
    volume         = "85",
    year           = "2025",
    pages          = "606",
    doi            = "10.1140/epjc/s10052-025-14062-8",
    reportNumber   = "CERN-EP-2024-023",
    eprint         = "2402.05858",
    archivePrefix  = "arXiv",
    primaryClass   = "hep-ex",
}

@Article{TOPQ-2020-20,
    author         = "{ATLAS Collaboration}",
    title          = "{Inclusive and differential cross-section measurements of \(t\bar{t}Z\) production in \(pp\) collisions at \(\sqrt{s} = 13\,\text{TeV}\) with the ATLAS detector, including EFT and spin-correlation interpretations}",
    journal        = "JHEP",
    volume         = "07",
    year           = "2024",
    pages          = "163",
    doi            = "10.1007/JHEP07(2024)163",
    reportNumber   = "CERN-EP-2023-252",
    eprint         = "2312.04450",
    archivePrefix  = "arXiv",
    primaryClass   = "hep-ex",
}

@Article{DAPR-2021-01,
    author         = "{ATLAS Collaboration}",
    title          = "{Luminosity determination in \(pp\) collisions at \(\sqrt{s} = 13\,\text{TeV}\) using the ATLAS detector at the LHC}",
    journal        = "Eur. Phys. J. C",
    volume         = "83",
    year           = "2023",
    pages          = "982",
    doi            = "10.1140/epjc/s10052-023-11747-w",
    reportNumber   = "CERN-EP-2022-281",
    eprint         = "2212.09379",
    archivePrefix  = "arXiv",
    primaryClass   = "hep-ex",
}

@Article{EGAM-2021-01,
    author         = "{ATLAS Collaboration}",
    title          = "{Electron and photon efficiencies in LHC Run~2 with the ATLAS experiment}",
    journal        = "JHEP",
    volume         = "05",
    year           = "2024",
    pages          = "162",
    doi            = "10.1007/JHEP05(2024)162",
    reportNumber   = "CERN-EP-2023-182",
    eprint         = "2308.13362",
    archivePrefix  = "arXiv",
    primaryClass   = "hep-ex",
}

@Article{EGAM-2021-02,
    author         = "{ATLAS Collaboration}",
    title          = "{Electron and photon energy calibration with the ATLAS detector using LHC Run~2 data}",
    journal        = "JINST",
    volume         = "19",
    year           = "2024",
    pages          = "P02009",
    doi            = "10.1088/1748-0221/19/02/P02009",
    reportNumber   = "CERN-EP-2023-128",
    eprint         = "2309.05471",
    archivePrefix  = "arXiv",
    primaryClass   = "hep-ex",
}

@Article{TOPQ-2021-08,
    author         = "{ATLAS Collaboration}",
    title          = "{Observation of four-top-quark production in the multilepton final state with the ATLAS detector}",
    journal        = "Eur. Phys. J. C",
    volume         = "83",
    year           = "2023",
    pages          = "496",
    doi            = "10.1140/epjc/s10052-023-11573-0",
    reportNumber   = "CERN-EP-2023-055",
    eprint         = "2303.15061",
    archivePrefix  = "arXiv",
    primaryClass   = "hep-ex",
    related        = "TOPQ-2021-08-err",
    relatedstring  = "Erratum:",
}

@Article{MUON-2022-01,
    author         = "{ATLAS Collaboration}",
    title          = "{Studies of the muon momentum calibration and performance of the ATLAS detector with \(pp\) collisions at \(\sqrt{s} = 13\,\text{TeV}\)}",
    journal        = "Eur. Phys. J. C",
    volume         = "83",
    year           = "2023",
    pages          = "686",
    doi            = "10.1140/epjc/s10052-023-11584-x",
    reportNumber   = "CERN-EP-2022-274",
    eprint         = "2212.07338",
    archivePrefix  = "arXiv",
    primaryClass   = "hep-ex",
}

@Article{SOFT-2022-02,
    author         = "{ATLAS Collaboration}",
    title          = "{Software and computing for Run~3 of the ATLAS experiment at the LHC}",
    journal        = "Eur. Phys. J. C",
    volume         = "85",
    year           = "2025",
    pages          = "234",
    doi            = "10.1140/epjc/s10052-024-13701-w",
    reportNumber   = "CERN-EP-2024-100",
    eprint         = "2404.06335",
    archivePrefix  = "arXiv",
    primaryClass   = "hep-ex",
}

@Article{TOPQ-2023-02,
    author         = "{ATLAS Collaboration}",
    title          = "{Measurement of high-mass \(t\bar{t} \ell^{+}\ell^{-}\) production and lepton flavour universality-inspired effective field theory interpretations at \(\sqrt{s}=13\,\text{TeV}\) with the ATLAS detector}",
    year           = "2025",
    reportNumber   = "CERN-EP-2025-053",
    eprint         = "2504.05919",
    archivePrefix  = "arXiv",
    primaryClass   = "hep-ex",
}

@Booklet{ATL-SOFT-PUB-2025-001,
    author         = "{ATLAS Collaboration}",
    title          = "{ATLAS Computing Acknowledgements}",
    howpublished   = "{ATL-SOFT-PUB-2025-001}",
    url            = "https://cds.cern.ch/record/2922210",
    year           = "2025",
}

@Report{ATLAS-TDR-19,
    author         = "{ATLAS Collaboration}",
    title          = "{ATLAS Insertable B-Layer: Technical Design Report}",
    type           = "ATLAS-TDR-19; CERN-LHCC-2010-013",
    year           = "2010",
    url            = "https://cds.cern.ch/record/1291633",
    related        = "ATLAS-TDR-19-addm",
    relatedstring  = "Addendum:",
}

@Article{CMS-TOP-21-001,
    author         = "{CMS Collaboration}",
    title          = "{Probing effective field theory operators in the associated production of top quarks with a \(Z\) boson in multilepton final states at \(\sqrt{s} = 13\,\text{TeV}\)}",
    journal        = "JHEP",
    volume         = "12",
    year           = "2021",
    pages          = "083",
    doi            = "10.1007/JHEP12(2021)083",
    eprint         = "2107.13896",
    archivePrefix  = "arXiv",
    primaryClass   = "hep-ex",
}

@Article{CMS-TOP-23-004,
    author         = "{CMS Collaboration}",
    title          = "{Measurements of inclusive and differential cross sections for top quark production in association with a \(Z\) boson in proton--proton collisions at \(\sqrt{s} = 13\,\text{TeV}\)}",
    journal        = "JHEP",
    volume         = "02",
    year           = "2025",
    pages          = "177",
    doi            = "10.1007/JHEP02(2025)177",
    reportNumber   = "CERN-EP-2024-251",
    eprint         = "2410.23475",
    archivePrefix  = "arXiv",
    primaryClass   = "hep-ex",
}

@Booklet{ATLAS-CONF-2014-018,
    author         = "{ATLAS Collaboration}",
    title          = "{Tagging and suppression of pileup jets with the ATLAS detector}",
    howpublished   = "{ATLAS-CONF-2014-018}",
    url            = "https://cds.cern.ch/record/1700870",
    year           = "2014",
}

@Booklet{ATL-PHYS-PUB-2011-014,
    author         = "{ATLAS Collaboration}",
    title          = "{Further ATLAS tunes of \textsc{Pythia}~6 and Pythia~8}",
    howpublished   = "{ATL-PHYS-PUB-2011-014}",
    url            = "https://cds.cern.ch/record/1400677",
    year           = "2011",
}

@Booklet{ATL-PHYS-PUB-2014-021,
    author         = "{ATLAS Collaboration}",
    title          = "{ATLAS Pythia~8 tunes to \(7~\text{TeV}\) data}",
    howpublished   = "{ATL-PHYS-PUB-2014-021}",
    url            = "https://cds.cern.ch/record/1966419",
    year           = "2014",
}

@Booklet{ATL-PHYS-PUB-2015-036,
    author         = "{ATLAS Collaboration}",
    title          = "{Properties of jets and inputs to jet reconstruction and calibration with the ATLAS detector using proton--proton collisions at \(\sqrt{s} = 13~\text{TeV}\)}",
    howpublished   = "{ATL-PHYS-PUB-2015-036}",
    url            = "https://cds.cern.ch/record/2044564",
    year           = "2015",
}

@Booklet{ATL-PHYS-PUB-2016-005,
    author         = "{ATLAS Collaboration}",
    title          = "{Modelling of the \(t\bar{t} H\) and \(t\bar{t}V (V = W, Z)\) processes for \(\sqrt{s} = 13~\text{TeV}\) ATLAS analyses}",
    howpublished   = "{ATL-PHYS-PUB-2016-005}",
    url            = "https://cds.cern.ch/record/2120826",
    year           = "2016",
}

@Booklet{ATL-SOFT-PUB-2018-002,
    author         = "{ATLAS Collaboration}",
    title          = "{The new Fast Calorimeter Simulation in ATLAS}",
    howpublished   = "{ATL-SOFT-PUB-2018-002}",
    url            = "https://cds.cern.ch/record/2630434",
    year           = "2018",
}
\clearpage
 
\begin{flushleft}
\hypersetup{urlcolor=black}
{\Large The ATLAS Collaboration}

\bigskip

\AtlasOrcid[0000-0002-6665-4934]{G.~Aad}$^\textrm{\scriptsize 104}$,
\AtlasOrcid[0000-0001-7616-1554]{E.~Aakvaag}$^\textrm{\scriptsize 17}$,
\AtlasOrcid[0000-0002-5888-2734]{B.~Abbott}$^\textrm{\scriptsize 123}$,
\AtlasOrcid[0000-0002-0287-5869]{S.~Abdelhameed}$^\textrm{\scriptsize 119a}$,
\AtlasOrcid[0000-0002-1002-1652]{K.~Abeling}$^\textrm{\scriptsize 55}$,
\AtlasOrcid[0000-0001-5763-2760]{N.J.~Abicht}$^\textrm{\scriptsize 49}$,
\AtlasOrcid[0000-0002-8496-9294]{S.H.~Abidi}$^\textrm{\scriptsize 30}$,
\AtlasOrcid[0009-0003-6578-220X]{M.~Aboelela}$^\textrm{\scriptsize 45}$,
\AtlasOrcid[0000-0002-9987-2292]{A.~Aboulhorma}$^\textrm{\scriptsize 36e}$,
\AtlasOrcid[0000-0001-5329-6640]{H.~Abramowicz}$^\textrm{\scriptsize 157}$
% ...
\bigskip
\\

% ...
$^{17}$Department for Physics and Technology, University of Bergen, Bergen; Norway.\\
% ...
$^{30}$Physics Department, Brookhaven National Laboratory, Upton NY; United States of America.\\
% ...
$^{36}$$^{(a)}$Facult\'e des Sciences Ain Chock, Universit\'e Hassan II de Casablanca;$^{(b)}$Facult\'{e} des Sciences, Universit\'{e} Ibn-Tofail, K\'{e}nitra;$^{(c)}$Facult\'e des Sciences Semlalia, Universit\'e Cadi Ayyad, LPHEA-Marrakech;$^{(d)}$LPMR, Facult\'e des Sciences, Universit\'e Mohamed Premier, Oujda;$^{(e)}$Facult\'e des sciences, Universit\'e Mohammed V, Rabat;$^{(f)}$Institute of Applied Physics, Mohammed VI Polytechnic University, Ben Guerir; Morocco.\\
% ...
$^{45}$Physics Department, Southern Methodist University, Dallas TX; United States of America.\\
% ...
$^{49}$Fakult\"{a}t Physik , Technische Universit{\"a}t Dortmund, Dortmund; Germany.\\
% ...
$^{55}$II. Physikalisches Institut, Georg-August-Universit\"{a}t G\"ottingen, G\"ottingen; Germany.\\
% ...
$^{104}$CPPM, Aix-Marseille Universit\'e, CNRS/IN2P3, Marseille; France.\\
% ...
$^{119}$$^{(a)}$New York University Abu Dhabi, Abu Dhabi;$^{(b)}$United Arab Emirates University, Al Ain; United Arab Emirates.\\
% ...
$^{123}$Homer L. Dodge Department of Physics and Astronomy, University of Oklahoma, Norman OK; United States of America.\\
% ...
$^{157}$Raymond and Beverly Sackler School of Physics and Astronomy, Tel Aviv University, Tel Aviv; Israel.\\
% ...
$^{a}$ Also at Affiliated with an institute formerly covered by a cooperation agreement with CERN.\\
$^{b}$ Also at An-Najah National University, Nablus; Palestine.\\
$^{c}$ Also at Borough of Manhattan Community College, City University of New York, New York NY; United States of America.\\
$^{d}$ Also at Center for Interdisciplinary Research and Innovation (CIRI-AUTH), Thessaloniki; Greece.\\
$^{e}$ Also at Centre of Physics of the Universities of Minho and Porto (CF-UM-UP); Portugal.\\
$^{f}$ Also at CERN, Geneva; Switzerland.\\
% ...

\end{flushleft}


%

%
%

%

%
%
%
%

\end{document}